\documentclass[12pt,nofootinbib]{revtex4}
\usepackage{mathrsfs}
\usepackage{verbatim}
\usepackage{amssymb}
\usepackage{graphicx}
\usepackage{subfigure}
\usepackage{color}
\usepackage{rotating}

\newcommand{\txtD}[2]{d #2 / d #1}
\newcommand{\txtpd}[2]{\partial #2 / \partial #1}

\newcommand{\vctr}[1]{\mathbf{#1}}

\newcommand{\D}[2]{\frac{d #2}{d #1}}
\newcommand{\pd}[2]{\frac{\partial #2}{\partial #1}}

\newcommand{\half}[0]{\frac{1}{2}}

\begin{document}


\title{A Periodic Table for Black Hole Orbits}


\author{Janna Levin${}^{*,!}$ and Gabe Perez-Giz${}^{**}$}
\affiliation{${}^{*}$Department of Physics and Astronomy, Barnard
College of Columbia University, 3009 Broadway, New York, NY 10027 }
\affiliation{${}^{!}$Institute for Strings, Cosmology and Astroparticle
  Physics, Columbia University, New York, NY 10027}
\affiliation{${}^{**}$Physics Department, Columbia University,
New York, NY 10027}

\widetext

\begin{abstract}
Understanding the dynamics around rotating black holes is imperative
to the success of the future gravitational wave
observatories. Although integrable in principle, test particle orbits
in the Kerr spacetime can also be elaborate, and while they have been
studied extensively, classifying their general properties has been a
challenge.  This is the first in a series of papers that adopts a
dynamical systems approach to the study of Kerr orbits, beginning with
equatorial orbits. 
We define a taxonomy of orbits that hinges on a
correspondence between periodic orbits and rational numbers.
The taxonomy defines the entire dynamics, 
including aperiodic motion, since every orbit is in or near the periodic set.
A remarkable implication of this periodic
orbit taxonomy is that the simple precessing ellipse familiar from
planetary orbits is not allowed in the strong-field regime. Instead,
eccentric orbits trace out precessions of multi-leaf clovers in the
final stages of inspiral.  Furthermore, for any black hole, there is
some point in the strong-field regime past which zoom-whirl behavior
becomes unavoidable.  Finally, we sketch the potential application of
the taxonomy to problems of astrophysical interest, in particular its
utility for computationally intensive gravitational wave calculations.
\end{abstract}

\maketitle

\vfill\eject
\section{Introduction}

When Einstein realized his General Theory of Relativity correctly
yielded the precession of the perihelion of Mercury, he had heart
palpitations, later writing to Ehrenfest \cite{folsing}, ``For some
days I was beyond myself with excitement.'' The anomalous precession
of the perihelion of Mercury was the only astronomical observation in
conflict with Newtonian gravity.  Even taking account of perturbations
from other planets, astronomers still saw Mercury's perihelion
overshoot its Keplerian target by an extra $43^{''}$/century, exactly
the amount predicted by General Relativity.

Just shy of a century later, we are on the brink of another important
test of General Relativity: the direct detection of gravitational
waves. Black hole binaries may be the most viable candidates for a
first direct detection.  Consequently, and interestingly, the first
gravitational waves observed will likely also probe relativistic
orbits, which display exotic behaviors beyond even Mercury's simple
precessing ellipse.  Relativistic orbital trajectories need not lie in
a plane.  They can also exhibit so-called ``zoom-whirl'' behavior
-- an extreme form of perihelion precession --
whirling around the central black hole
before zooming out quasi-elliptically \cite{{barackcutler},{gk}}.
Nevertheless, even though 
orbits of a non-spinning test particle around a spinning
black hole have been studied since Carter proved their integrability
in the 1970's \cite{carter}, a language for making simple, general
claims about their properties has been elusive.

In this paper, we introduce a powerful taxonomy that defines the full
range of orbital dynamics in the equatorial plane of a Kerr black hole.
We use this scheme to illustrate behaviors that run counter even to our
relativistic intuitions. In the strong-field
regime, precessing elliptical orbits such as Mercury's are excluded.
Instead, at close separations, eccentric orbits trace out precessions
of patterns best described as multi-leaf clovers. We can also
demarcate a region in orbital parameter space where the aforementioned
zoom-whirl behavior is not merely prevalent but unavoidable, the size
of that region increasing as the spin of the central black hole
increases.

\centerline{\bf Not all orbits are created equal}

Our taxonomy emphasizes a dynamically special set of orbits -- the
periodic orbits that return exactly to their initial conditions
after a finite time.  First widely touted by Poincar\'e \cite{poinc},
who suggested that the general behavior of any classical system could
be gleaned from a study of repeating motions, periodic orbits have
played a crucial role in the treatment of some difficult problems in
celestial mechanics, including the motions of planetary satellites,
the long term stability of the solar system, and motion in galactic
potentials.  
In contrast, periodic orbits in relativistic
astrophysical systems like compact object binaries have gone largely
unexamined\footnote{While some authors
\cite{{chandra1},{cont1},{cont2},{cont3},{moeckel},{cornishfrankel},{bc}}
have explored inherently
relativistic periodic motion, the systems studied have been more
mathematically informative than astrophysically descriptive.},
typically with the disclaimer that because they have measure zero in
the space of all possible orbits, these closed orbits merit
correspondingly little attention.  Particularly in the case of test
particle motion in the Schwarzschild and Kerr spacetimes -- the only
analytically soluble relativistic orbital systems -- mapping out the
properties of a particular measure zero set has seemed unnecessary.

By contrast, we will side with Poincar\'e. To paraphrase Orwell, while
all measure zero sets are equal, some are more equal than others.  For
instance, circular orbits receive special attention, even
though they are also a measure zero set, because they have two special
dynamical features:
\begin{enumerate}
  \item circular orbits are easy to handle, and 
  \item  some orbits look like small
perturbations to circular ones.
\end{enumerate}
As a result, an analysis of circular orbits reveals fairly detailed
information about nearby low-eccentricity orbits with relatively low
overhead.  Most relativity texts derive the famed precession of the
perihelion of Mercury in precisely this way \cite{wald}. Of course,
since most orbits are not close to circular ones, circular
orbits alone do not encode the entirety of black hole orbital
dynamics.

\begin{figure}
  \vspace{-20pt}
  \centering
\hspace{-30pt} 
 \includegraphics[width=0.45\textwidth]{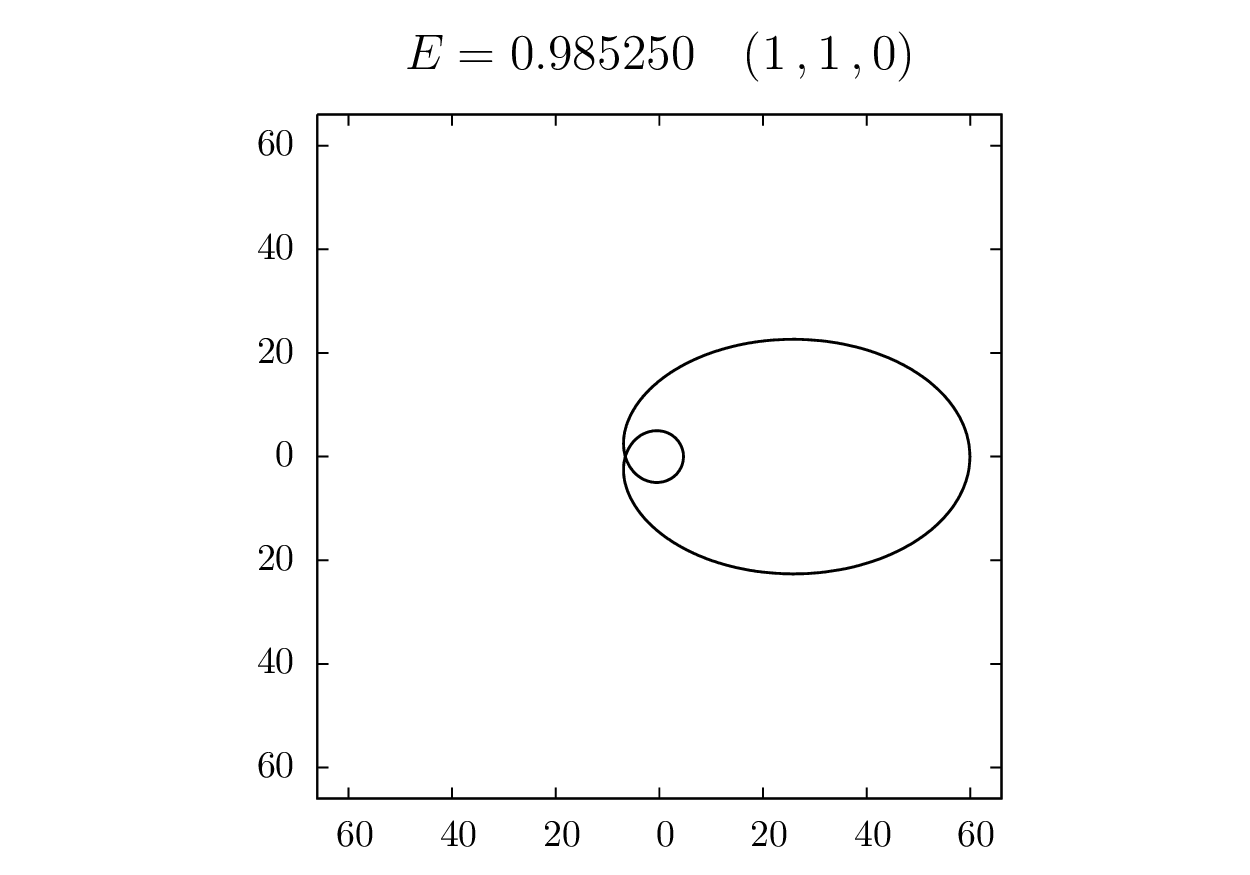}
\hspace{-80pt}
  \includegraphics[width=0.45\textwidth]{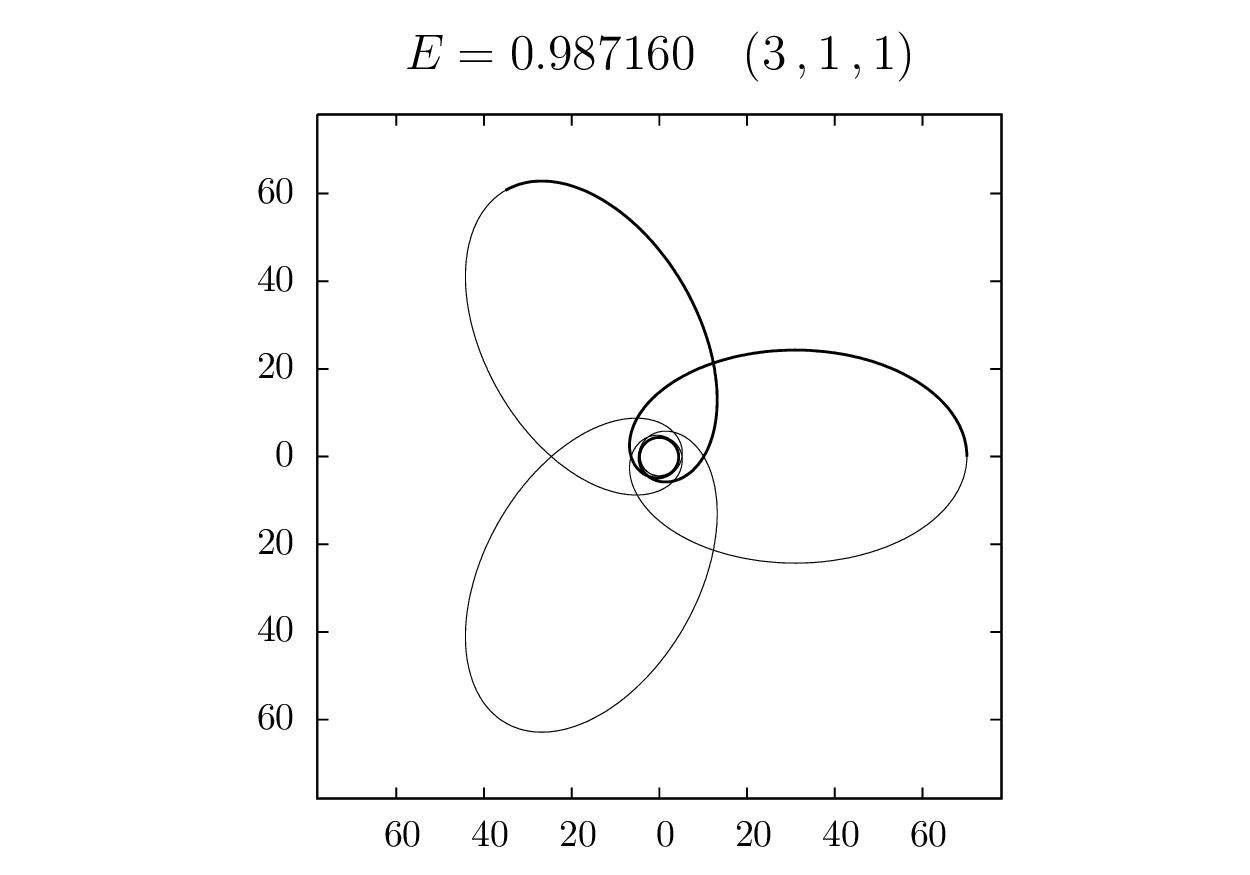} 
\hspace{-80pt}
  \includegraphics[width=0.45\textwidth]{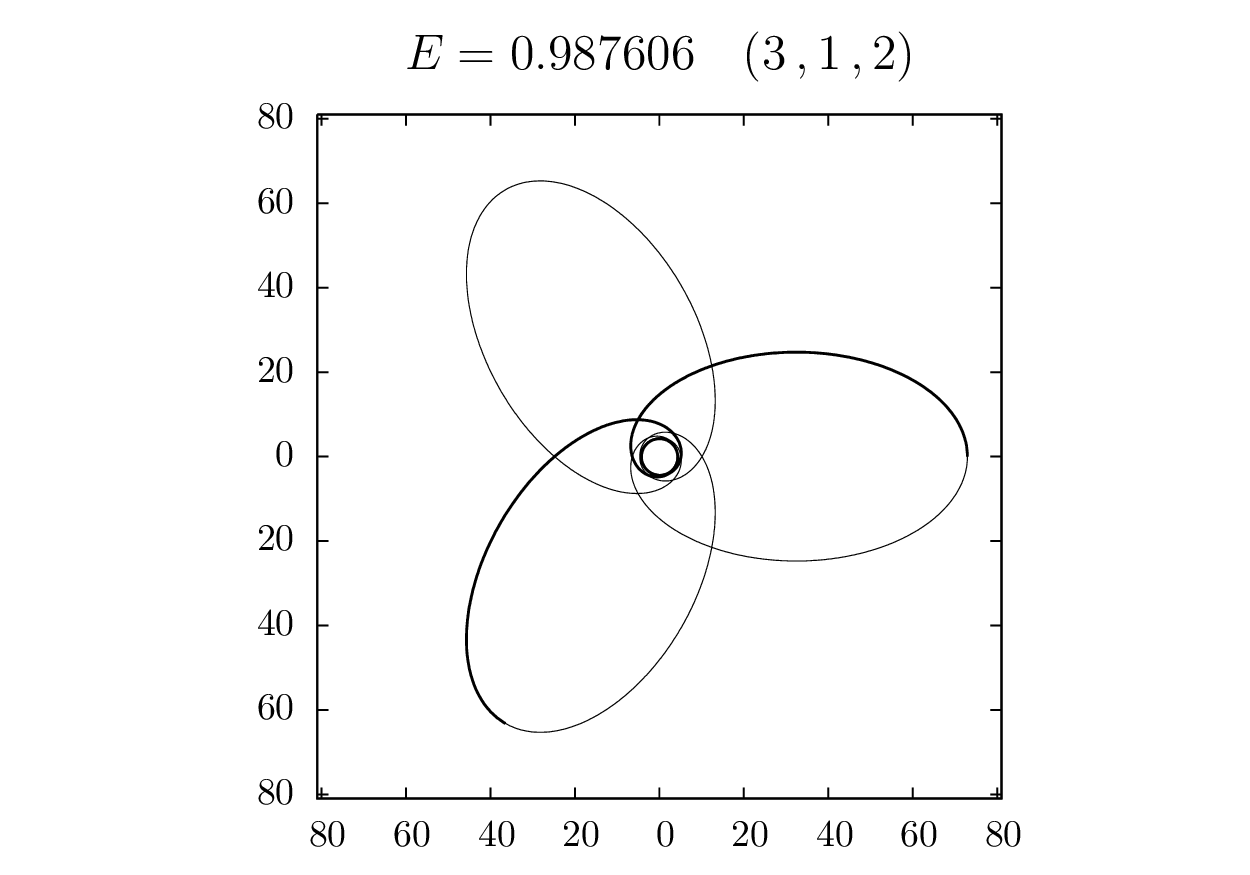}
\hfill
\\
\hspace{-26pt}
  \includegraphics[width=0.45\textwidth]{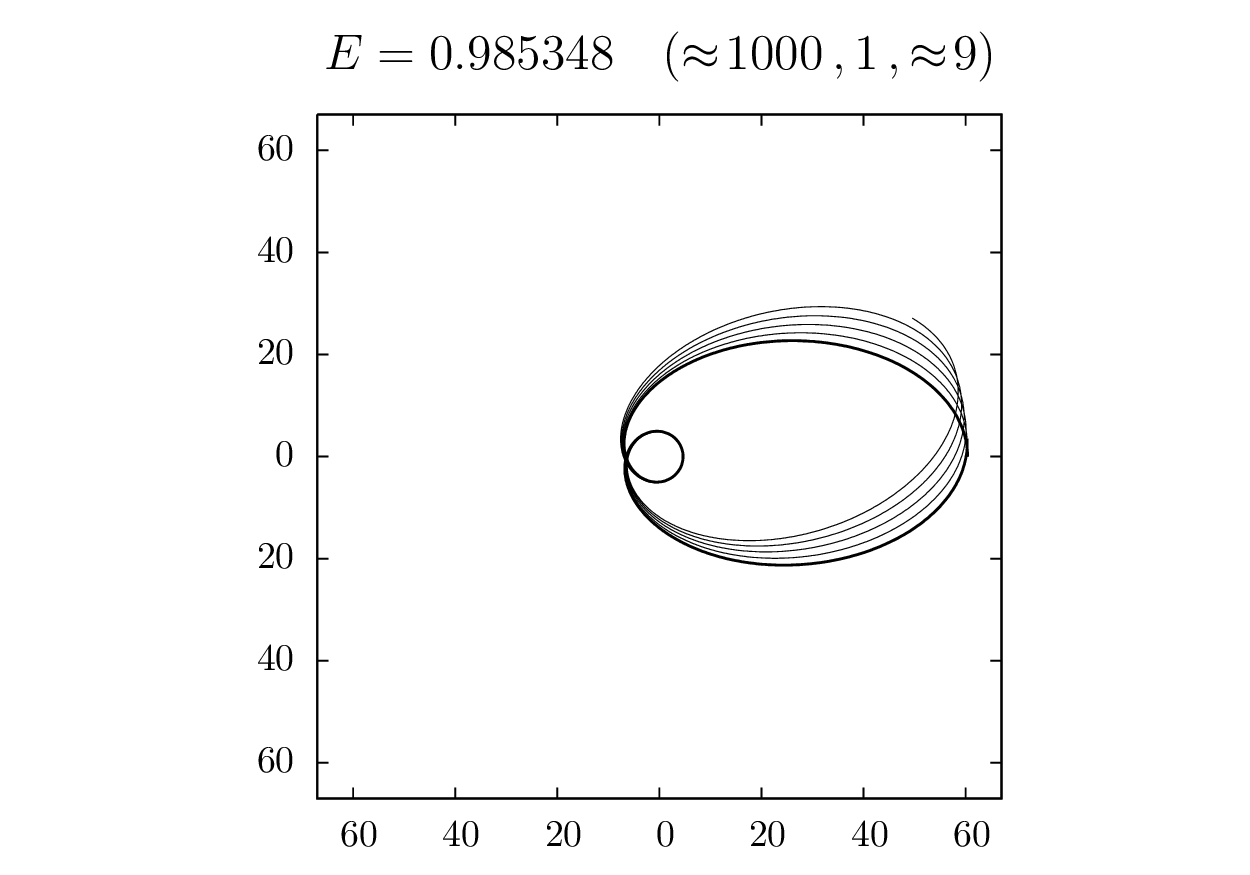}
\hspace{-80pt}
  \includegraphics[width=0.45\textwidth]{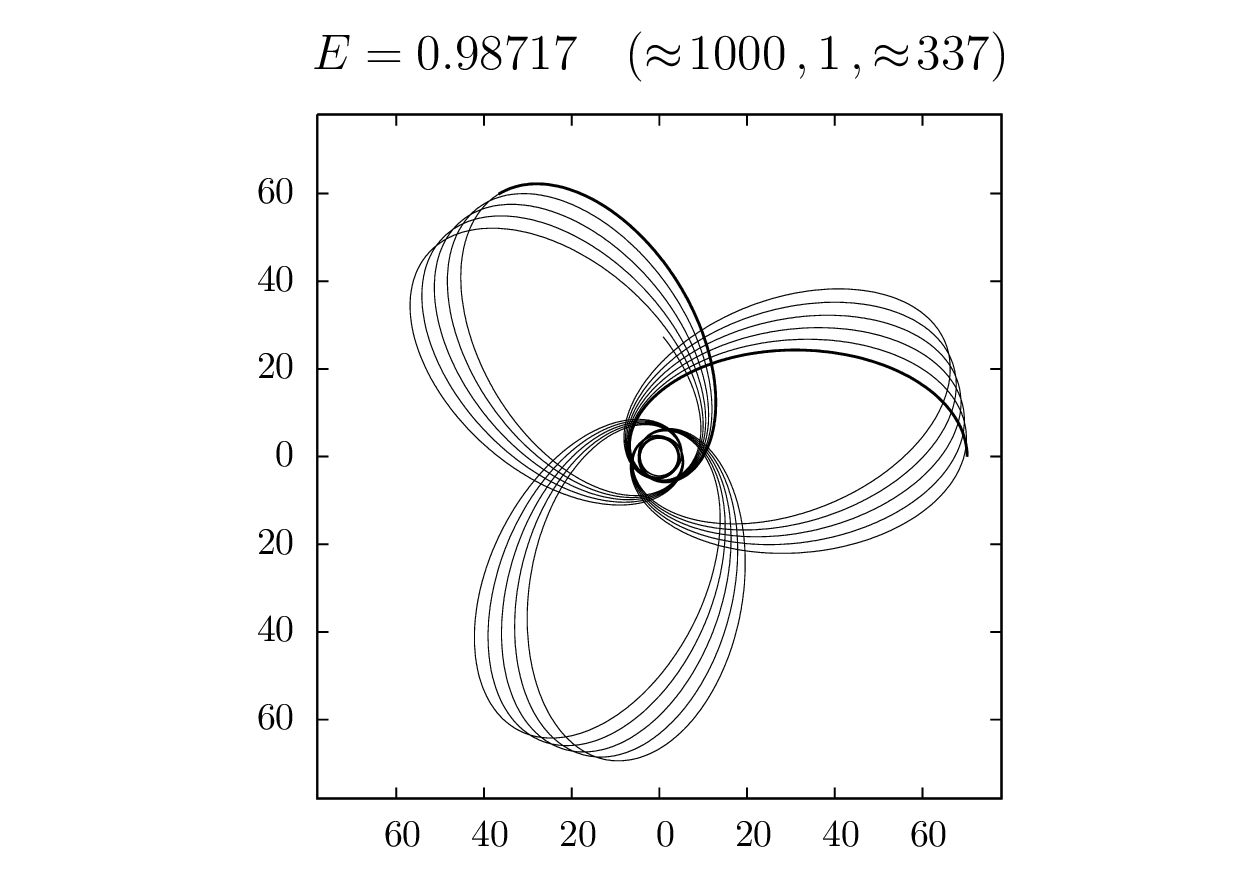}
\hspace{-80pt}
  \includegraphics[width=0.45\textwidth]{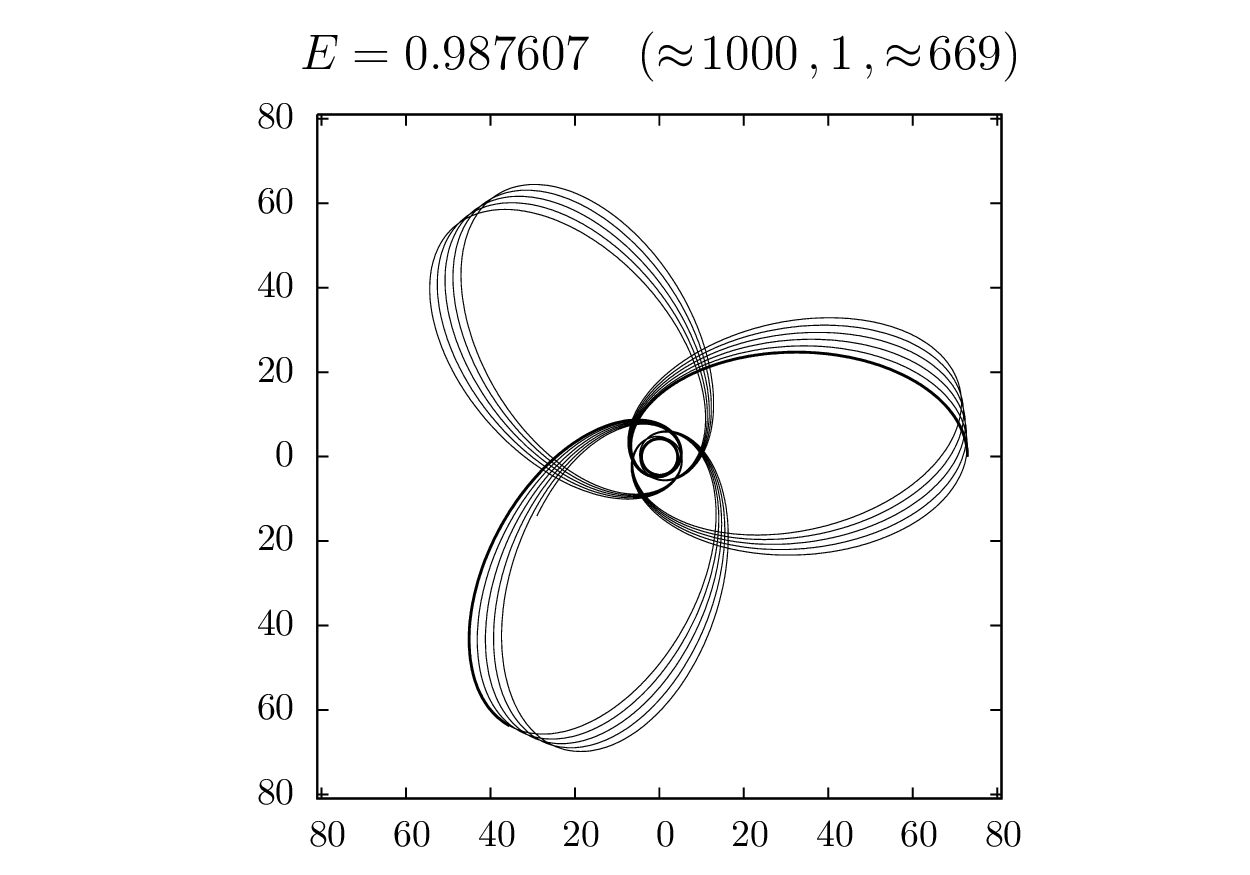}
\hfill
\\
\hspace{-24pt}
  \includegraphics[width=0.45\textwidth]{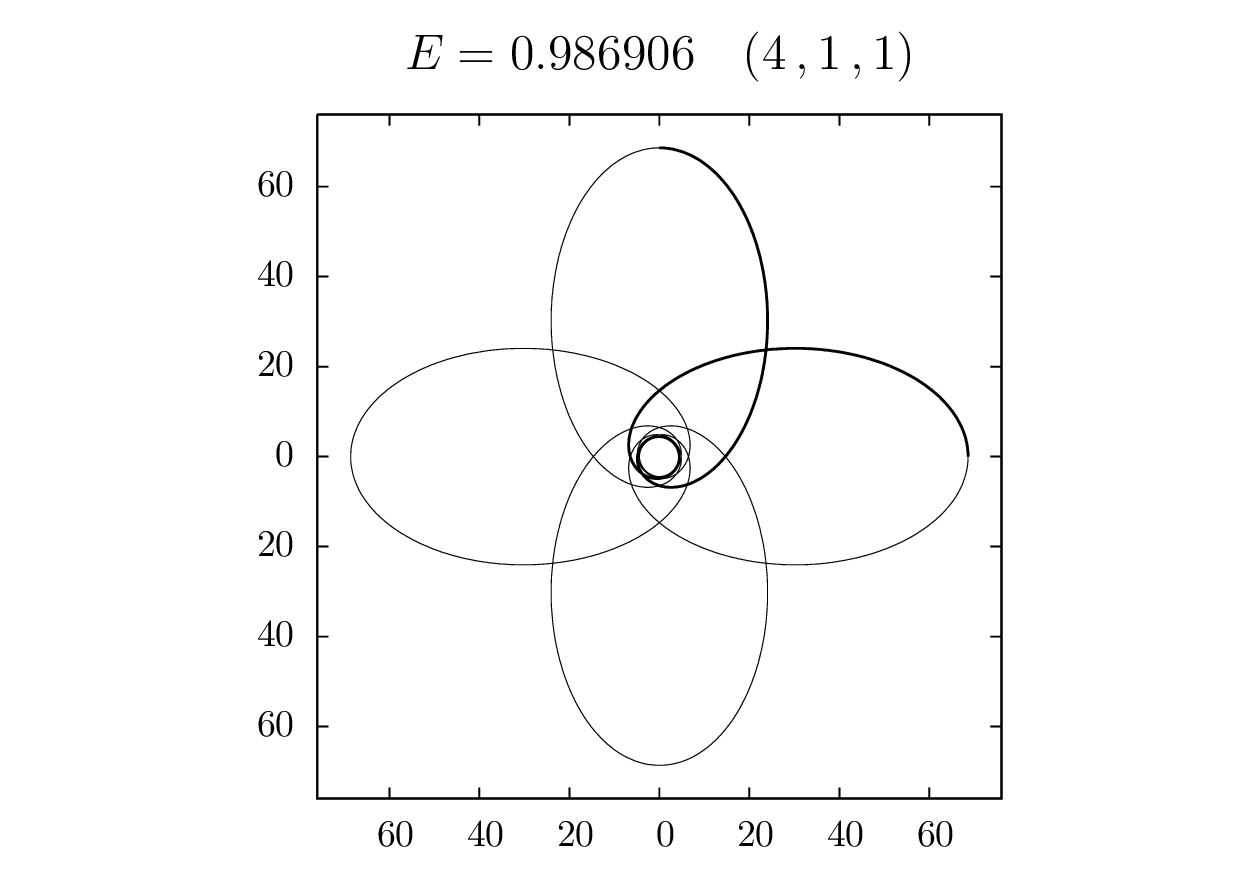}
\hspace{-80pt}
  \includegraphics[width=0.45\textwidth]{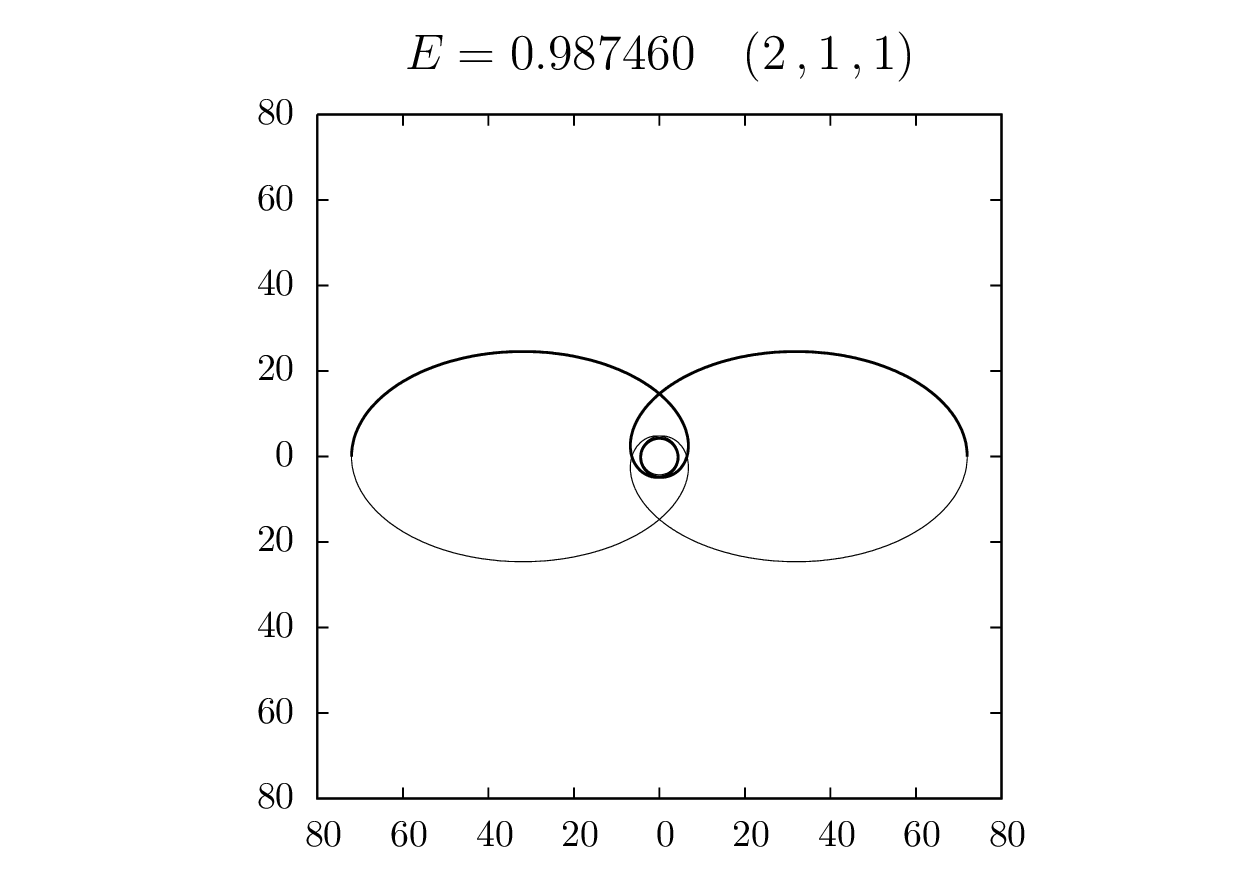}
\hspace{-80pt}
  \includegraphics[width=0.45\textwidth]{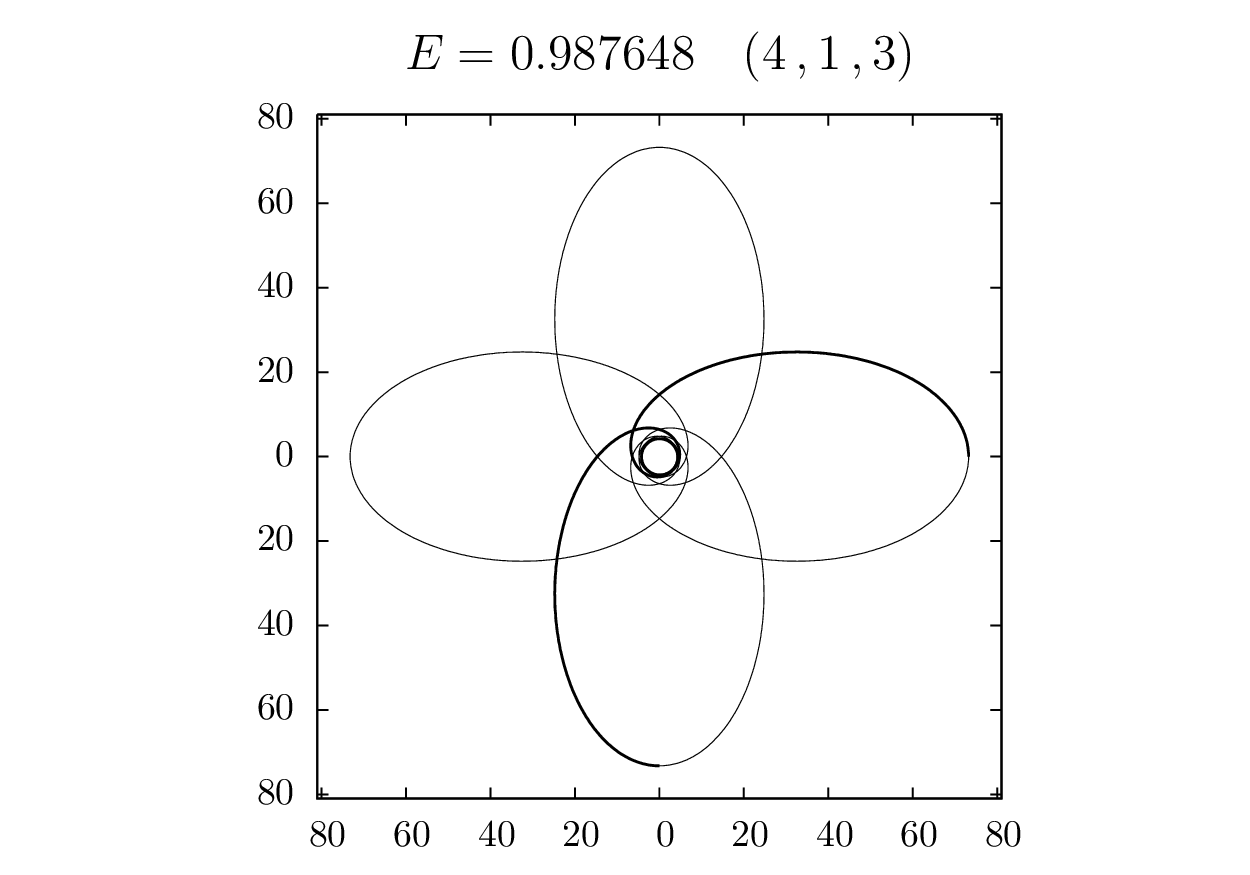}
\hfill
\\
\hspace{-22pt}
  \includegraphics[width=0.45\textwidth]{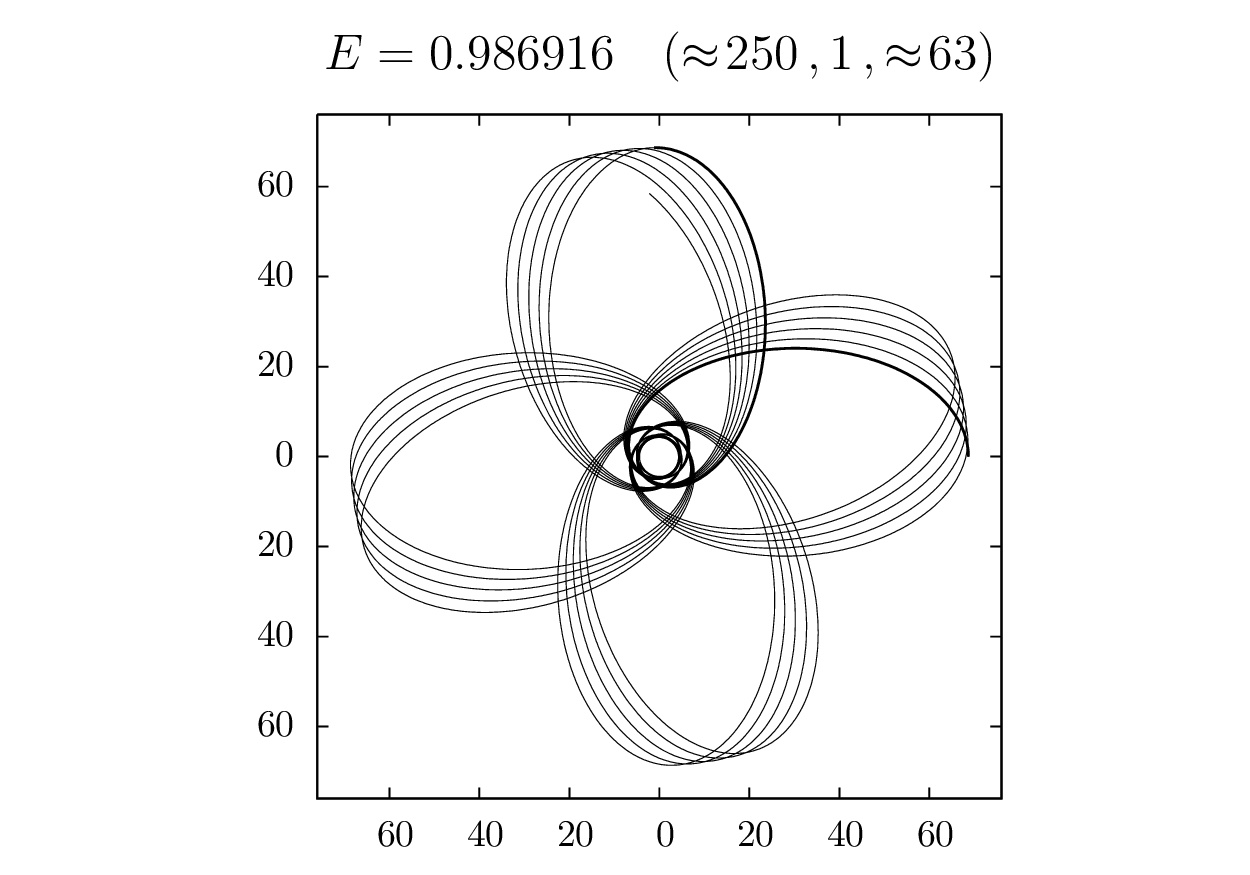}
\hspace{-80pt}
  \includegraphics[width=0.45\textwidth]{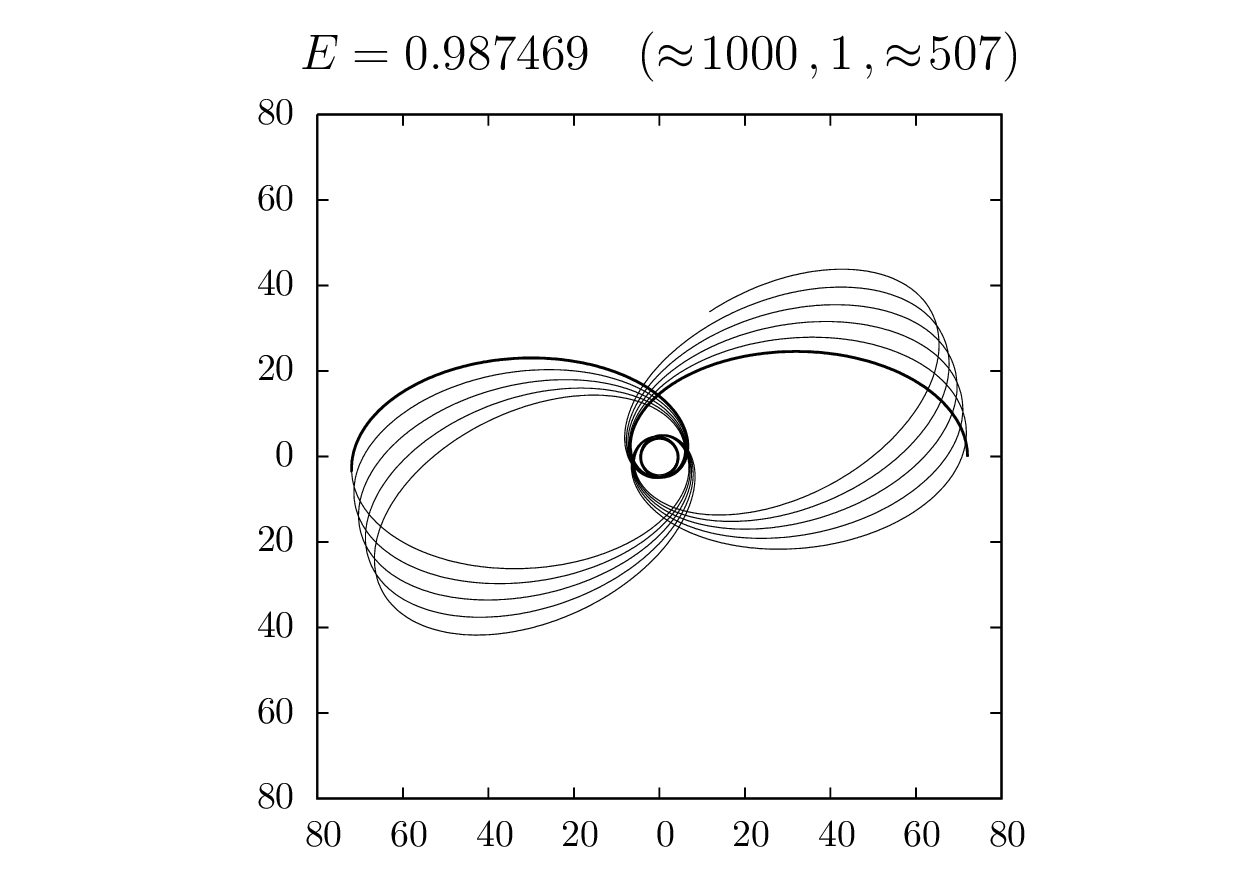}
\hspace{-80pt}
  \includegraphics[width=0.45\textwidth]{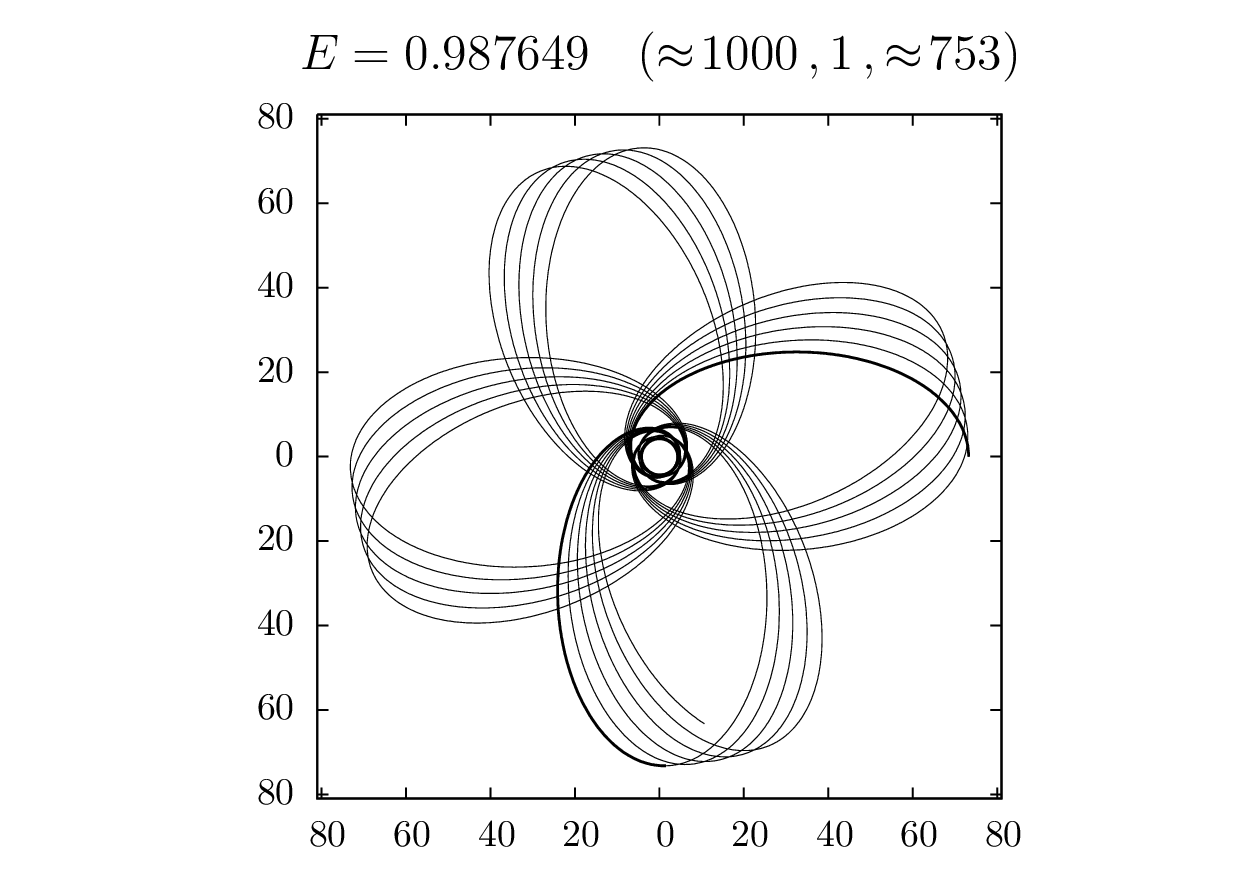}
\hfill
  \caption{This figure is a preview of figure \ref{twelveS}. Rows 1 and
  3 show exactly periodic orbits. Rows 2 and 4 show nearby aperiodic
  orbits.}
  \label{skeleton}
\end{figure}

Remarkably, periodic orbits do.
In 
fact they have even greater dynamical power than circular orbits because
\begin{enumerate}
  \item periodic orbits are also easy to handle, and
  \item \emph{all} generic orbits look like small
  perturbations to periodic ones.
\end{enumerate}
The latter fact, first noted by Poincar\'e, stems from a beautiful
correspondence between periodic orbits and the rational numbers.  The
density of rationals on the number line thus implies a
corresponding density of periodic orbits in the space of all possible
orbits, so that any generic orbit can be viewed as an arbitarily small
deviation from some exactly periodic counterpart.

The result is a highly geometric skeleton of periodic orbits in terms
of which the properties of even generic, aperiodic orbits can be
described.  Figure\ \ref{skeleton} offers a preview of the anatomy of
this skeleton.  Ignoring for the moment the details of these orbits,
all of which are explained in the body of the paper, we can still see
at a glance the two special dynamical properties discussed above.
Rows 1 and 3 show a set of exactly periodic orbits.  Besides being
visually elegant, those correspond to a 
rational number according to a scheme detailed in \S \ref{tax}.  Just below
each periodic orbit is a generic aperiodic orbit.  Notice that each
aperiodic orbit looks like a slow precession not of an ellipse but of
the periodic orbit immediately above.  Additionally, each such orbit
can be 
assigned an approximating rational.  
What's more, we can describe all black hole orbital dynamics by such a
periodic table. 

This paper is the first in a series that realizes Poincar\'e's dictum
for equatorial orbits in the Kerr spacetime\footnote{For clarity of
exposition, we relegate the dynamically more involved nonequatorial
Kerr orbits to another work.}.  After defining and filling
out this periodic skeleton for equatorial orbits (\S \ref{tax}), we extract its
dynamical consequences (\S \ref{dynamics}), and explore its potential
applications to a 
host of astrophysical problems (\S \ref{util}). In particular,
we
outline how periodic orbits might facilitate the execution
of the computationally intensive task of calculating the gravitational
waveforms from extreme mass ratio inspirals
\cite{{barackcutler},{gk},{hopman},{hughesconf2},{narayan},{dh},{deh}}.

Although important and interesting in its own right, for simplicity
of presentation we
relegate to an appendix the Hamiltonian formalism we use to identify
periodic orbits 
in the Kerr system. For the body of the paper, the
reader need only know that $a$ is the spin parameter for each central
black hole and that every orbit is specified by the energy $E$ and
angular momentum $L$ of a test-particle as measured by an observer at
infinity. 
As detailed in appendix \ref{kerreqs}, we work in units in which $a$,
$E$, $L$, and all coordinates and frequencies are dimensionless.

\section{Taxonomy of periodic orbits}
\label{tax}

The goal of this section is to detail a system for indexing all closed
orbits around a black hole with a triplet of integers $(z,w,v)$.  The
reason behind the choice of symbols will become clear shortly.  The
scheme is topological, and our approach is to establish the connection
between a given periodic orbit and its $(z, w, v)$ label visually.  We
then establish the relationship between a given periodic orbit and a
specific rational number in two ways: first based on topological
features of the orbit, and then based on frequencies associated with
its radial and azimuthal motions.

Finally, there is the matter of how we know that any particular
periodic orbit we reference even exists. Answering this rather
important question is the central objective of the dynamical section
of the paper, \S \ref{dynamics}.

\subsection{The essential taxonomy}
\label{essential}

We have a simple topological method for identifying all of the closed
orbits  
for a given angular momentum $L$ around a given black hole (fixed by
the spin $a$). Each
periodic orbit traces out a finite number of leaves before
closing. We will call the number of leaves $z$ for
zoom.\footnote{
Consider again the periodic orbits in rows 1 and 3 of figure
\ref{skeleton}.  An immediately recognizable integer associated with
each such orbit is the number of leaves it traces out before returning
to its starting point. Each such leaf corresponds to the
quasi-elliptical ``zooming'' behavior mentioned in the introduction.}

\begin{figure}
  \vspace{-20pt}
\hspace{-400pt}
  \centering
\hfill
\includegraphics[width=100mm]{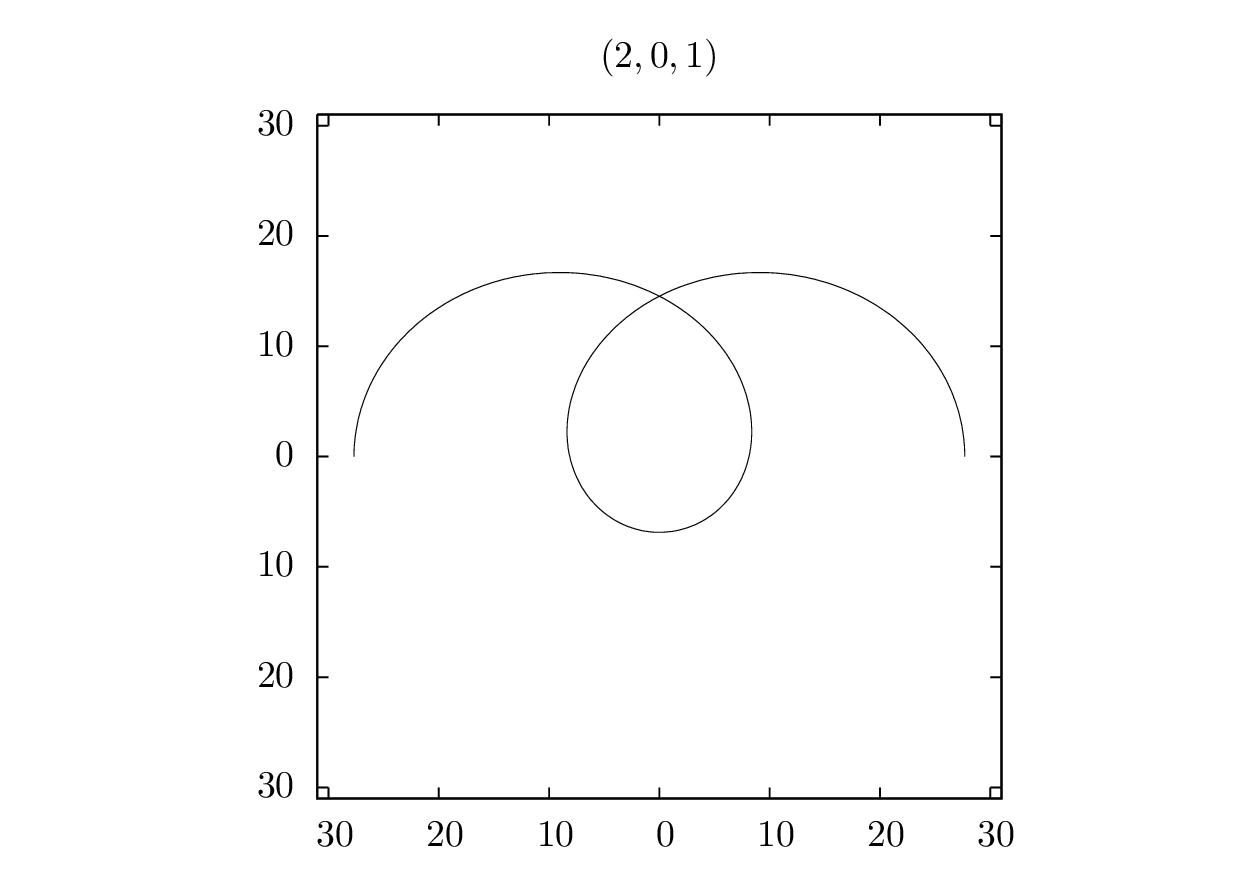}
\hspace{-80pt}
\includegraphics[width=100mm]{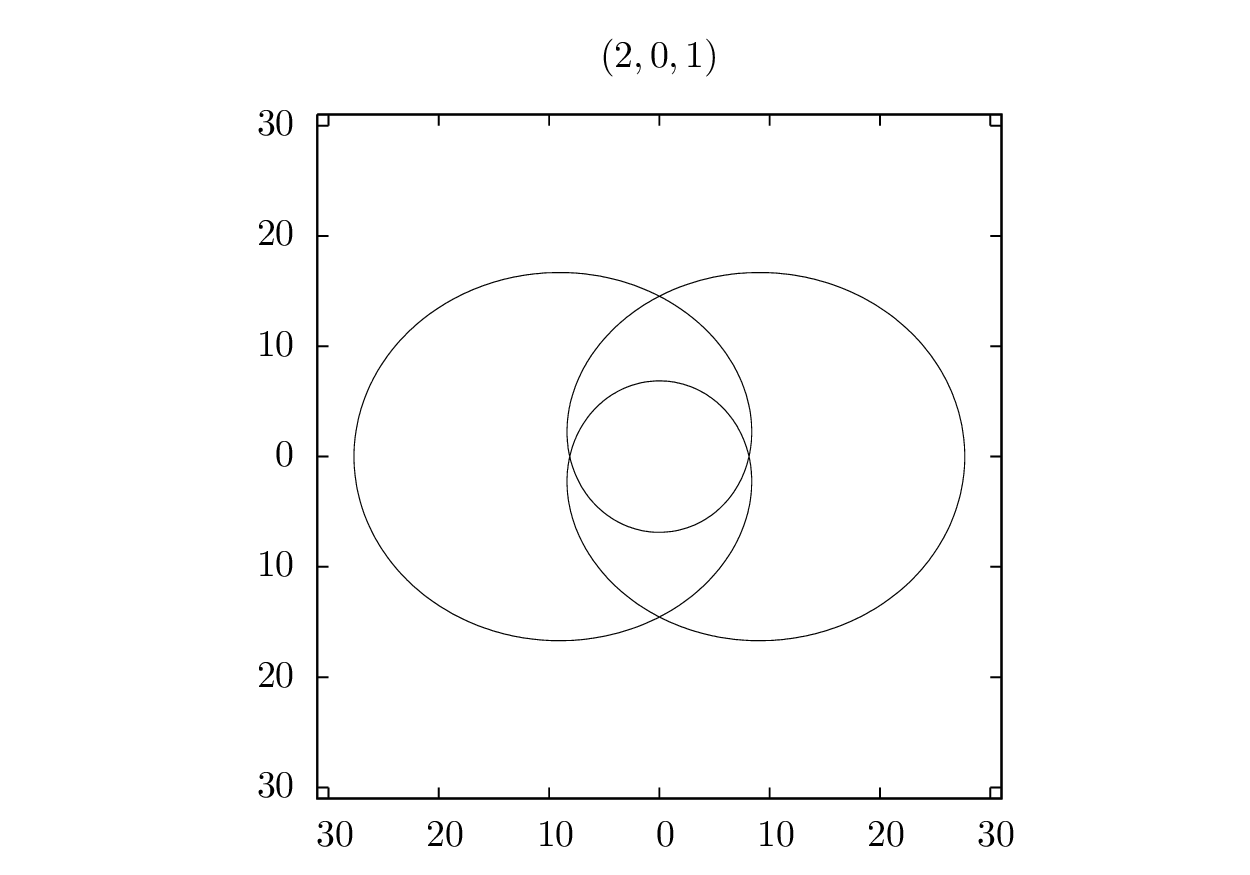}
\hfill
  \caption{Left: Half of the $(z=2,w=0,v=1)$ periodic orbit. Right:
  The full $(z=2,w=0,v=1)$ closed orbit. The orbit has $a=0,
L=3.980393$, and $  E=0.973101$.}  
  \label{(2,0,1)}
\end{figure}

Figure \ref{(2,0,1)} shows a $z=2$ orbit in the right panel, with a
single radial cycle of the same orbit (from one apastron to periastron
to the next apastron) shown in the left panel.  Notice from the figure
that the accumulated angle from apastron to apastron, $\Delta \varphi_r$,
is $3\pi$. In a complete orbit, the accumulated angle is $\Delta \varphi=
z\Delta \varphi_r=6\pi$.

This is not, however, the only kind of $z=2$ orbit. Figure
\ref{(2,1,1)} shows another 2-leaf orbit that whirls around the center
an additional $2\pi$ before zooming out to apastron again.  In fact,
there is a 2-leaf orbit that whirls $4\pi$ longer around the center
before returing to apastron, and in general one that whirls $2\pi w$
longer around the center before returning to apastron, for any
positive integer $w$.
Therefore, we will distinguish orbits by their number of
whirls, $w$, as well as by $z$.

\begin{figure}
  \vspace{-20pt}
\hspace{-500pt}
  \centering
\hfill
\includegraphics[width=100mm]{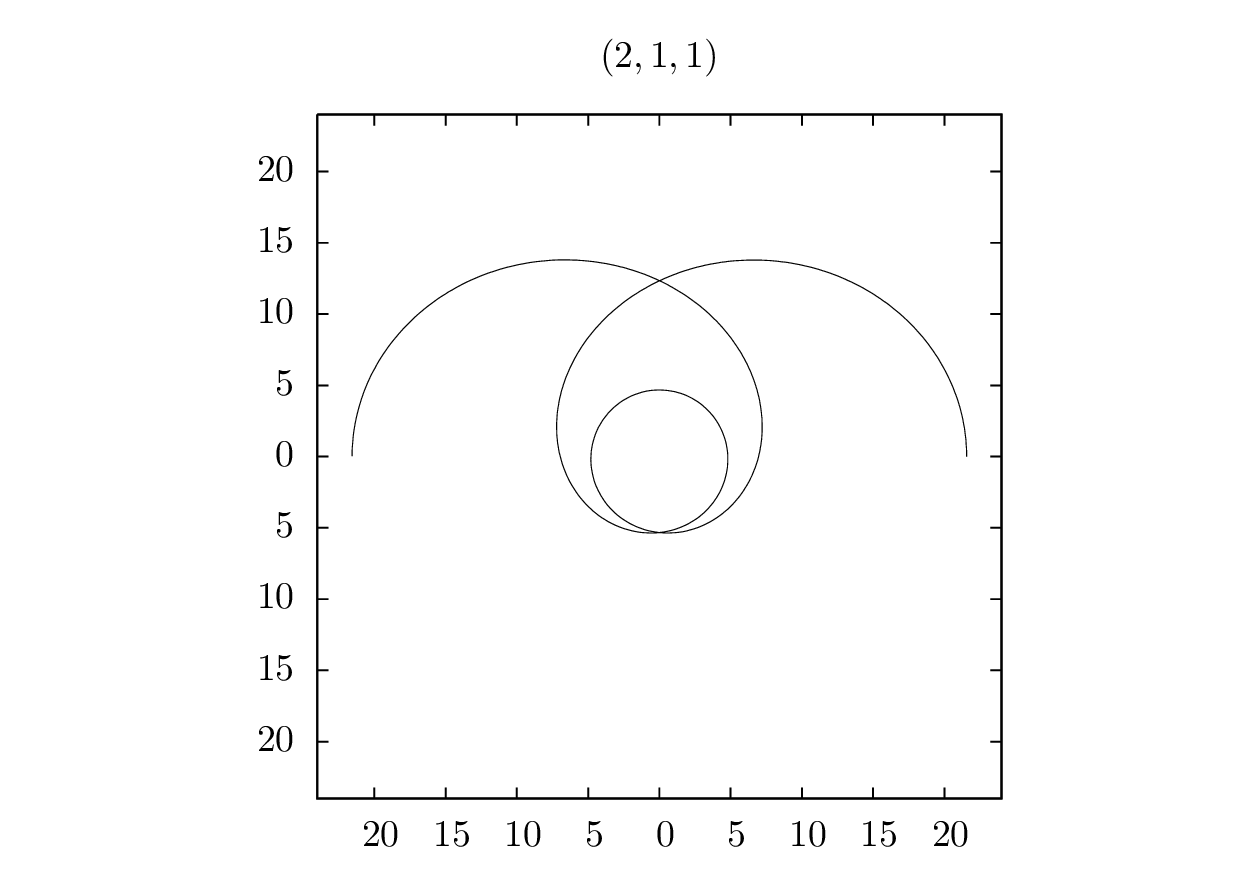}
\hspace{-100pt}
\includegraphics[width=100mm]{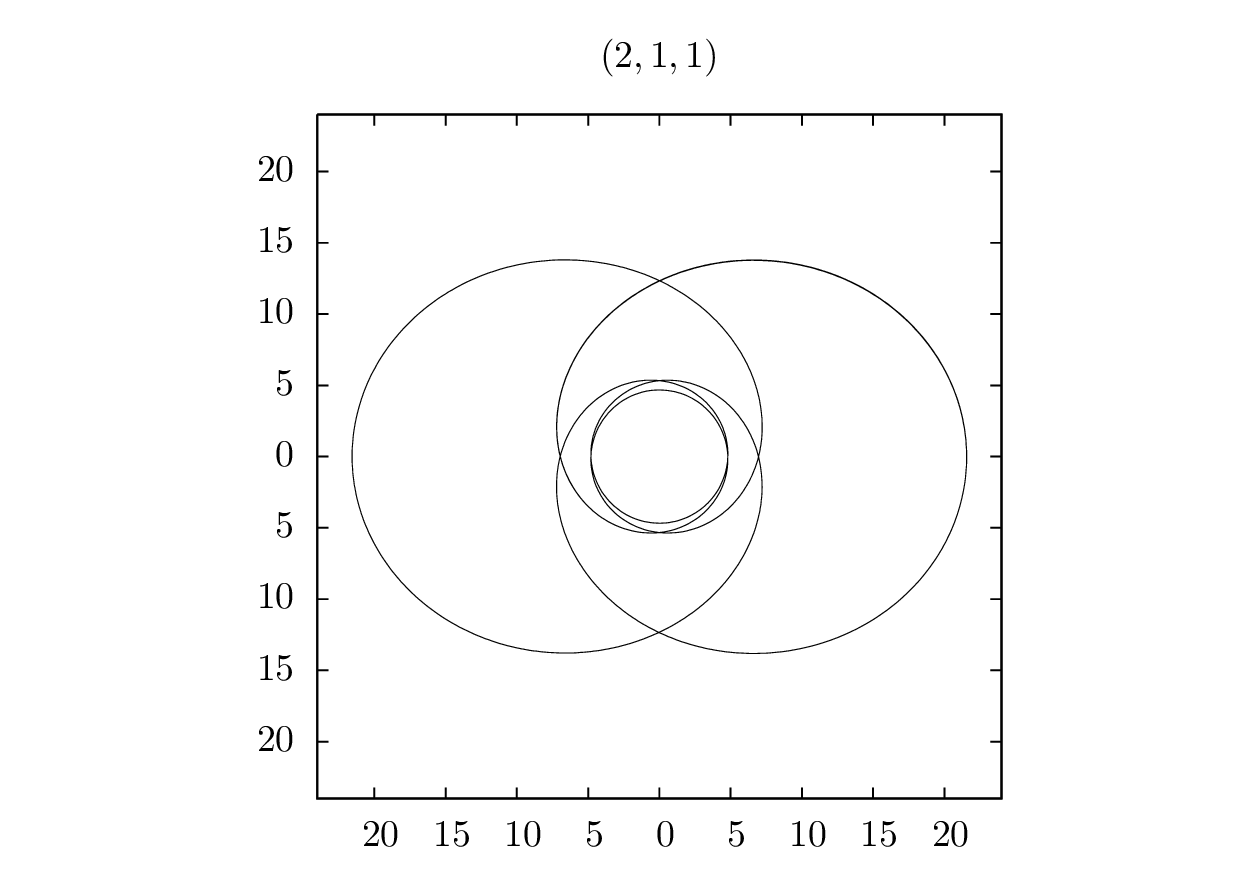}
\hfill
  \caption{Left: Half of the $(z=2,w=1,v=1)$ periodic orbit. Right:
  The full $(z=2,w=1,v=1)$ closed orbit. The orbital parameters are 
$a=0, L=3.718679$, and $E=0.966555$.}
  \label{(2,1,1)}
\end{figure}

Still, $z$ and $w$ alone are not sufficient to specify the geometric
features of a periodic orbit.  To see this, note that the successive
apastra of a periodic orbit with $z>2$ form the vertices of a regular
polygon.  We will label the vertices of these polygons with a third
integer $v$, counting the starting apastron of the orbit as $v=0$ and
increasing in the same rotational sense as the orbit (counterclockwise
for prograde orbits, clockwise for retrograde orbits), as shown in
figure \ref{(4,1,1)}.  Now, given any $z>2$, an orbit might move from
the starting apastron immediately to the next vertex in the
polygon. Such an orbit will be labelled $v=1$. However, an orbit with
the same $(z,w)$ might skip the next neighbor vertex.  We will assign
that orbit a $v$ of 2.  In general, a periodic orbit with a given
$z>2$ can skip any number of vertices less than $z$ when moving
between successive apastra.  All orbits will therefore be specified by
$(z,w,v)$ where $v$ indicates the first vertex hit by the orbit after
$v=0$, and where $v$ has the range
\begin{equation}
1\le v\le z-1 \quad \quad .
\label{vlimold}
\end{equation}
The orbit on the left of figure \ref{(4,1,1)} for instance is
a $(4,1,1)$ while
the orbit drawn on the right is a $(4,1,3)$.
We can still use eqn.\ (\ref{vlimold}) for the $z=2$ leaf orbits, despite
the fact that two points do not trace out a polygon, so that
$v=1$. Following this rubric,
the orbit of figure \ref{(2,0,1)} is a $(2,0,1)$ and that of
figure \ref{(2,1,1)} is a $(2,1,1)$.

\begin{figure}
  \vspace{-20pt}
\hspace{-500pt}
  \centering
\hfill
  \includegraphics[width=100mm]{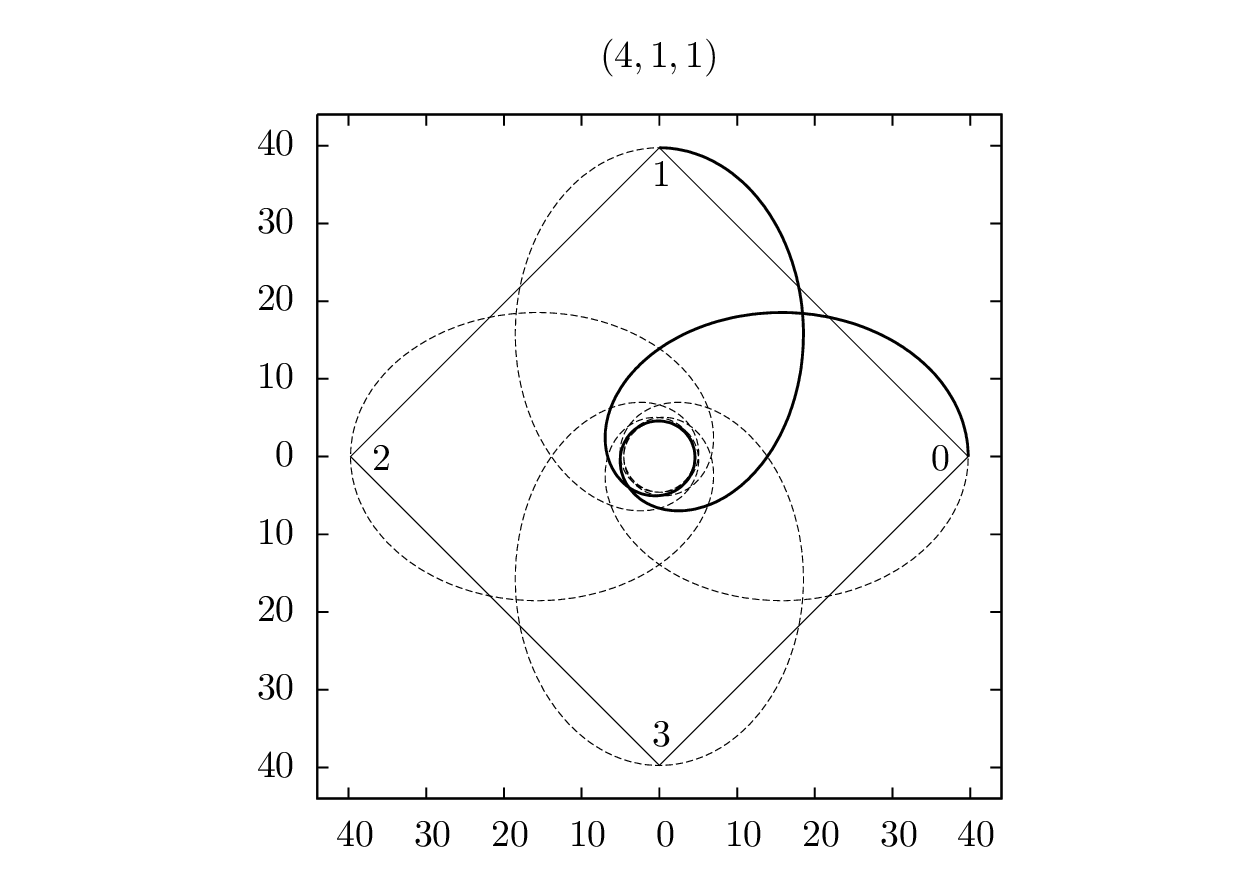}
\hspace{-100pt}
  \includegraphics[width=100mm]{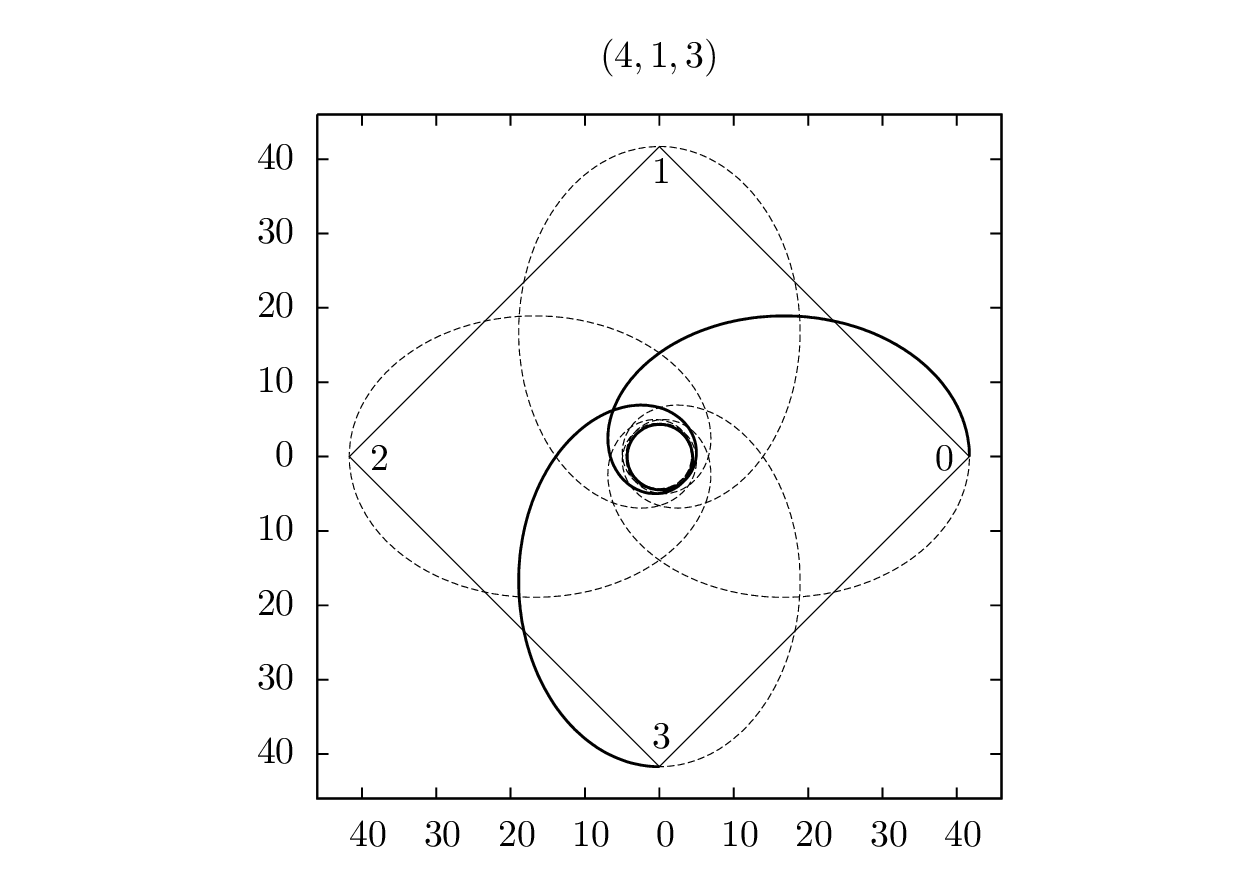}
\hfill
  \caption{4-leafed orbits with no whirls. Left: Leaves are traced out
  in sequential order for the $(z=4,w=1,v=1)$ closed orbit. Right:
  Leaves are traced out of order for the $(z=4,w=1,v=3)$ closed orbit.
The orbital parameters are 
$a=0, L=3.834058$ for both. The energy of the leftmost orbit is
  $E=0.979032$
and the energy of the rightmost orbit is $E=0.979842$.}
  \label{(4,1,1)}
\end{figure}

\begin{figure}
  \vspace{-20pt}
  \centering
\hspace{-20pt}
  \includegraphics[width=100mm]{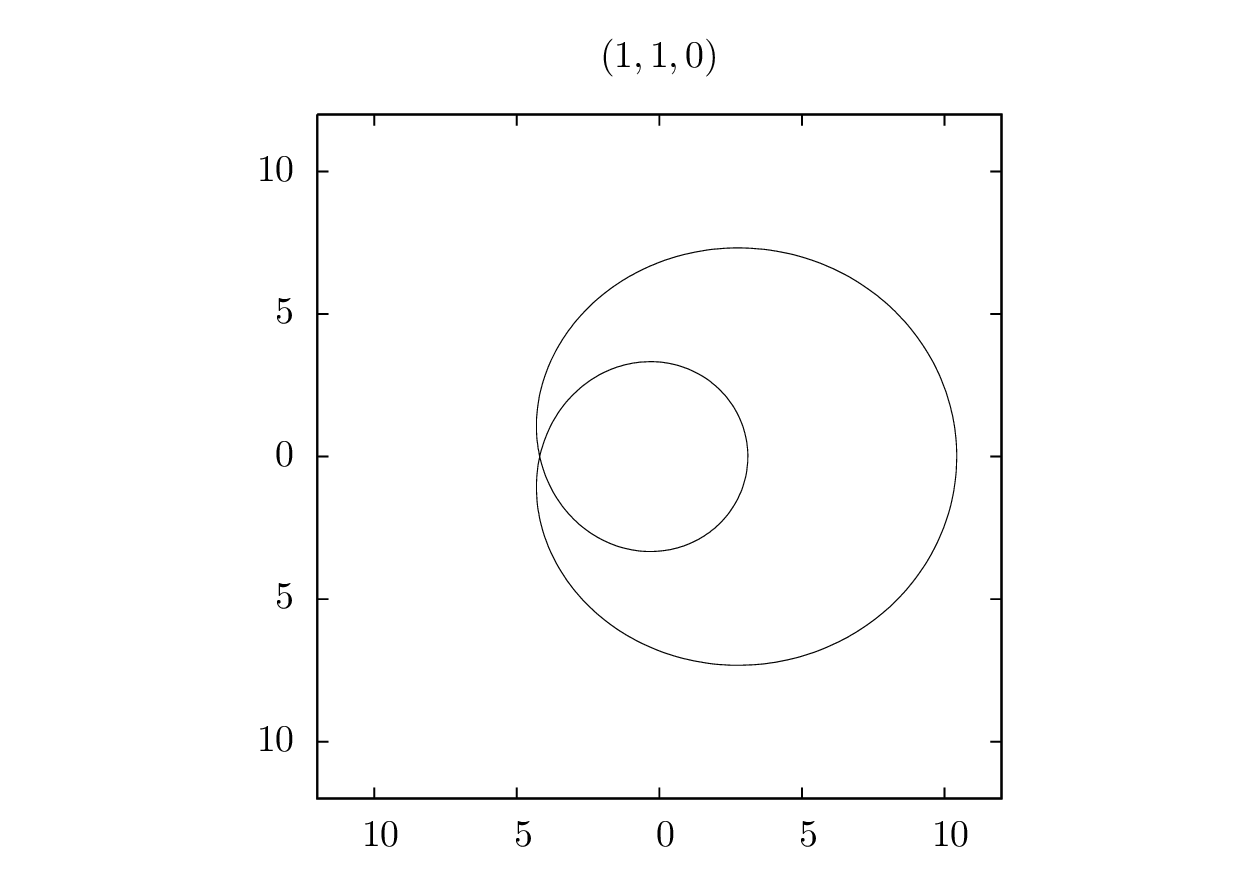}
\hspace{-100pt}
  \includegraphics[width=100mm]{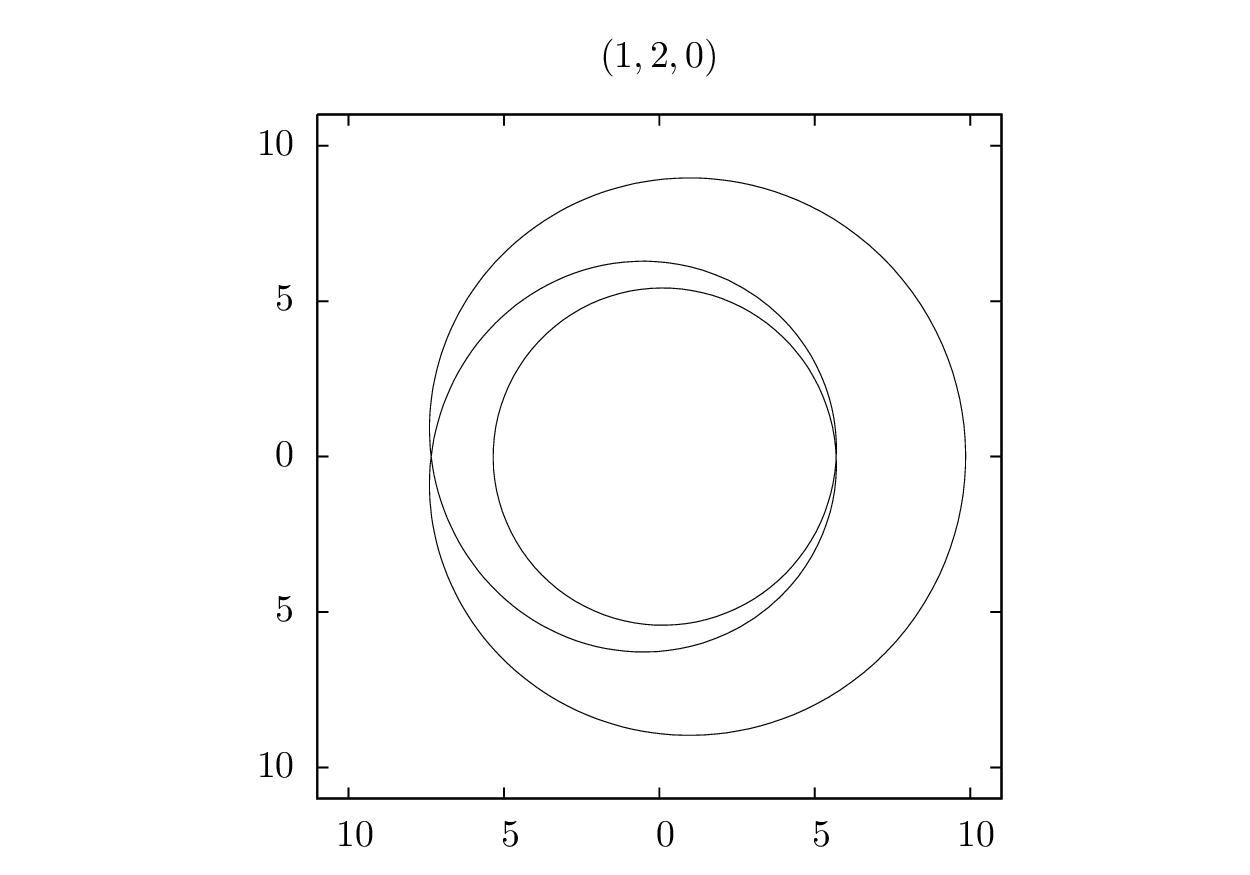}
\hfill
  \caption{Left: The full $(z=1,w=1,v=0)$ closed orbit. The orbital
    parameters are  
$a=0, L=2.714326$, and $E=0.932703$.
Right: The
  full $(z=1,w=2,v=0)$ closed orbit.
The orbital parameters are 
$a=0, L=3.535534$, and $E=0.948491$.
}
  \label{z1}
\end{figure}

Single leaf orbits, such as those shown in figure \ref{z1}, need
separate discussion.  Instinctively, we want to assign them a $z$
value of 1, but 
then the $v$ restricted by the range in (\ref{vlimold}) would be
undefined.  We can handle this problem by assigning
such orbits $v=0$, since successive apastra for these orbits
are actually the same single apastron.  That would lead us to modify
the allowed range of $v$ to
\begin{eqnarray}
  1 \leq v \leq z - 1 & , &\mathrm{if}\, z > 1 \nonumber \\
  v = 0 & , & \mathrm{if}\, z = 1 \quad .
\label{vlimnew}
\end{eqnarray}

There is another degeneracy to address.  For a given $w$, some $(z,v)$
pairs describe the same orbit.  For instance, the $(4,1,2)$ orbit
closes after only two leaves and is identical with the $(2,1,1)$ with
the same orbital parameters.  We remove this degeneracy by requiring
that $z$ and $v$ be relatively prime.  The allowed $v$ are then
\begin{eqnarray}
  1 \leq v \leq z - 1 & , &\mathrm{if}\, z > 1,\, z \bmod v = 0\nonumber \\
  v = 0 & , & \mathrm{if}\, z = 1 \quad .
\label{vlim}
\end{eqnarray}
Having pruned those orbits out of any counting, the $(z,w,v)$ label
now uniquely specifies the topological features of a closed orbit.

\subsection{Periodic orbits and rational numbers}
\label{find}

Our $(z,w,v)$ taxononmy naturally forges the relationship between
rational numbers and periodic orbits.  Note first that any positive
rational number $q$ can be written in the form
\begin{equation}
  q = s + \frac{m}{n} \quad ,
\end{equation}
where $s \geq 0$ is the integer part and $m/n$ is the fractional part,
with $m$ and $n$ relatively prime integers satisfying
\begin{equation}
1\le m\le n-1 \quad \quad .  
\end{equation}
Of course, these are exactly the conditions that $w, v$ and $z$
satisfy.  Since every periodic orbit corresponds to a $(z,w,v)$
set in our scheme, we can therefore associate a rational number
\begin{equation}
  q \equiv w + \frac{v}{z}
  \label{qdef}
\end{equation}
to every periodic orbit.  This association
reflects the 
physical observation that
for a periodic orbit, the accumulated azimuth between successive
apastra must be given by
\begin{equation}
\Delta \varphi_r=2\pi \left(1+w+\frac{v}{z}\right )=\frac{\Delta \varphi}{z}
\quad ,
\label{dphir}
\end{equation}
where the total accumulated angle $\Delta \varphi$ in one full orbital
period is $z\Delta \varphi_r$.

There is another sense in which we can associate a rational number to
a periodic orbit.  Every eccentric equatorial orbit has 2 associated
orbital frequencies: a radial frequency
\begin{equation}
\omega_r=\frac{2\pi}{T_r} \quad\quad ,
\end{equation} 
where $T_r$ is the (coordinate) time elapsed during one radial
cycle (not the total period of an orbit), and an angular frequency
\begin{equation}
\omega_\varphi=\frac{1}{T_r} \int_0^{T_r}\frac{d\varphi}{dt} dt =
\frac{\Delta\varphi_r}{T_r}  \quad\quad
\end{equation}
corresponding to the time-averaged value of $\txtD{t}{\varphi}$ over one
radial period.  For a generic orbit, the ratio of those frequencies
can be arbitrary, but for a periodic orbit, it must satisfy
\begin{equation}
  \frac{\omega_\varphi}{\omega_r} = \frac{\Delta\varphi_r}{2\pi} =
  1 + w +\frac{v}{z} = 1 + q \quad .
\end{equation}
The periodic orbits, then, are those whose fundamental orbital
frequencies are rationally related.

It is worth noting that the orbital frequencies $\omega_r$ and
$\omega_\varphi$ are the same frequencies that arise in an action-angle
Hamiltonian formulation of the dynamics \cite{schmidt}.  We draw the
connection explicitly in appendix \ref{actionangle}.

\subsection{Circular orbits}
\label{circs1}

Circular orbits, while strictly periodic, do not fit into our indexing
scheme.  This is not surprising, since circular orbits close
compulsorily rather than by the tuning of any orbital parameters.  Put
in other terms, a circular orbit has only a single rotational
frequency $\omega_\varphi$; there is no additional librational frequency
to which it must relate rationally in order to close.  Circular orbits
exhibit a periodicity of a different flavor, and no association to
rational numbers is required.  They are, in a sense, periodic by
default.

Perhaps unexpectedly, there is nonetheless a natural way to fit the
\emph{stable} circular orbits into our taxonomy.  Consider, for a
given black hole spin $a$, all the orbits with a given angular
momentum $L$.  For such a fixed $a, L$ pair, the frequencies
$\omega_\varphi$ and $\omega_r$ vary continuously\footnote{In fact, as
we'll see in Sections \ref{zooS} and \ref{zooK}, they vary
monotonically with energy and eccentricity at fixed $a, L$.} with
eccentricity.  Not surprisingly, the zero eccentricity limit of
$\omega_\varphi$ is just the $\txtD{t}{\varphi}$ of the stable circular
orbit.  Additionally, as detailed in appendix \ref{circs}, the zero
eccentricity limit of $\omega_r$ is not zero but rather the frequency
of radial oscillations for small perturbations of the circular orbit.
In this limiting sense, then, we can assign every stable circular
orbit an effective
\begin{equation}
  q_c = \lim_{e \to 0}\left ( \frac{\omega_\varphi}{\omega_r} -
  1\right )
\label{qcirc}
\end{equation}
and thus an effective $(z,w,v)$ triplet.

As shown in Appendix \ref{circs}, $q_c$ grows monotonically with
decreasing $r_c$, the radius of the stable circular orbit.  In fact,
it turns out that for a given $a$, every 
positive rational $q$ corresponds to some stable circular orbit, with
$q_c \to 0$ as the circular radius $r_c \to \infty$ and $q_c \to
\infty$ as $r_c \to r_{ISCO}$, the radius of the innermost stable
circular orbit.  At low $r$, $q_c$ will eventually go
above 1 so that $w_c > 1$.  Additionally, the corresponding $z_c$ will
sometimes be small ($z \lesssim 10$).  Both these facts
will leave their footprint
on orbits in the strong-field regime.

\subsection{Precession of periodic orbits}
\label{precess}

Our taxonomy focuses on the set of measure zero closed orbits.
A generic orbit will experience an accumulated
angle that is
irrational so the orbit precesses and never closes. However,
since the rationals are dense on the number line,
any generic orbit can
be approximated arbitrarily closely by some rational. Therefore,
  although it seems that we are describing a special set of extremely
  rare orbits, we might as well be describing every orbit. The
  potential power of this taxonomy 
lies in this simple observation.

\begin{figure}
  \vspace{-20pt}
\hspace{-500pt}
  \centering
\hfill
  \includegraphics[width=100mm]{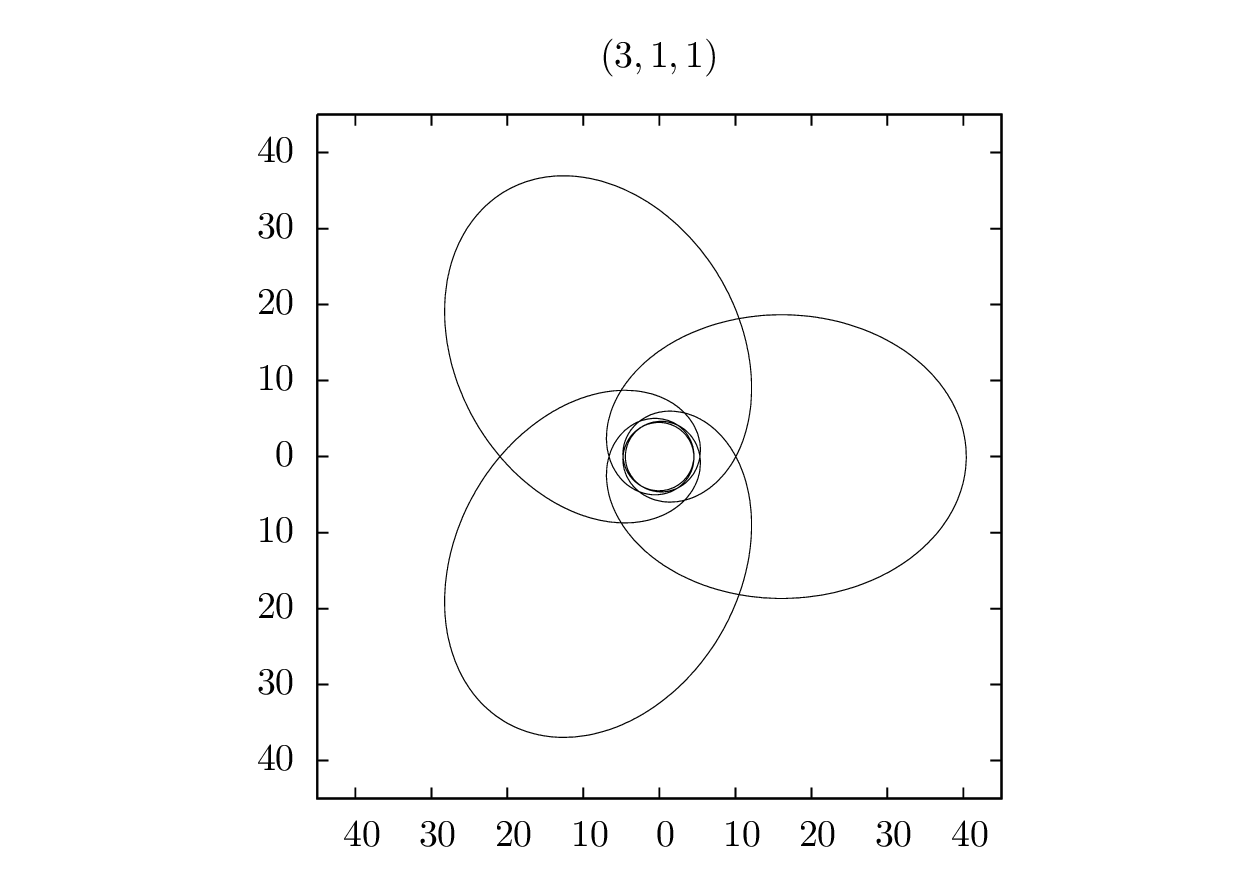}
\hspace{-100pt}
  \includegraphics[width=100mm]{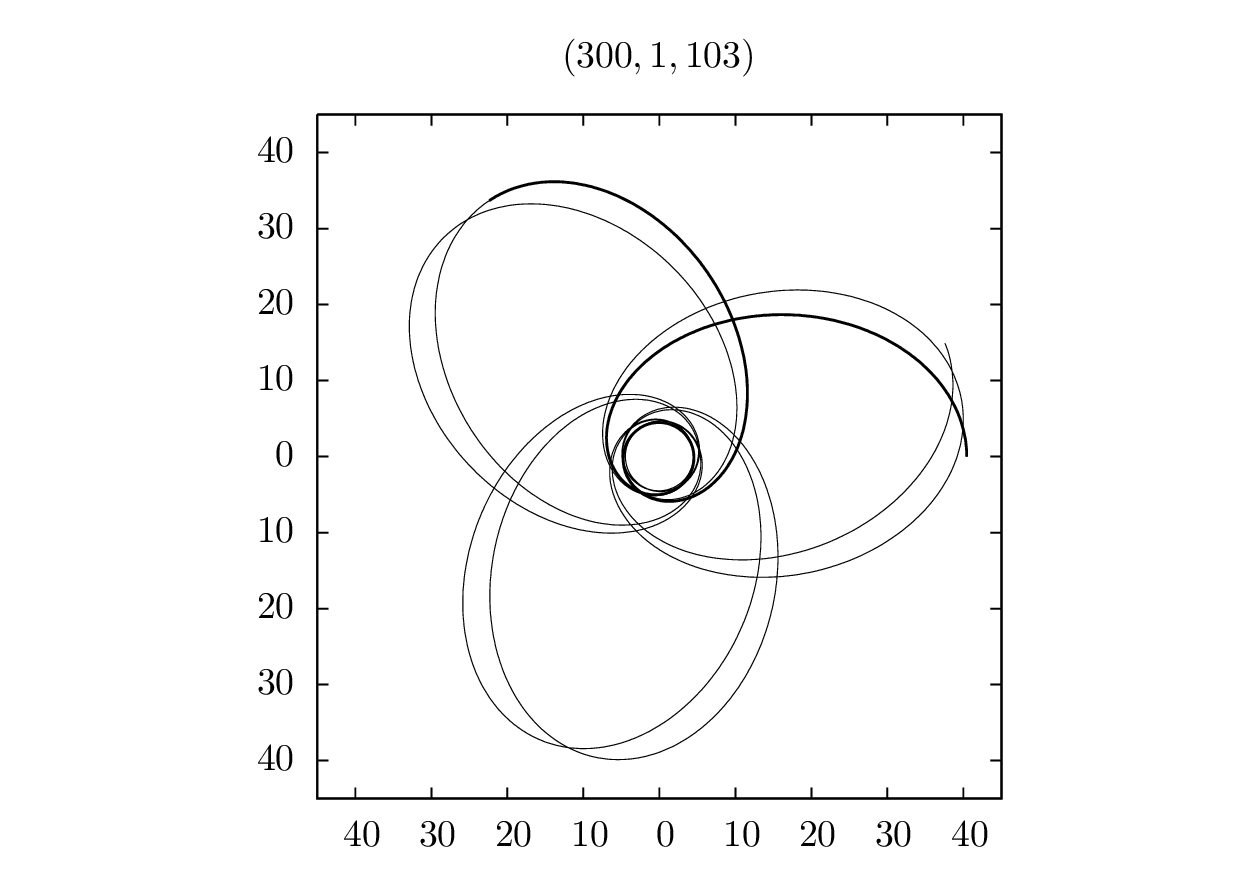}
\hfill
  \caption{Left: The closed $(z=3,w=1,v=1)$ orbit. right: The closed
  $(z=300,w=1,v=103)$ orbit looks like a precession of the
  $(z=3,w=1,v=1)$ orbit. Only 6 of the 300 leaves are shown. The first
  segment from apastron to apastron is emphasized in bold.
The orbital parameters are 
$a=0, L=3.834058$ for both. The leftmost orbit has energy $E=0.979304$ and the
rightmost orbit has energy $E=0.979331$.  \label{(3,1,1)}}
\end{figure}

To illustrate the approximation of a generic orbit by periodic ones,
consider the $(3,1,1)$ closed multi-leaf orbit on the left of figure 
\ref{(3,1,1)}. There are aperiodic orbits which lie close
to this orbit. For instance consider an orbit that accumulates a
slightly greater
angle in one cycle from apastron to apastron, $w+v/z
=1+1/3+\delta$ where the irrational $\delta<<1$. It's
three leaf pattern will precess, 
never closing. The precessing
orbit could be 
approximated by the periodic orbit $(3,1,1)$. The smaller the
precession, the better the approximation. Or, we could do
better. For the sake of 
argument, let's approximate the irrational drift by the rational
$\delta\approx 1/100$ and write  
$w+v/z=1+1/3+1/100$ as
$w+v/z=1+103/300$. We can now approximate the precessing orbit by the
$(300,1,103)$ multi-leafed orbit on the right of figure \ref{(3,1,1)}. 
Our rational approximate is a
300-leaf orbit. Since $v=103$, it skips 102 leaves in the
pattern each time it moves out to apastron so that it appears to
trace out
a precessing three leaf pattern. The periodic approximate does
actually close but only
after tracing all 300 leaves.

We emphasize that the approximation of
generic orbits by a nearby periodic
is a reality foisted on us by  finite precision, both
numerically and observationally. 
Every computer program truncates numbers to some finite
precision and in so doing explicitly outputs a rational.
Every measurement performed delivers an observable to
some finite precision and in so doing explicitly delivers all
data as rationals.
Likewise, every calculated aperiodic black hole
orbit will necessarily be indistinguishable from some periodic one.
We have no choice about this.
The taxonomy embraces
that fact and allows us to estimate the rational approximate deliberately
rather than inadvertently through the restrictions of finite precision.

\subsection{Not all orbits allowed}
\label{notall}

An obvious question to ask is whether all periodic orbits are allowed in
every Kerr system. In other words, is every rational in the set 
$0<q<\infty$ 
allowed? The answer will turn out to be a very interesting
no.

\section{Black Hole orbital dynamics}
\label{dynamics}

Indeed, not all periodic orbits are allowed in a given Kerr system.
We now show that
the rational numbers occur in the
range
\begin{equation}
q_{c}
\le q
\le
q_{max} \quad \quad ,
\label{lims}
\end{equation}
bounded below by the $q_c$ of the stable circular orbit and
above by the $q_{max}$ of the maximum energy bound orbit.\footnote{To be
  definite, we mean {\it non-plunging}, bound orbit.}
Periodic orbits corresponding to all rational
numbers in this range populate the phase space.
The limits vary not only for
different black hole spins, but also for different values
of the constant angular momentum $L$, as we will demonstrate
in \S \ref{zooS} and \S \ref{zooK}.

Equation (\ref{lims}) turns out to be profound. 
The crucial insight that allows us to determine the admissable periodic orbits
is the observation that 
$\Delta \varphi_r$, and therefore the associated rational $q=w+v/z$,
increase monotonically with energy.
We
extract the dynamical consequences of this fact,
contrasting the weak and strong-field behaviors.

\subsection{The Schwarzschild zoo}
\label{zooS}

\begin{figure}
\centering
{

\vspace{-20pt}
    \includegraphics[width=10cm]{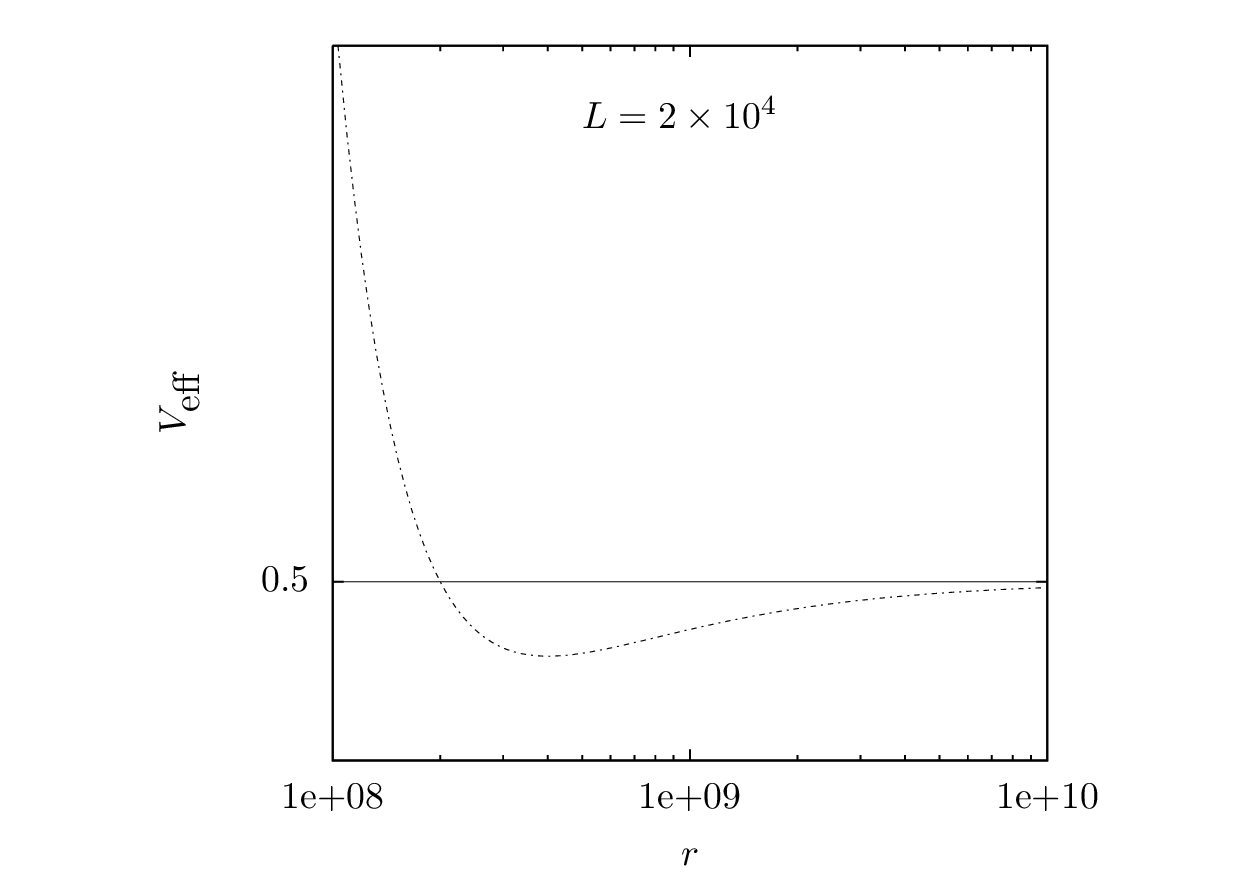}

}
\caption{The effective radial potential $V_{\textrm{eff}}(r)$
for very large $r$ (plotted on a $\ln r$ scale) in the vicinity of
solar system values. 
The horizontal line at $V_{\textrm{eff}} = 1/2$ corresponds to particle
  energy $E = 1$.
\label{Vsolar}}
\end{figure}

\begin{figure}
\centering
\subfigure[The effective radial potential $V_{\textrm{eff}}(r)$ as a
  function of $r $ on a log scale
for $L$ just above $L_{IBCO}$. 
The horizontal line at $V_{\textrm{eff}} = 1/2$ corresponds to particle
  energy $E = 1$. The $y$-axis ranges from $0.5+ 9\times 10^{-9}$ down
  to $0.5-3\times 10^{-9}$.
]
{
  \label{Vabove}
\vspace{-20pt}
    \includegraphics[width=10cm]{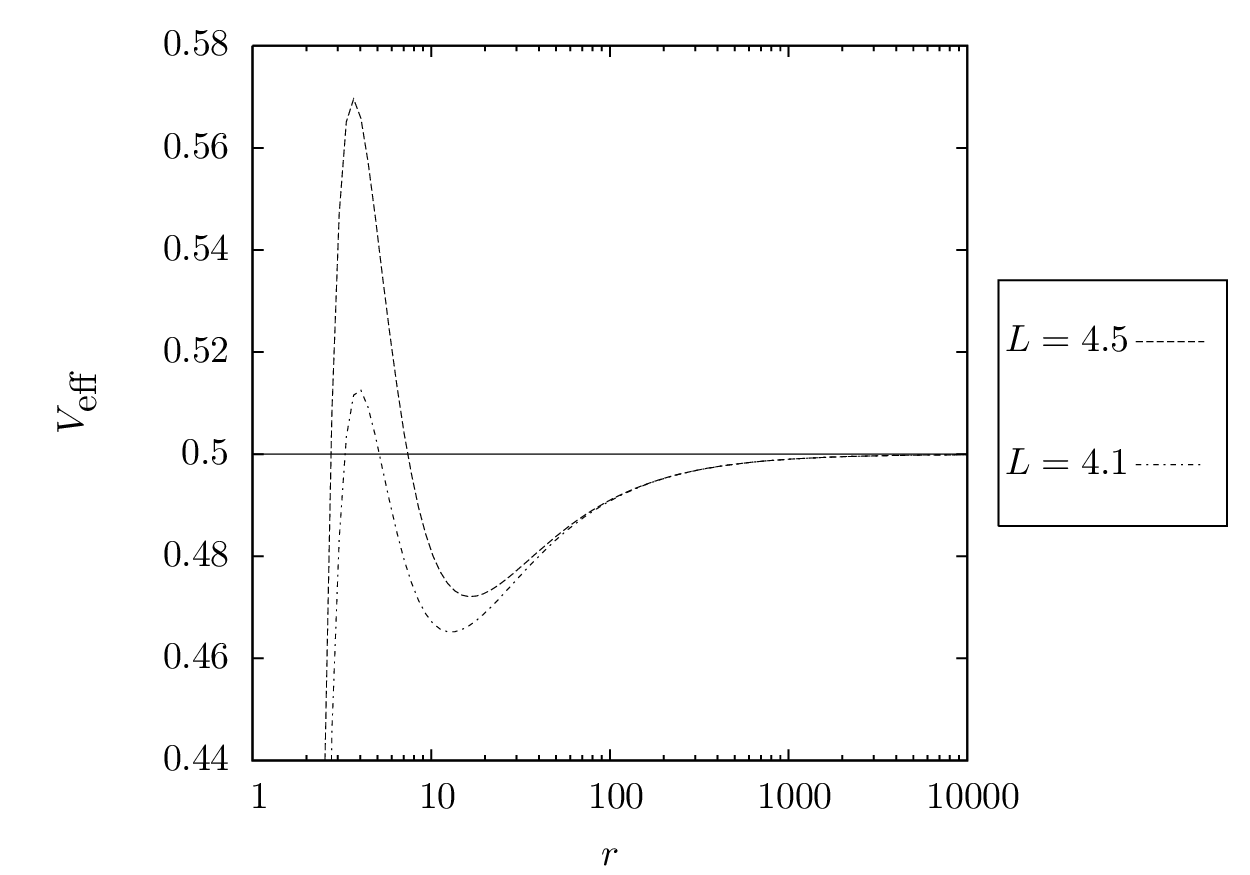}
}
 \\
\centering
\hspace{-95pt}
\subfigure[$w+v/z$ versus energy.] 
{
 \label{wvzEabove}
    \includegraphics[width=12.4cm]{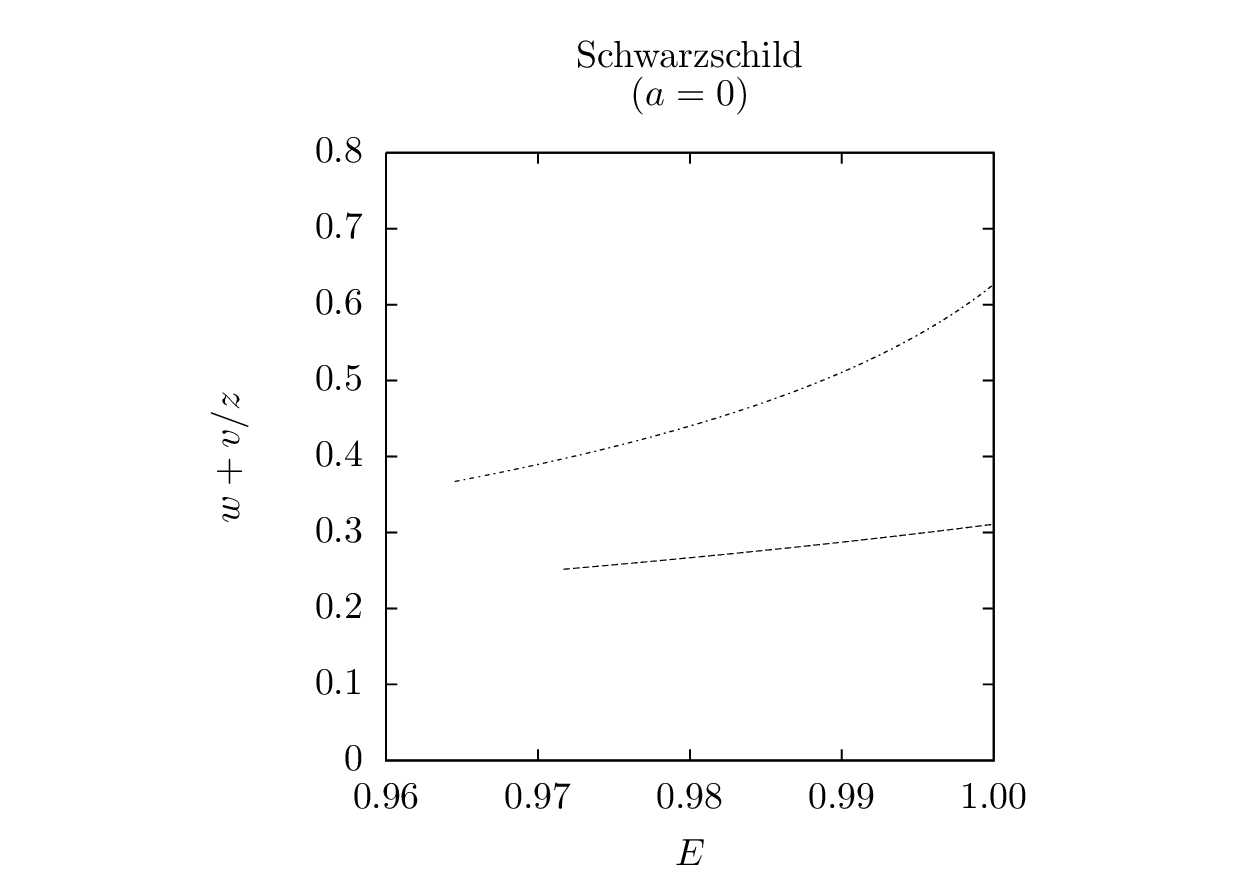}
}
\centering
\hspace{-165pt}
\subfigure[$w+v/z$ versus eccentricity.] 
{
 \label{wvzeccabove}
    \includegraphics[width=12.4cm]{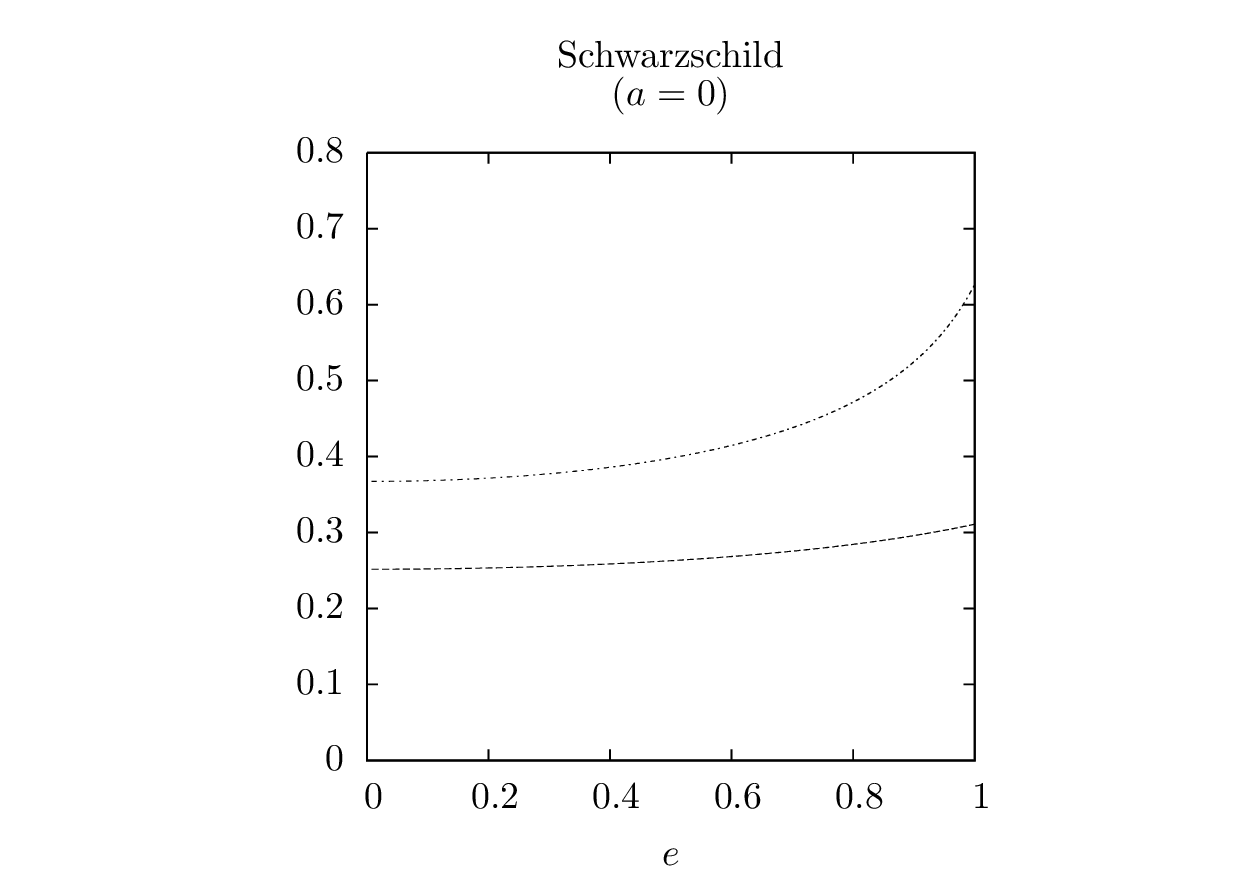}
}
\hspace{-60pt}
\caption{Schwarzschild, values of $L$ just above $L_{IBCO}$. In terms of accumulated angle, $(\Delta \varphi_r/2\pi)-1=w+v/z$.}  \label{above}
\end{figure}

\begin{figure}
\centering
\subfigure[The effective radial potential $V_{\textrm{eff}}(r)$
  bounded on the top by $L_{IBCO}=4$ and on the bottom by $L_{ISCO}=\sqrt{12}$.
]
{
  \label{SchVeffFig}
\vspace{-20pt}
    \includegraphics[width=10cm]{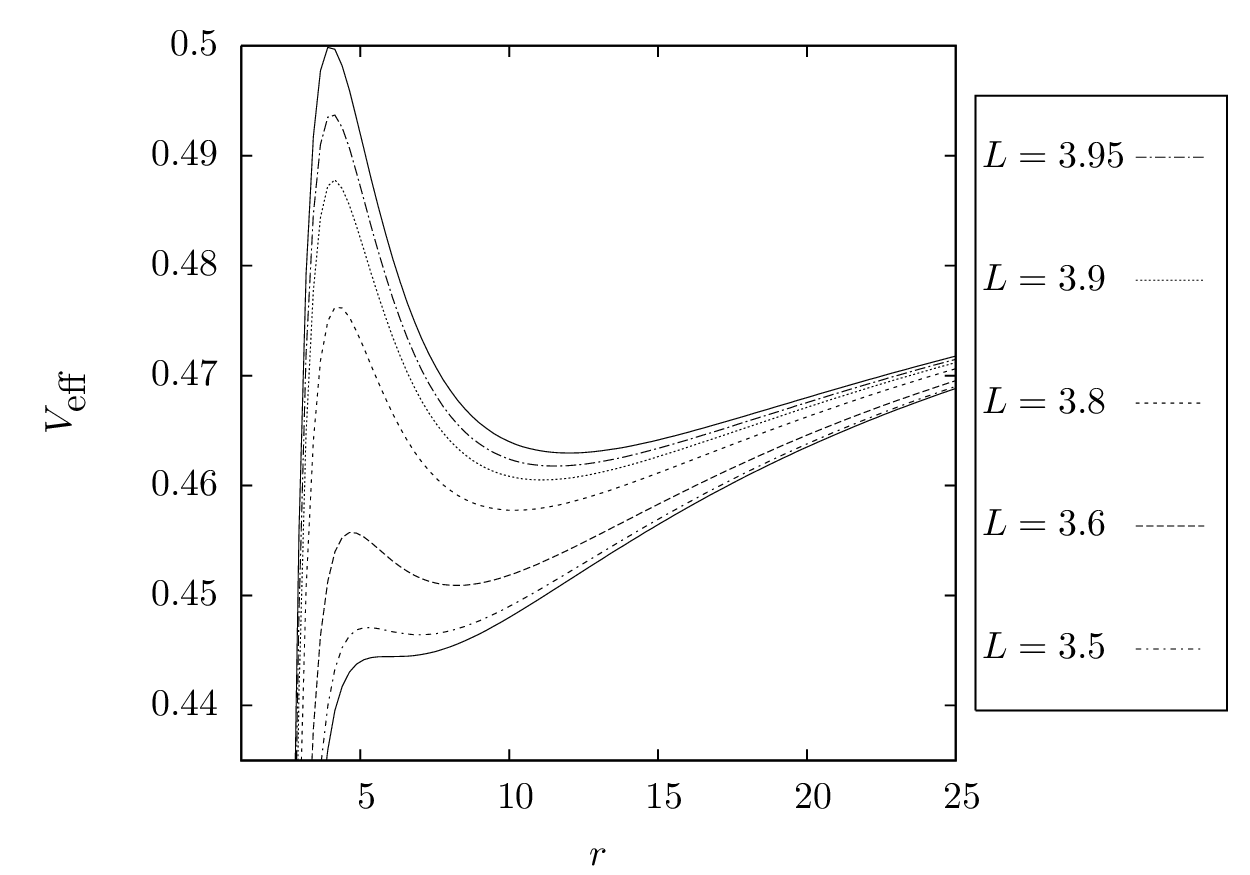}
}
 \\
\centering
\hspace{-95pt}
\subfigure[$w+v/z$ versus energy.] 
{
 \label{wvzES}
    \includegraphics[width=12.4cm]{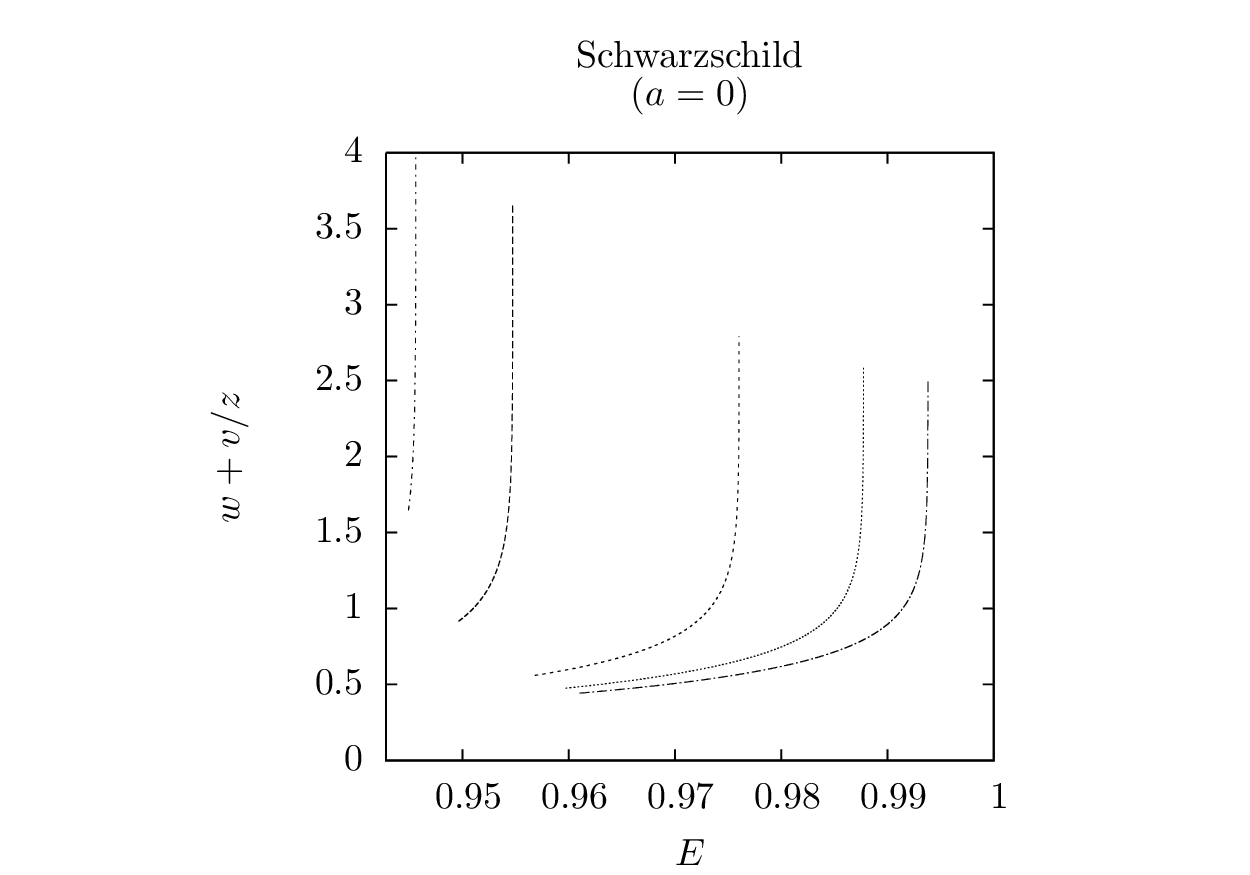}
}
\centering
\hspace{-165pt}
\subfigure[$w+v/z$ versus eccentricity.] 
{
 \label{wvzeccS}
    \includegraphics[width=12.4cm]{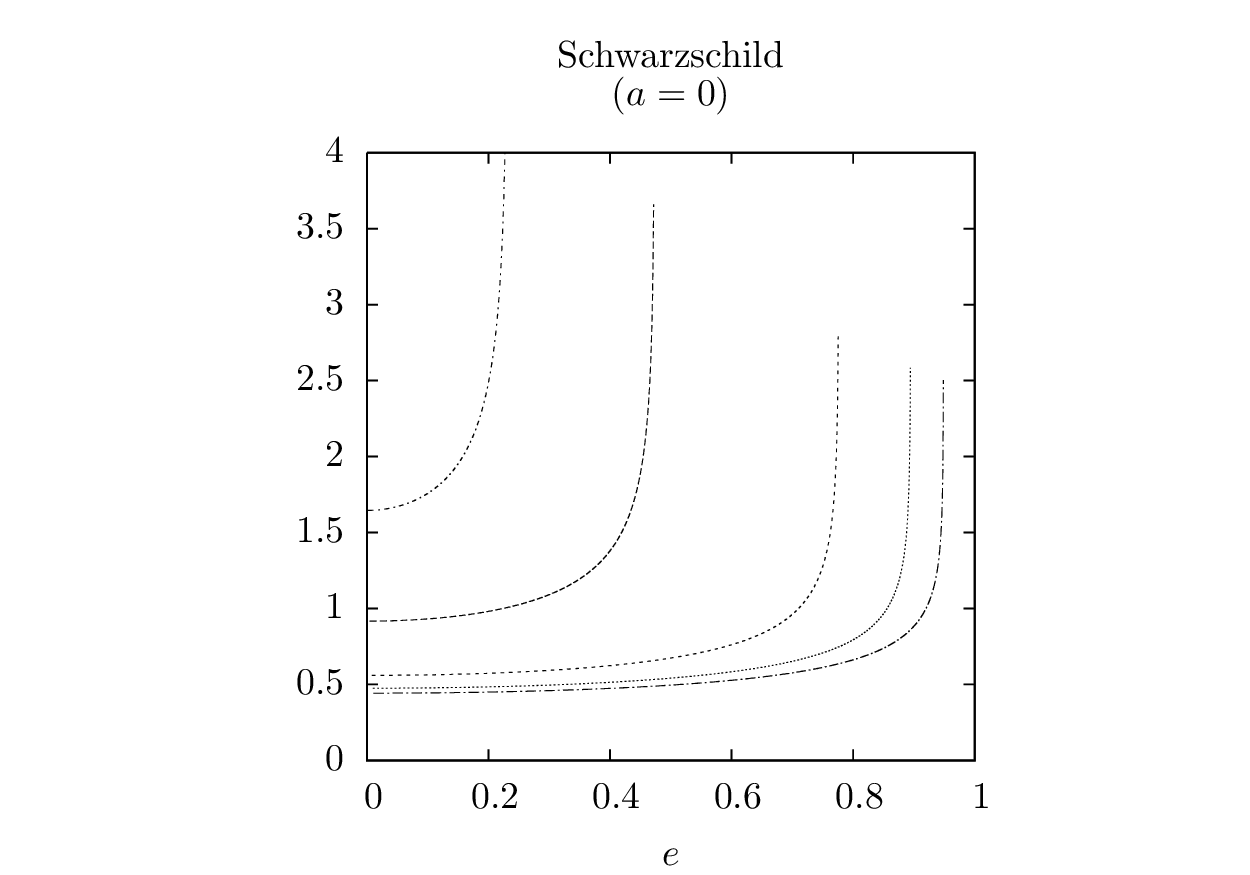}
}
\hspace{-60pt}
\caption{Schwarzschild for $L_{ISCO}< L<L_{IBCO}$. In terms of accumulated angle, $(\Delta \varphi_r/2\pi)-1=w+v/z$.}  \label{sch}
\end{figure}

To appreciate the significance of observation (\ref{lims}), 
we first consider Schwarzschild ($a=0$) orbits
because they admit a formal effective potential formulation that 
lends clarity to the discussion.
Schwarzschild orbits
can be described as one-dimensional motion in an effective
potential \cite{wald}:
\begin{equation}
\frac{1}{2}\dot r^2+V_{\rm eff}=\varepsilon_{\rm eff}
\label{schw}
\end{equation}
where
\begin{equation}
  \varepsilon_{\rm eff} =\half E \, , \quad V_{\rm eff} = \half
  -\frac{1}{r}+\frac{L^2}{2r^2}-\frac{L^2}{r^3} \quad\quad .
\label{veff}
\end{equation}
Since the shape of the potential is fixed by the value of $L$, a given
$V_{\rm eff}$ snapshot is tantamount to a snapshot of the family of
orbits with the same angular momentum.
Consider the effective potentials of figures  \ref{Vsolar}, \ref{Vabove},
and \ref{SchVeffFig}, which show $V_{\rm eff}$'s ranging from large
$L$ to low $L$, or equivalently 
successive $V_{\rm eff}$'s from the weak-field to the strong-field
regimes.
Clearly, the stable circular orbit always has the
lowest energy for a given $L$. If, as we will show, $\Delta
\varphi_r=2\pi(q+1)$ 
increases monotonically with energy, it must also be the case that
$q=w+v/z$ increases monotonically. It follows immediately that eqn.\
(\ref{lims}) is true. We verify these claims for the Schwarzschild
spacetime in this section.

Spinning black hole
spacetimes do not admit a comparable 
one-dimensional effective potential
description. However, all the qualitative features of the
Schwarzschild 
analysis
survive and, in fact, all the interesting dynamical
effects become more pronounced. We outline the dynamical results for the
spinning black hole spacetime in \S \ref{zooK}.


We begin in the weak-field regime at large radii (large
$L$) and will watch the dynamics evolve as we move into the
strong-field regime at close separations (low $L$). 
Consider, for
instance, solar system values, so that $L$ is in the range of Mercury's
angular momentum $L\sim 2\times 10^4$.
Such a potential at large $r$ is shown in
figure \ref{Vsolar}. There is effectively a large centrifugal barrier
at low $r$. The energy of the bound orbits ranges from
$E=1$ for the 
eccentricity $e=1$ trajectory down to the stable circular orbit
$e=0$. (Following
the standard convention in the literature we define eccentricity as
\begin{equation}
e = \frac{r_a-r_p}{r_a+r_p} \quad\quad ,
\end{equation}
where $r_a$ is the apastron and $r_p$ is the perihelion.)

All of the bound orbits with this $L$ are well described 
as minute precessions of an ellipse. In our taxonmy, these orbits
correspond to $(z,0,1)$ with a very high number of
zooms $z$
($q=w+v/z<<1$).
Mercury, for instance, precesses roughly
$43^{\prime \prime}$/century. With an orbital period of $88$ days,
this comes to roughly $0.1^{\prime\prime}$/orbit, which is $\sim
360^{o}/(3600^2)$. Mercury, therefore, is close to a periodic orbit
with $z=3600^2=1.296\times 10^7$ leaves and no whirls, which always
advances to the next available apastron  
 -- that is, a $(1.296\times 10^7,0,1)$.
Even the perfectly periodic $(1.296\times 10^7,0,1)$ will give
the appearance of a simple precessing ellipse.

In figure 
\ref{above}, we move into the intermmediate-field regime and show the
potentials for angular momenta $L=4.1$ and $L=4.5$, with
the centrifugal barrier better resolved to
show the unstable $E>1$ circular orbit.

The key conclusions for the black hole dynamics comes from the simple
figures \ref{wvzEabove} and \ref{wvzeccabove}.
Some salient features turn out to be universal for
all $L$. Most important, figures \ref{wvzEabove} and \ref{wvzeccabove} show
that, for a fixed $L$, $q=w+v/z$ increases monotonically with both $E$
and $e$ and terminates at some finite value. 
As per the comments at the beginning of \S \ref{dynamics}, this
justifies eqn.\ (\ref{lims}).

There are other interesting dynamical features evident from these
figures alone.
For instance, we see that
no orbits in this intermediate $L$ range whirl since $q=w+v/z<1$
and therefore $w=0$ over the 
entire range of eccentricities. Incidentally, the same thing is true
for the planetary $L$ (whose $q$ versus $e$ plot we omitted because it
was graphically indistinguishable from the $q=0$ axis). This
explains why we do not see zoom-whirl behavior in the solar system, or
for that matter, even for Schwarzschild orbits whose periastra are as
close to the 
black hole as $r\approx 8$.

For $L=4.5$,
the allowed closed 
orbits correspond to rational numbers bounded by the range
\begin{equation}
\frac{1}{4} \lesssim q\lesssim \frac{3}{10}
\quad \quad ,
\end{equation}
where we have approximated the limits by the nearest low $v,z$
option ($w=0$). It follows that all orbits for $L=4.5$ will look like
precessions of $(4,0,1)$ not quite going over to precessions of
$(3,0,1)$ at the upper limit as they march through the rationals with
energy increasing to $E=1$. 
This is dramatically different from Mercury-type motion. These are not
precessing ellipses but are precessions of a four-leaf clover. 

As the angular momentum drops even further to $L=4.1$, the
allowed closed
orbits correspond to rational numbers bounded by the range
\begin{equation}
\frac{3}{8} \lesssim q \lesssim \frac{3}{5}
\quad \quad ,
\end{equation}
where we have approximated the limits by the nearest low $v,z$
option. Orbits will pass through $3/8,2/5,3/7,1/2,4/7,$ to $3/5$ and
all the rationals in between as the energy increases from that of the
stable circular to $E=1$. 


The zoo becomes more crowded as the angular momentum drops to the
point that the energy of the unstable circular orbit drops below 1.
The top darkened curve in figure \ref{SchVeffFig} is the potential for
$L=L_{IBCO} = 4$, the angular momentum of the innermost bound circular
orbit (IBCO).  The bottom darkened curve is the potential for
$L=L_{ISCO}=\sqrt{12}$, the angular momentum of the innermost stable
circular orbit (ISCO).  For every $L_{ISCO} < L < L_{IBCO}$, several
of which figure \ref{SchVeffFig} shows, there is an energetically
bound ($E < 1$) unstable circular orbit.  Accompanying
each bound unstable circular orbit is an eccentric orbit called a
homoclinic orbit.  These orbits are so central to the
discussion of strong-field dynamics that we now briefly outline how
they fit into the $(z,w,v)$ taxonomy before continuing with the rest
of the dynamical discussion.

\begin{figure}
  \vspace{-20pt}
  \centering
  \includegraphics[width=0.5\textwidth]{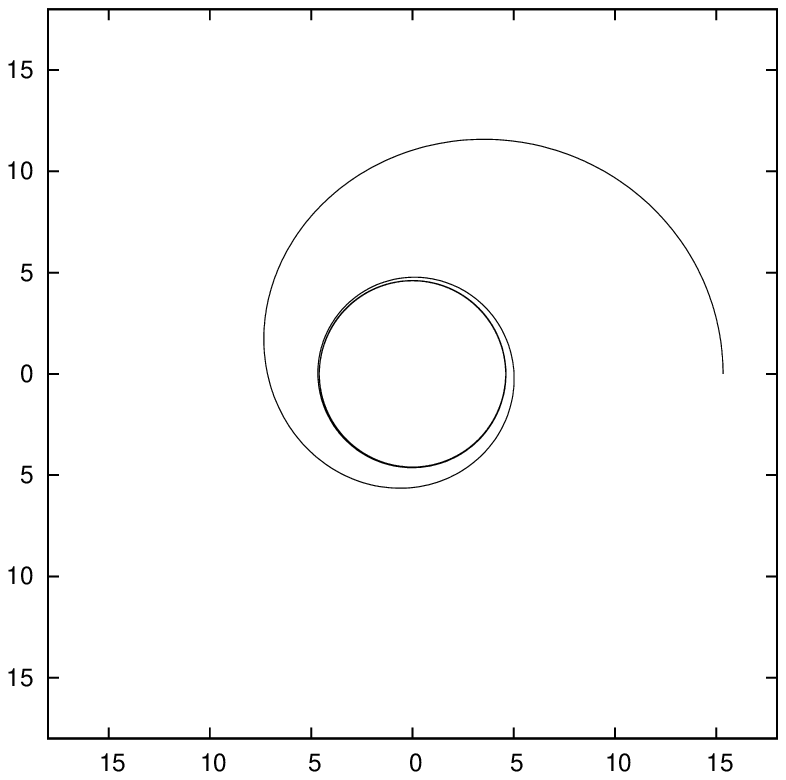}
  \caption{A homoclinic orbit $(1,\infty,0)$ for $L_{ISCO}<L<L_{IBCO}$.
The orbital parameters are 
$a=0, L=3.636619$, and $E=0.958373$.}
  \label{(1,inf,0)}
\end{figure}

A homoclinic orbit asymptotically recedes from a periodic orbit (in
this case, an unstable circular orbit) in the infinite past, zooms out
to some finite apastron, and then asymptotically approaches the same
periodic orbit in the infinite future.  Though eccentric, each
homoclinic orbit has the same energy $E_u$ and angular momentum $L_u$
as the unstable circular orbit to which it asymptotes.  When it
exists, a homoclinic orbit forms the boundary (for a given $L$)
between orbits that plunge over the top of the potential ($E > E_u$)
and orbits that do not ($E < E_u$).

To see the role of homoclinic orbits in our taxonomy, consider the
behavior of a homoclinic orbit that starts at apastron, such as the
one in figure \ref{(1,inf,0)}.  As the particle climbs toward the
maximum of the effective potential, it asymptotes toward the circular
orbit located there and executes an infinite number of whirls. Because
it never returns to 
apastron, the
homoclinic orbit is not strictly periodic. Nevertheless, it 
is appropriate to think of this single leaf orbit as the $w \to
\infty$ limit of the progression of the $(1,w,0)$ orbits. We thus
label each homoclinic orbit as $(1,\infty,0)$, so that each has an
associated rational $q = \infty$.

It is perhaps surprising that the $q=\infty$ orbits are bound and have
eccentricities $e < 1$.  More specifically, homoclinic orbits with $L
\to L_{IBCO}$ have eccentricities $e \to 1$, while those with $L \to
L_{ISCO}$ have $e \to 0$.  In fact, the ISCO \emph{is} just the zero
eccentricity homoclinic orbit.  Properties of homoclinic orbits beyond
what is needed for the current discussion, including analytic
expressions for their trajectories, can be found in Refs.\
\cite{{bc},{ush}}.

Since the homoclinic orbit is the maximum energy bound orbit (for a
given $L$, it
defines the upper range of allowed periodic orbits, a la eqn.\
(\ref{lims}). Therefore $q_{max}=\infty$ for all $L_{ISCO}<L<L_{IBCO}$.
As always, the lower bound is greater than zero because $q$ terminates at a
finite value.

To see the consequences of this fact,
consider the $L=3.9$ line in figure \ref{wvzES}, which shows a $q_c$
of approximately
$0.475\approx 19/40$. By eqn.\
(\ref{lims}), we have the bound
\begin{equation}
\frac{19}{40}\le 
q \le \infty
\quad\quad 
\end{equation}
for $L=3.9$.
We can 
approximate the lower limit better by going to higher $v,z$, but
these values 
are sufficient to make a point: 
there are no Mercury type motions.
Rather, small perturbations of the circular orbit can be
approximated by a periodic orbit with 40 leaves that jumps 
19 apastra away 
every radial cycle. Such an orbit will look very
much like a precession of a two-leaf orbit. 
Again, the one-leaf, zero-whirl orbit is forbidden.
The
$(1,0,0)$ corresponds to a $q=0$ and is not reached by any $L=3.9$
orbit as figure \ref{wvzES} shows.

\begin{figure}
  \centering
\hspace{-26.5pt} 
 \includegraphics[width=0.425\textwidth]{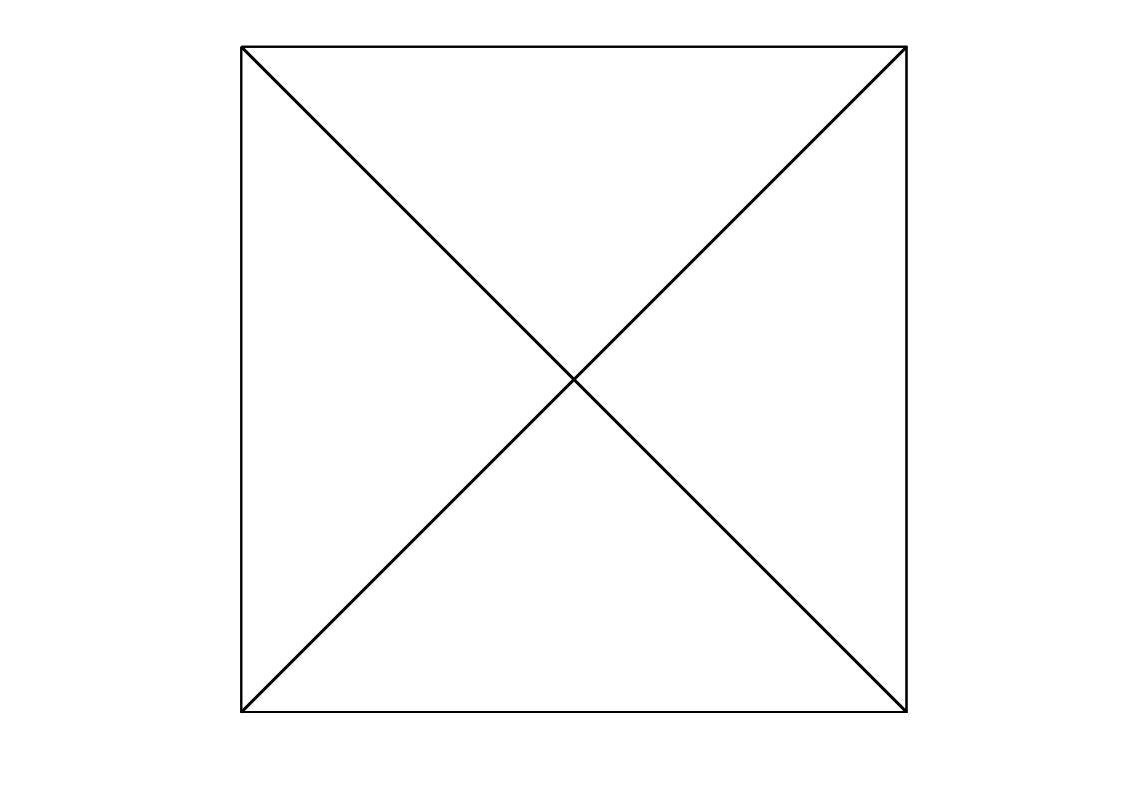}
\hspace{-63.5pt}
  \includegraphics[width=0.425\textwidth]{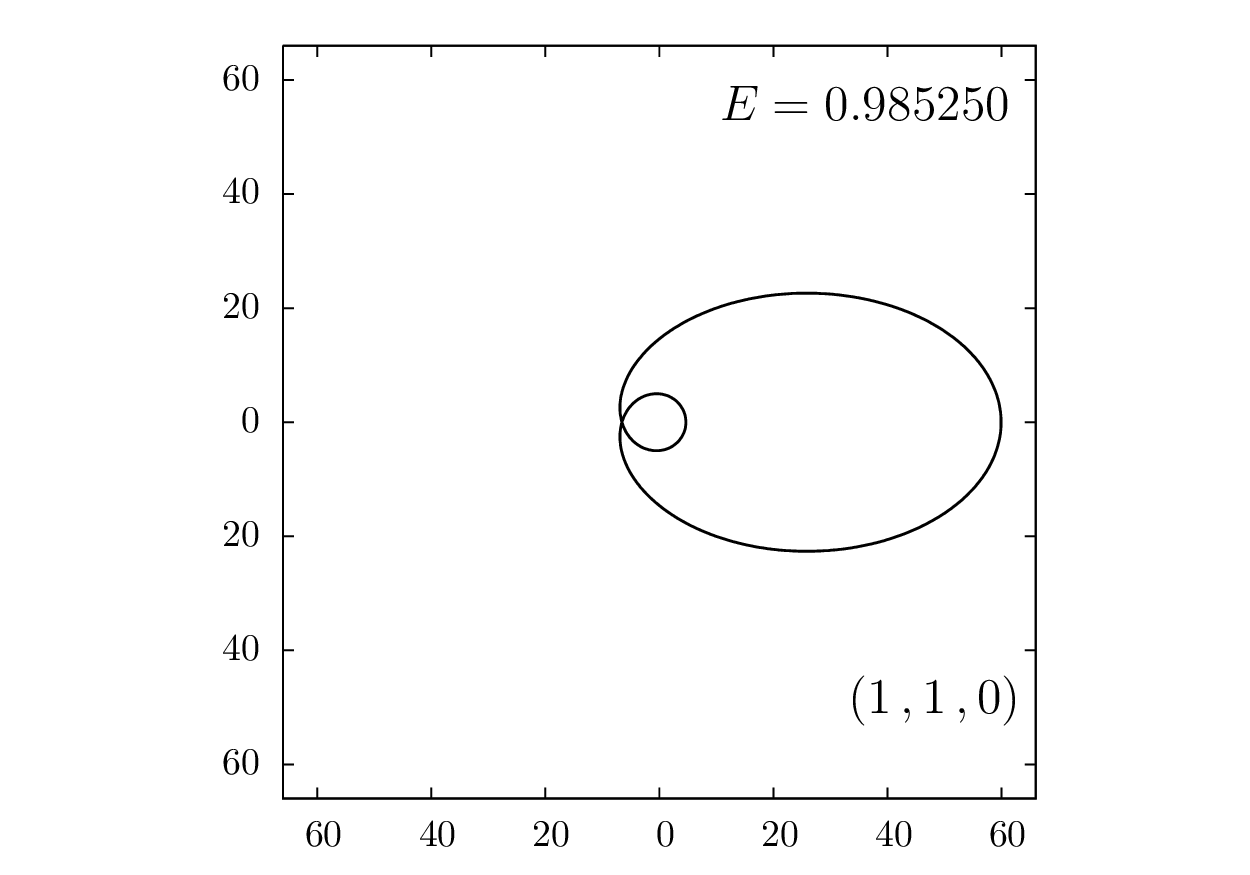}
\hspace{-60pt}
  \includegraphics[width=0.425\textwidth]{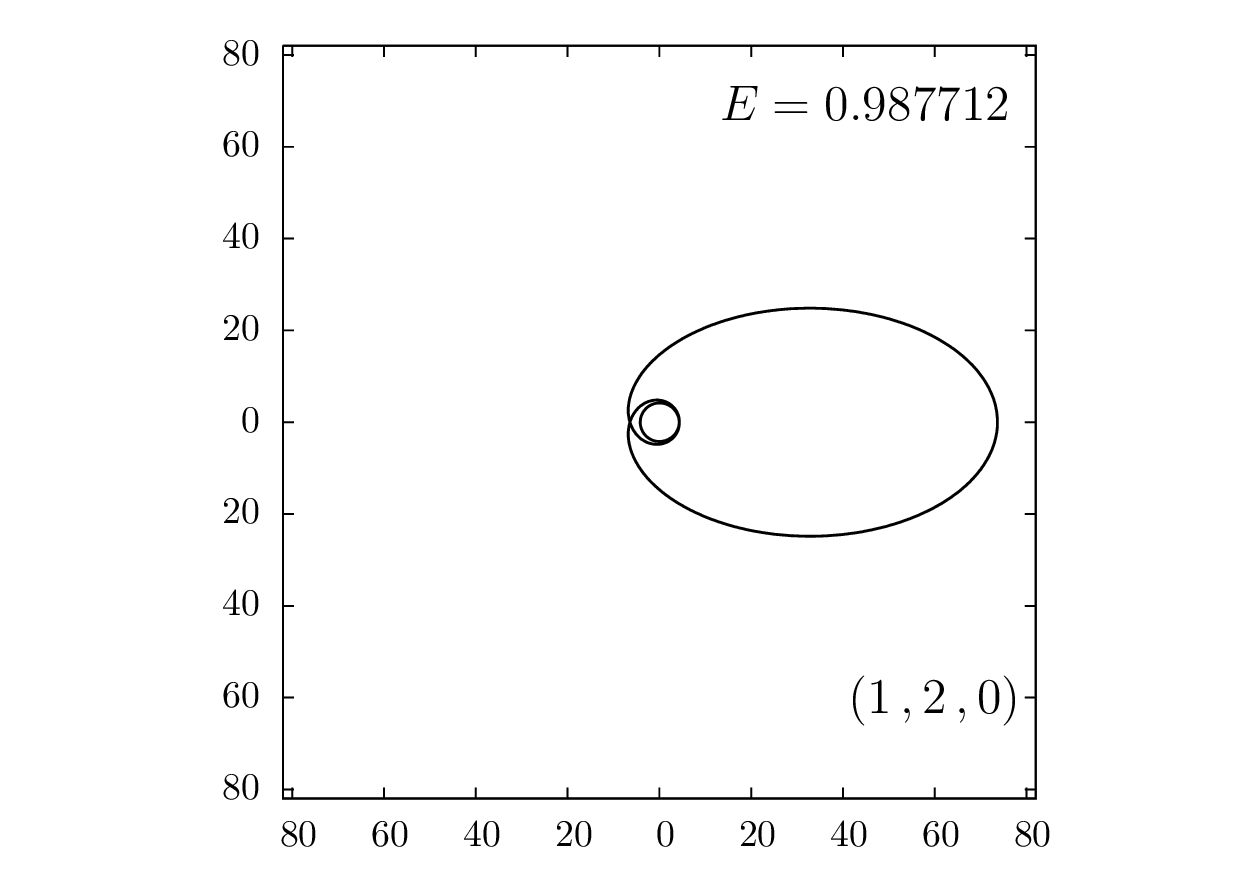}
\hfill
\\
\hspace{-26.5pt}
  \includegraphics[width=0.425\textwidth]{pplots/a0.0/L3.9/bigx.eps}
\hspace{-63.5pt}
  \includegraphics[width=0.425\textwidth]{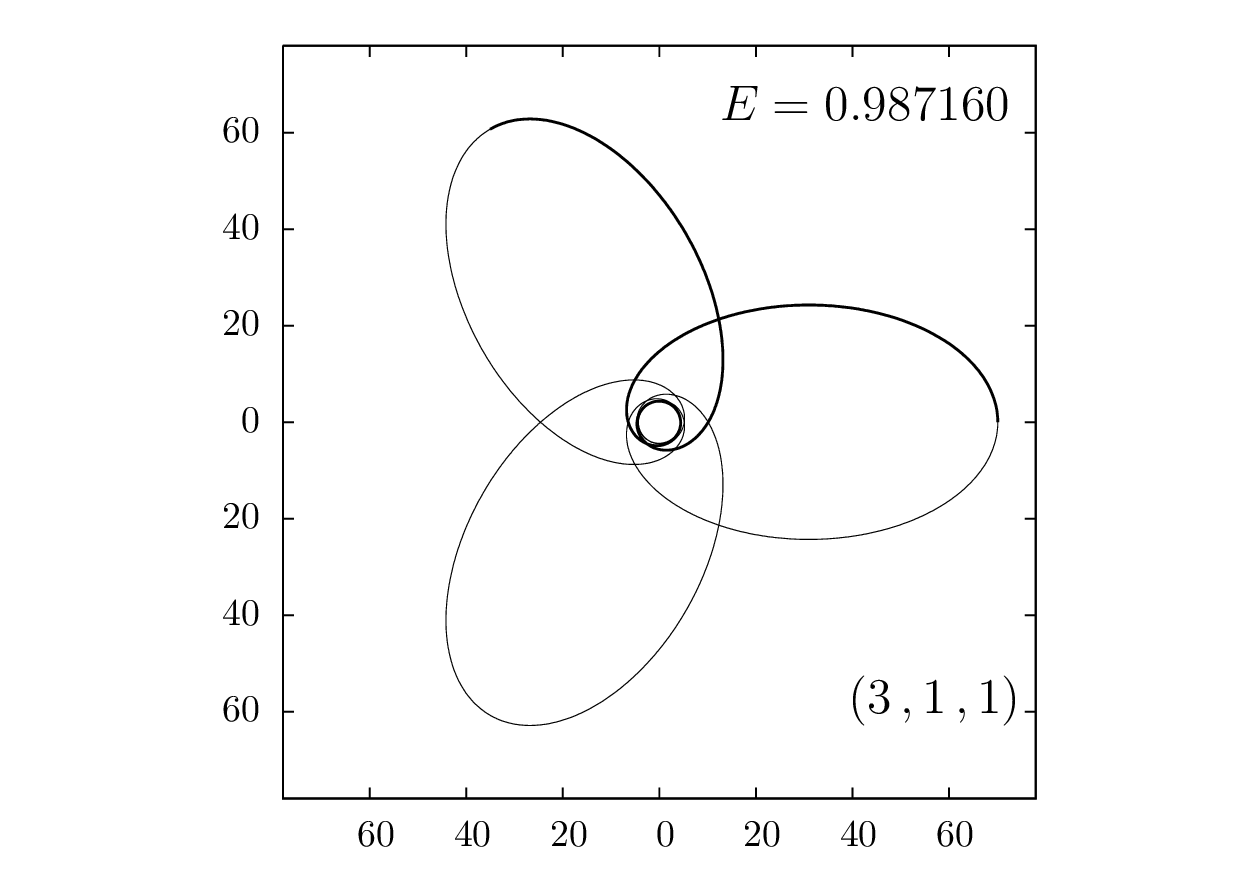}
\hspace{-60pt}
  \includegraphics[width=0.425\textwidth]{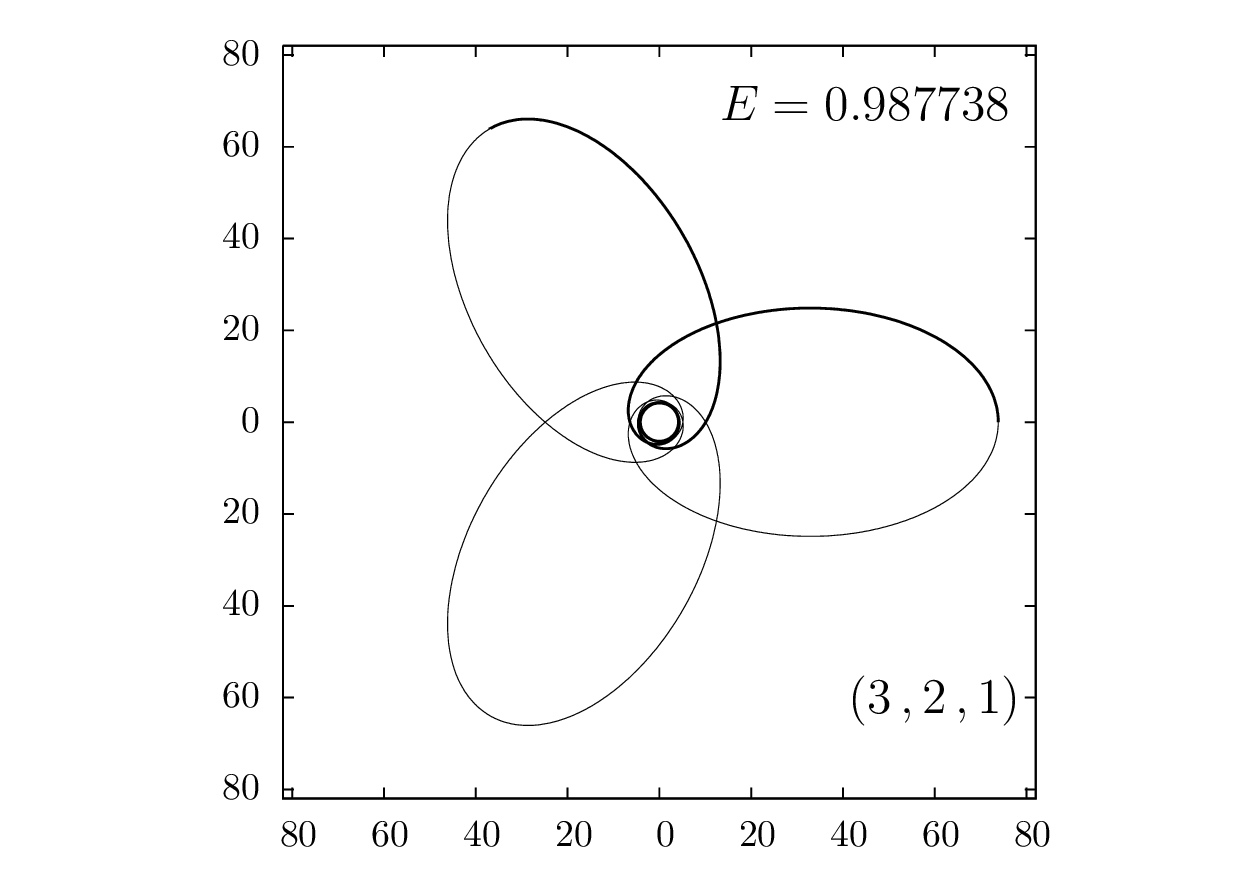}
\hfill
\\
\hspace{-30pt}
  \includegraphics[width=0.425\textwidth]{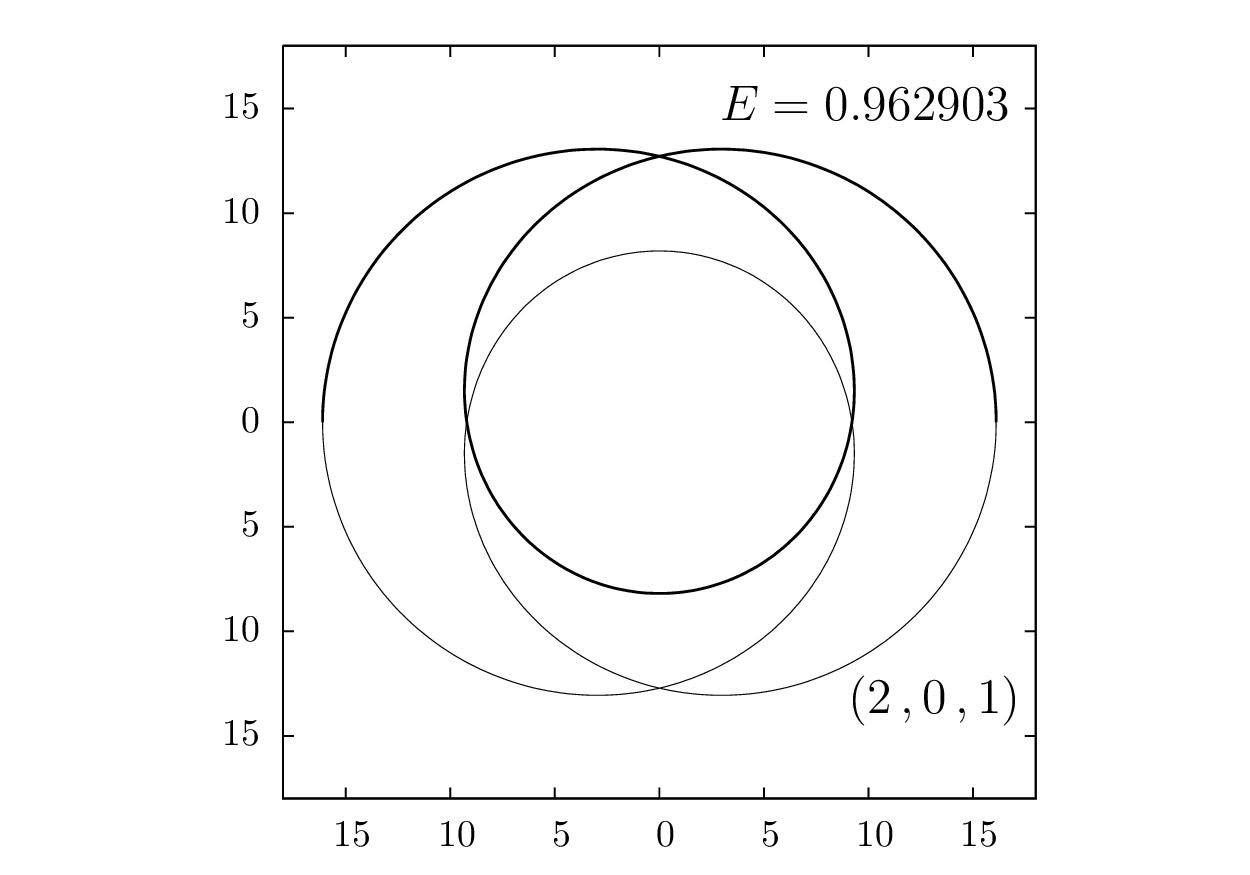}
\hspace{-60pt}
  \includegraphics[width=0.425\textwidth]{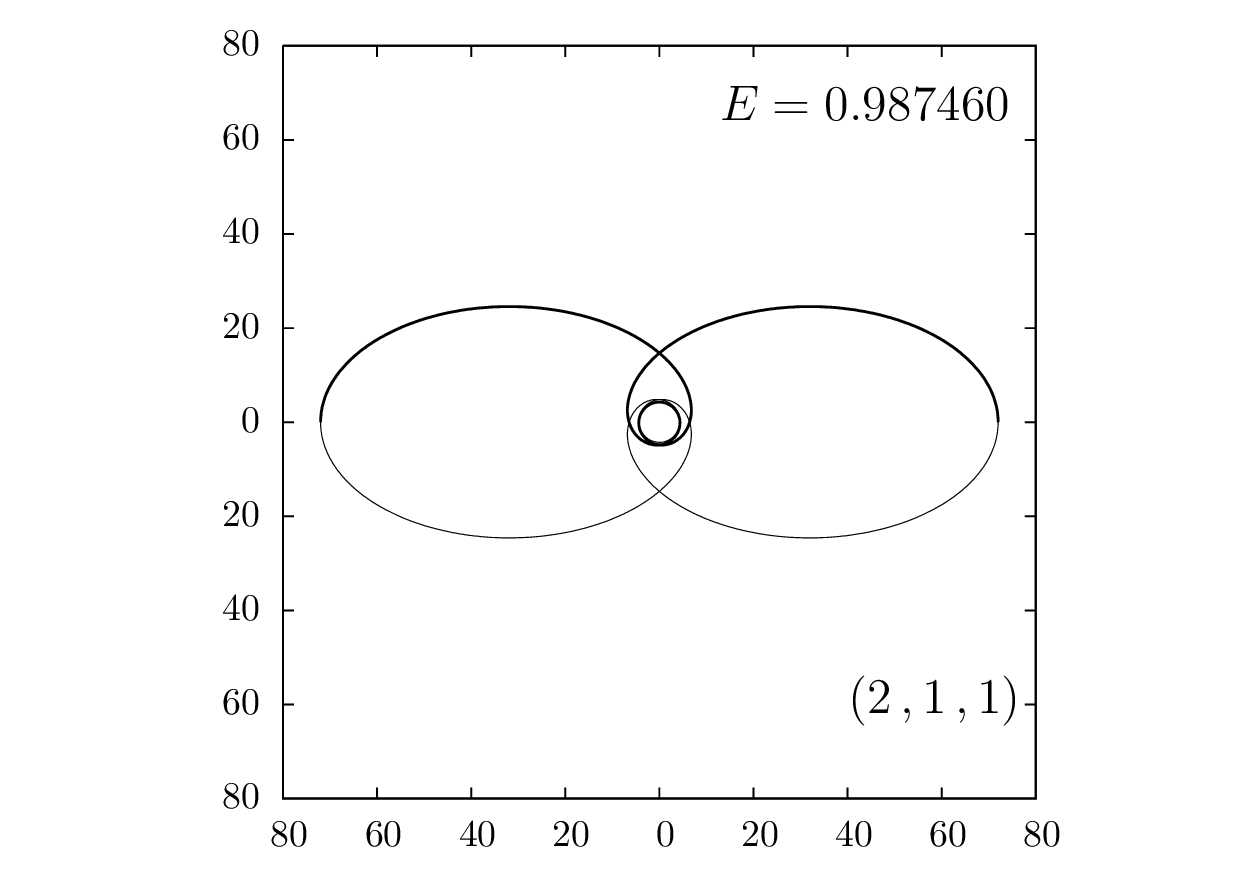}
\hspace{-60pt}
  \includegraphics[width=0.425\textwidth]{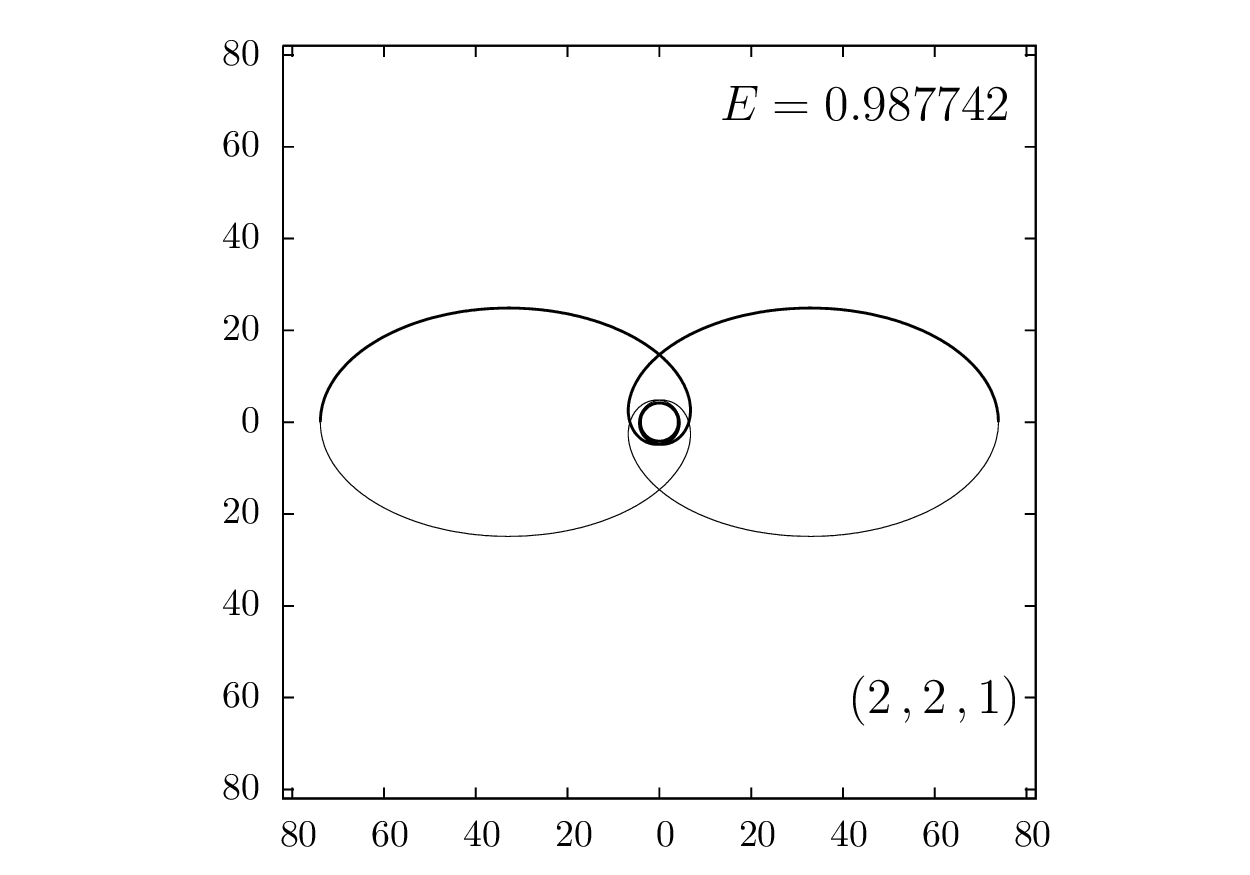}
\hfill
\\
\hspace{-30pt}
  \includegraphics[width=0.425\textwidth]{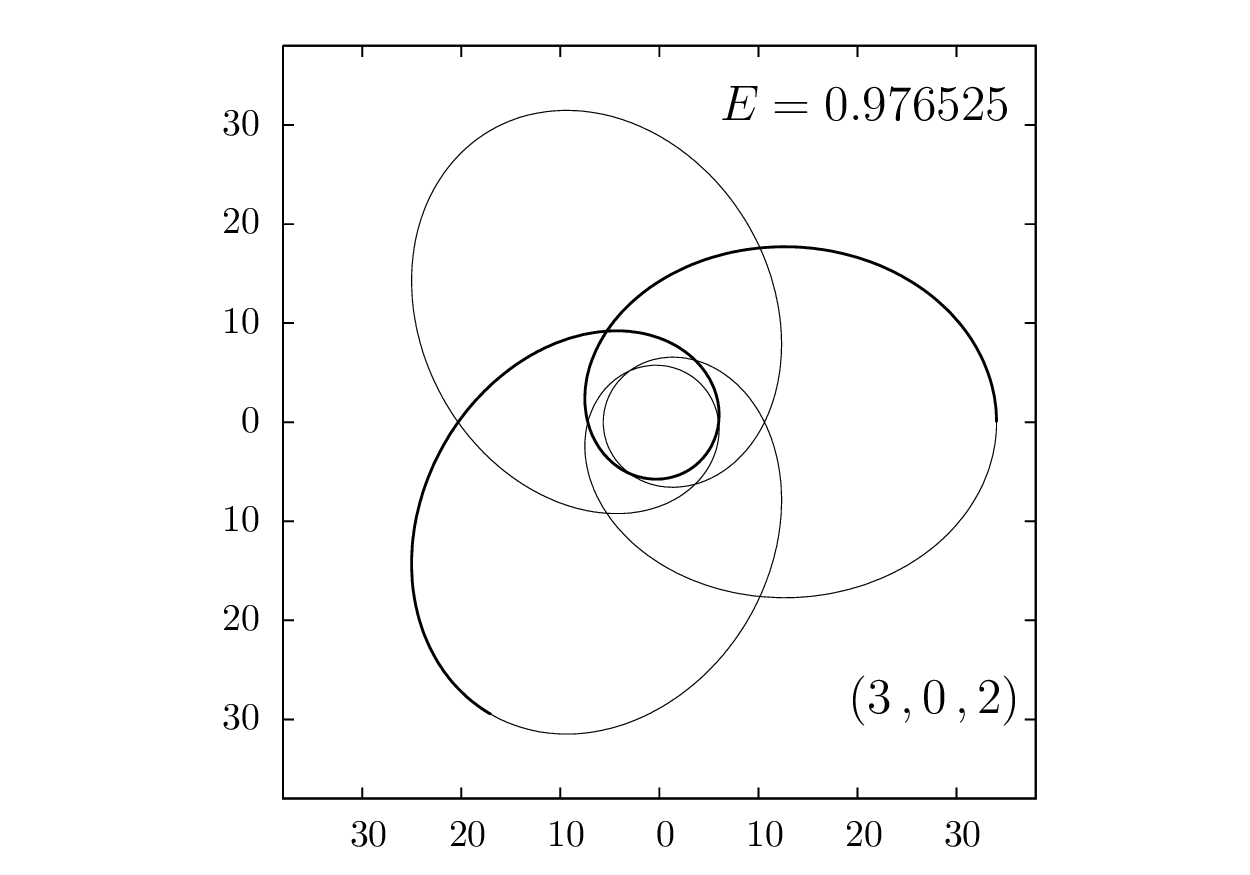}
\hspace{-60pt}
  \includegraphics[width=0.425\textwidth]{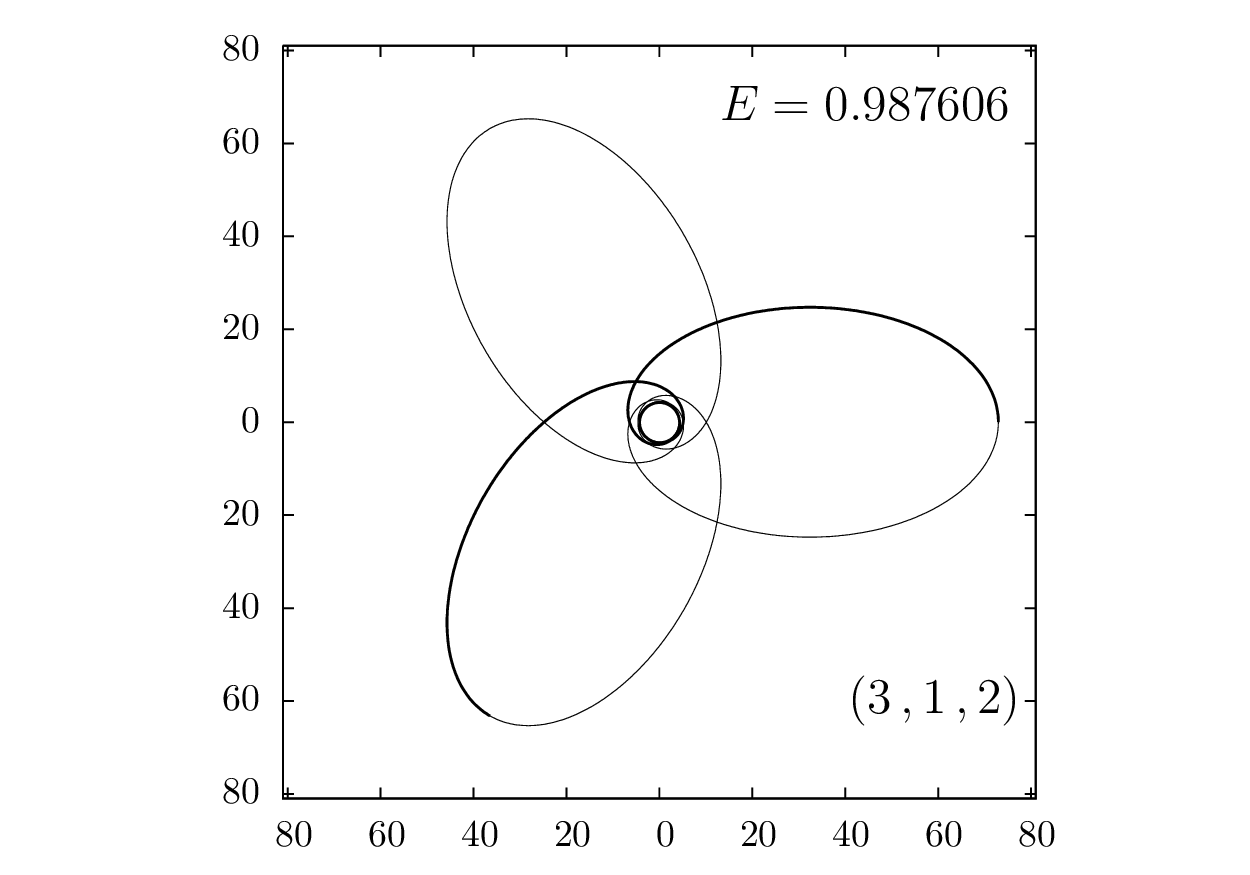}
\hspace{-60pt}
  \includegraphics[width=0.425\textwidth]{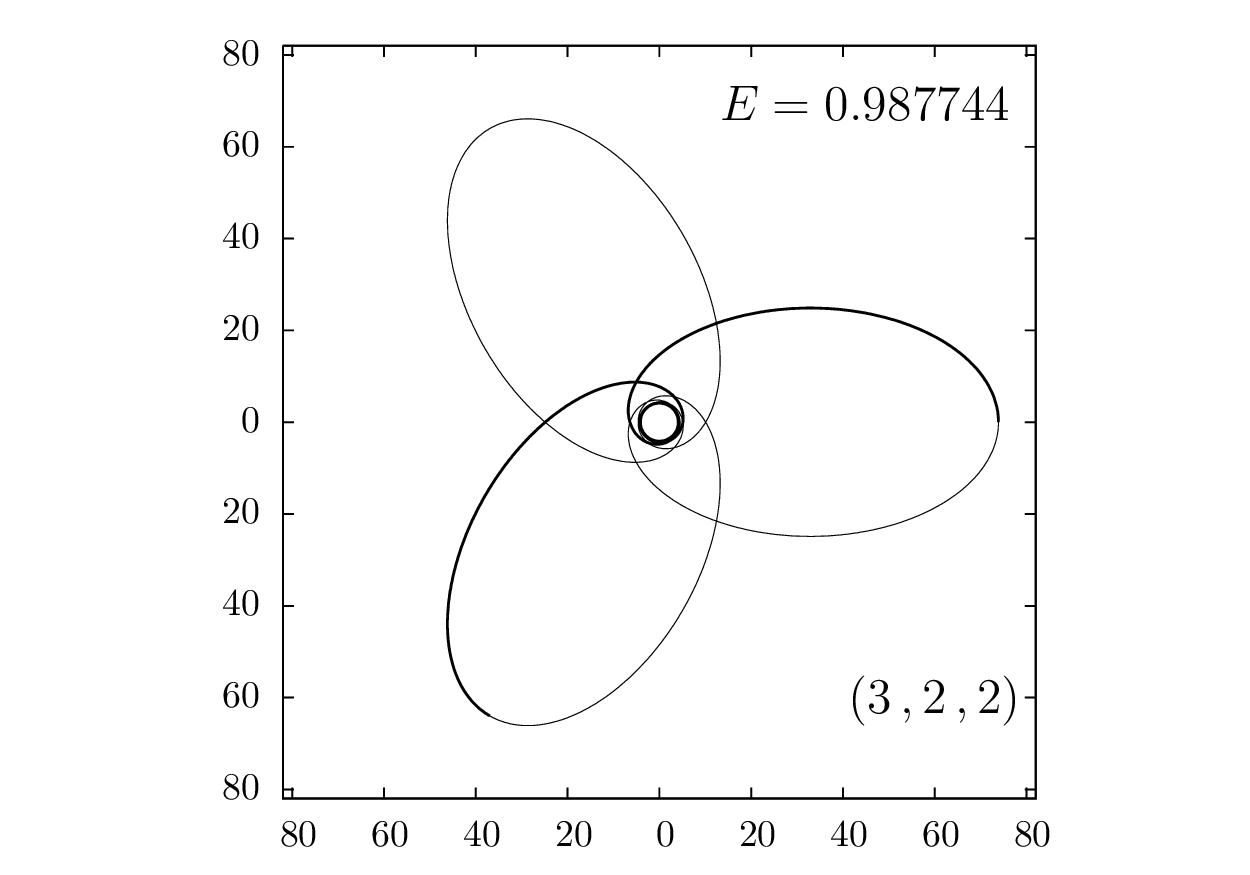}
  \caption{All $z=1,2,3$ orbits with $w=0$ for the first column, $w=1$
  for the middle column, and $w=2$ for the last column. All orbits lie
  on the $L=3.9$ line of {figure \ref{sch}}. Norbits
  increase in energy from top to bottom and left to right. The first
  radial cycle is emphasized in bold for each orbit. Notice the first
  and second entry in the first column are blank, indicating the
  inaccessibility of the $(1,0,0)$ and $(3,0,1)$ orbits.}
  \label{tenS}
\end{figure}

\begin{figure}
  \vspace{-20pt}
  \centering
\hspace{-30pt} 
 \includegraphics[width=0.45\textwidth]{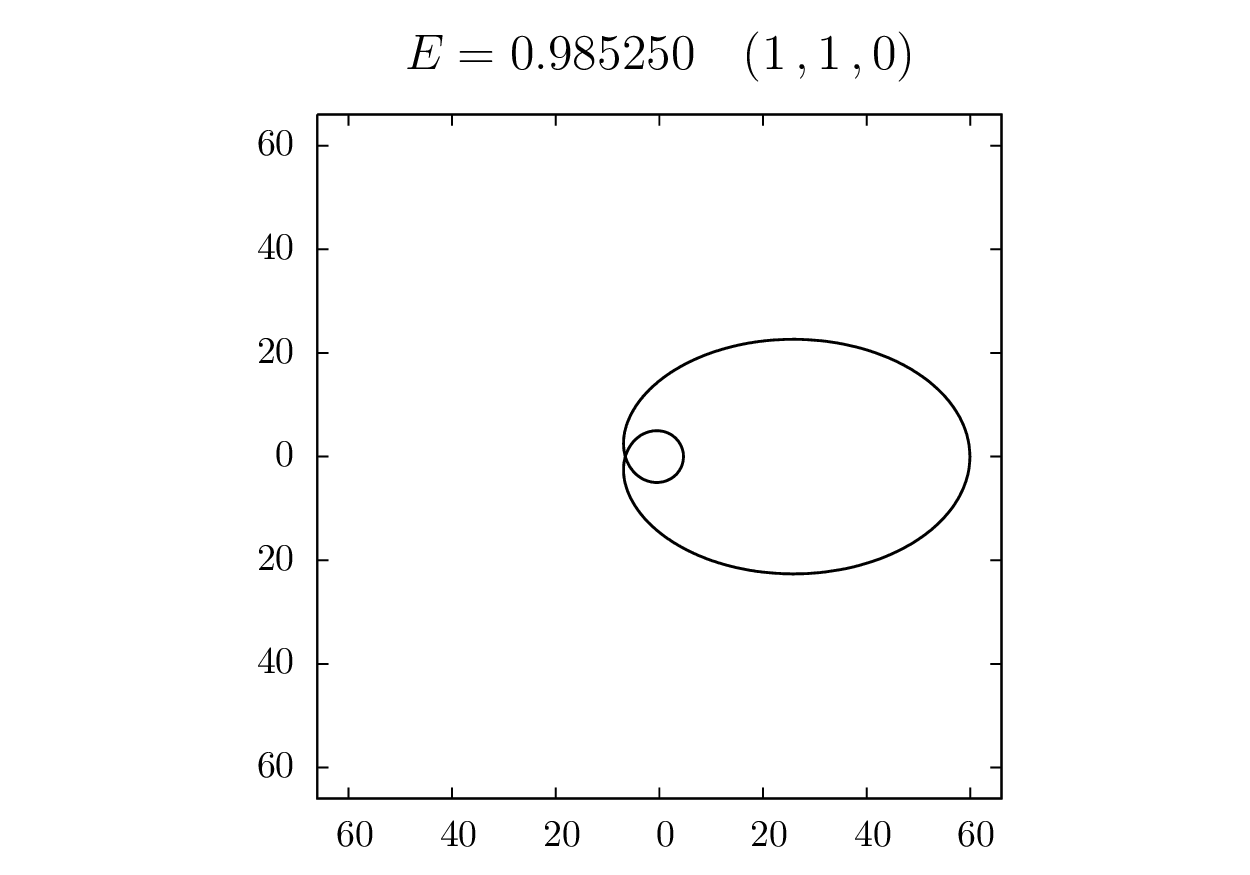}
\hspace{-80pt}
  \includegraphics[width=0.45\textwidth]{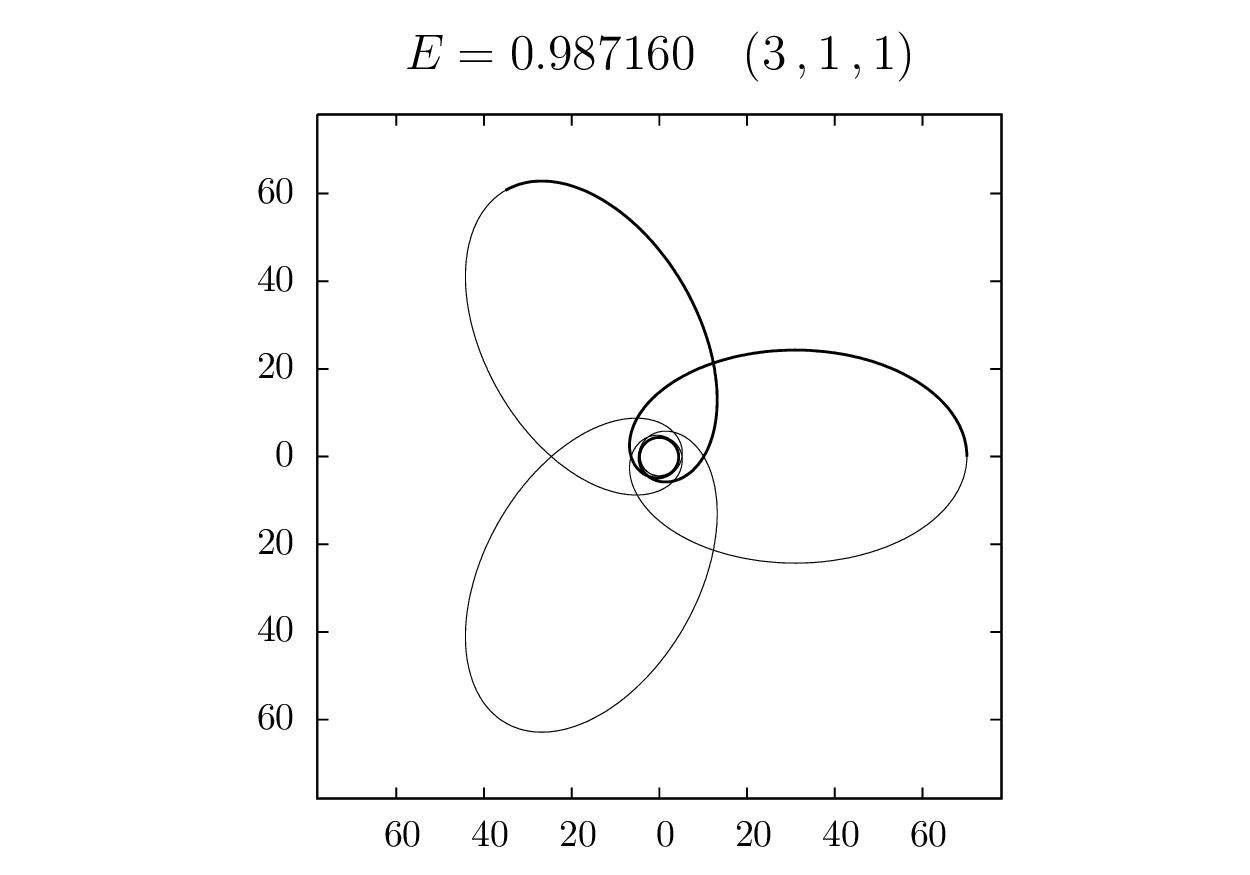} 
\hspace{-80pt}
  \includegraphics[width=0.45\textwidth]{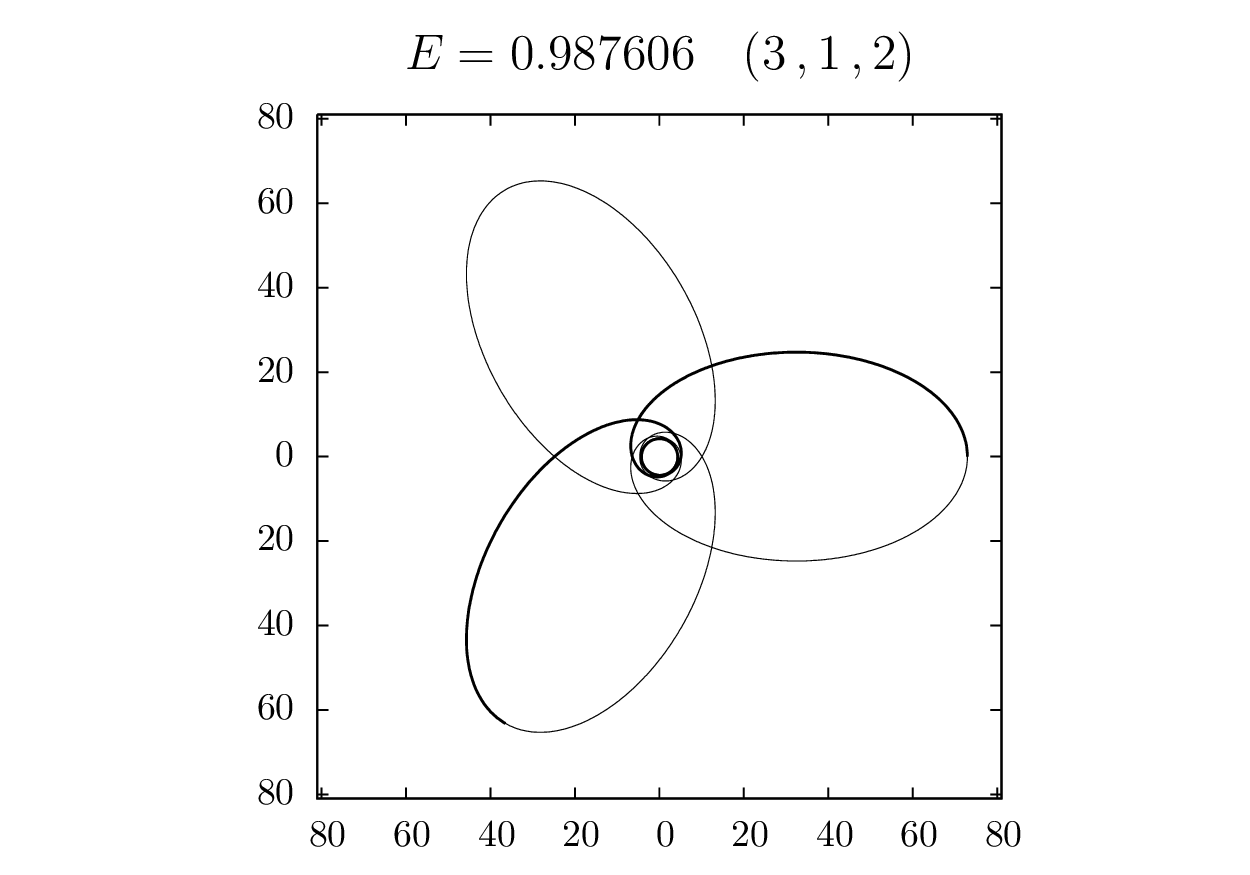}
\hfill
\\
\hspace{-26pt}
  \includegraphics[width=0.45\textwidth]{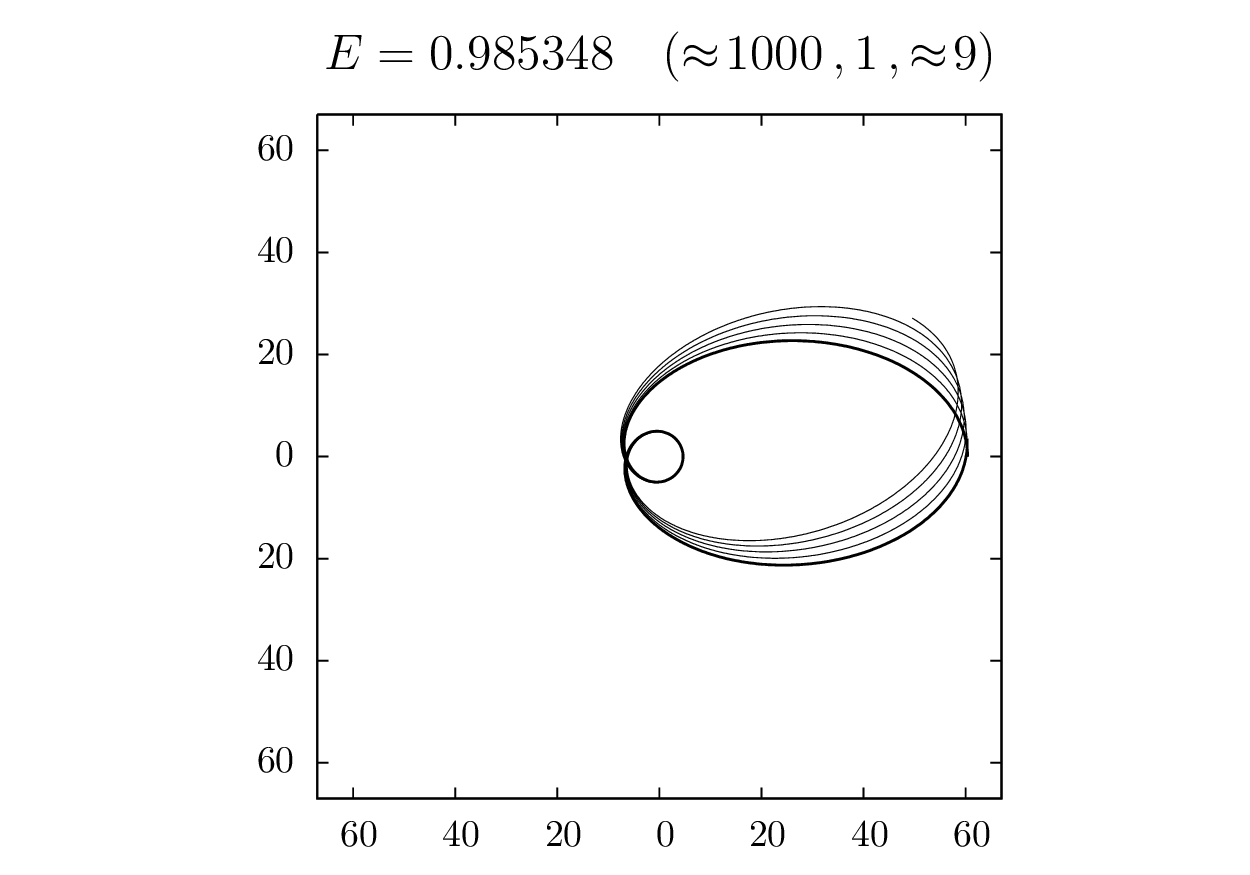}
\hspace{-80pt}
  \includegraphics[width=0.45\textwidth]{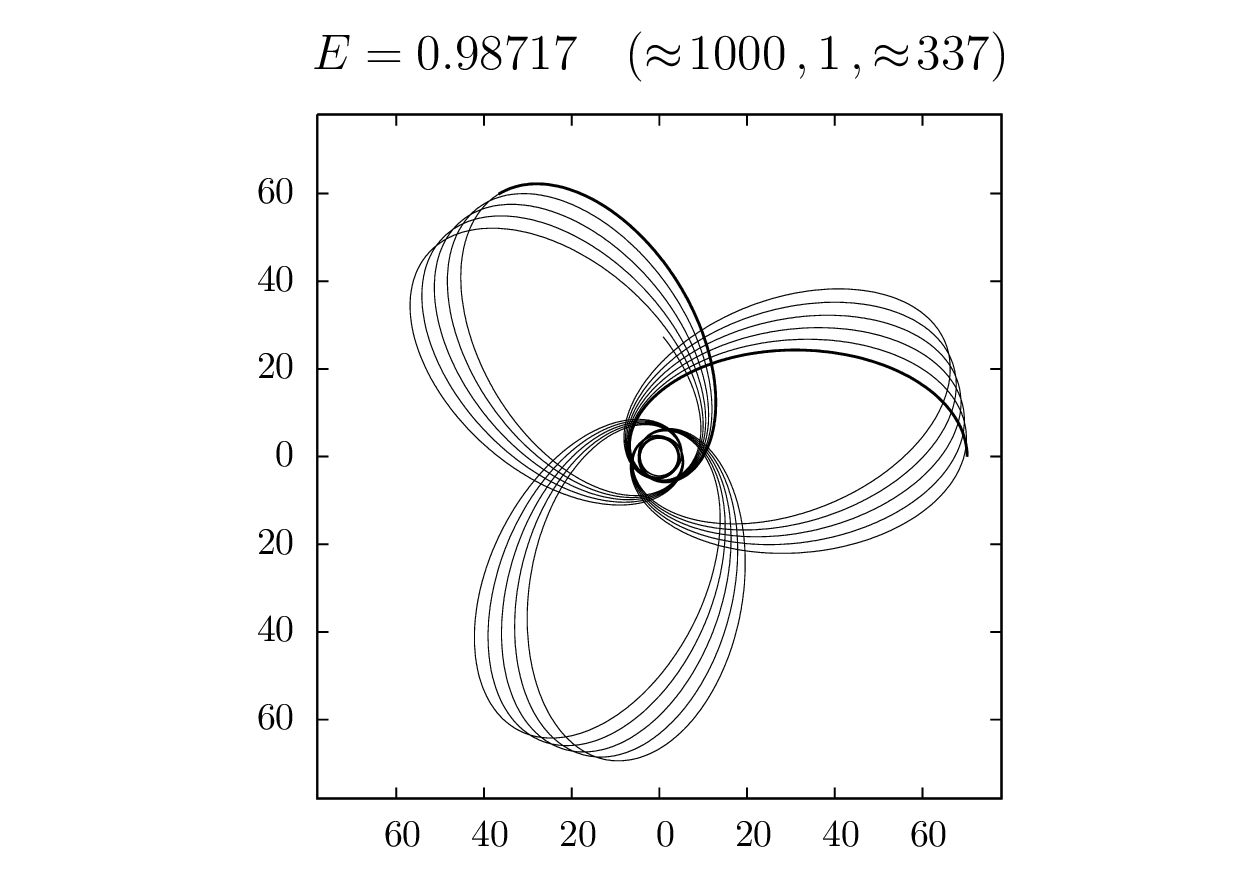}
\hspace{-80pt}
  \includegraphics[width=0.45\textwidth]{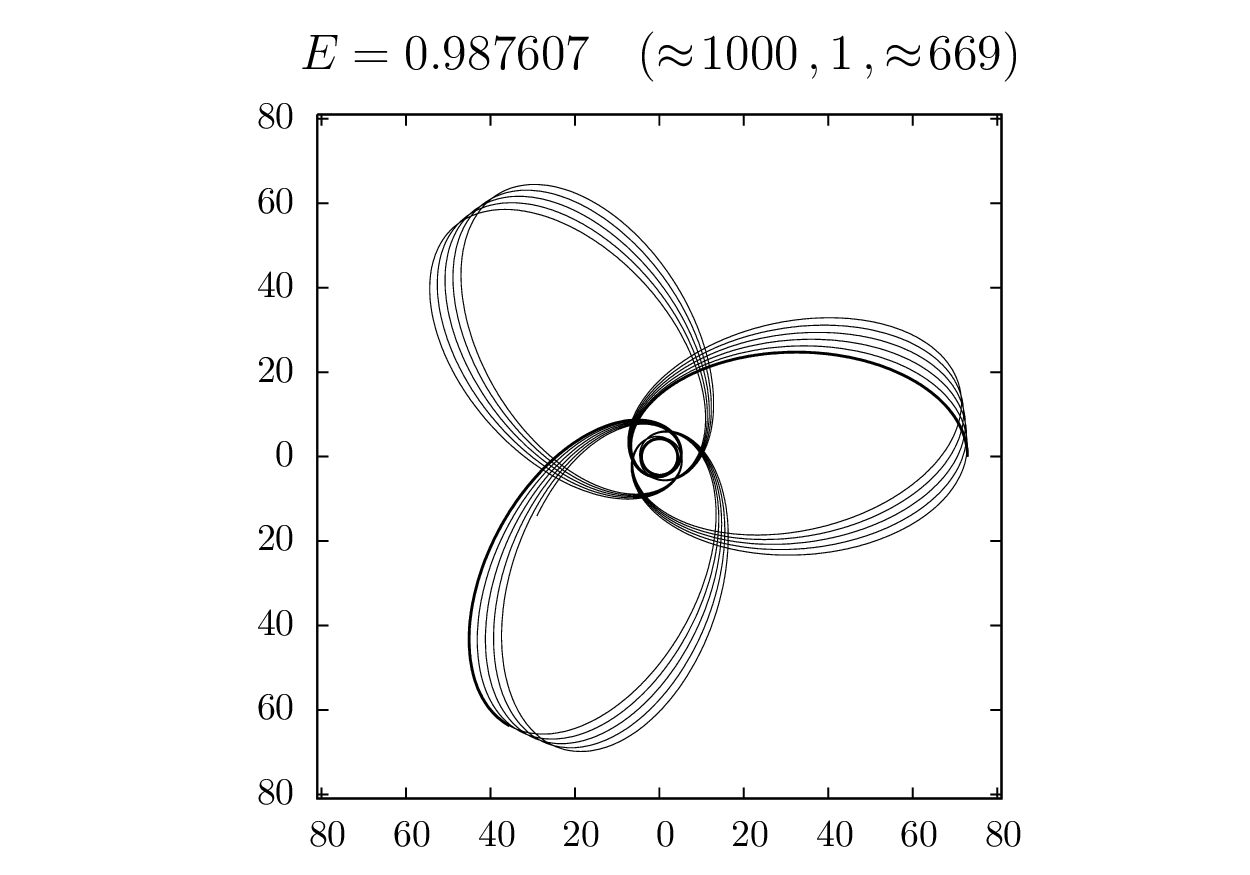}
\hfill
\\
\hspace{-24pt}
  \includegraphics[width=0.45\textwidth]{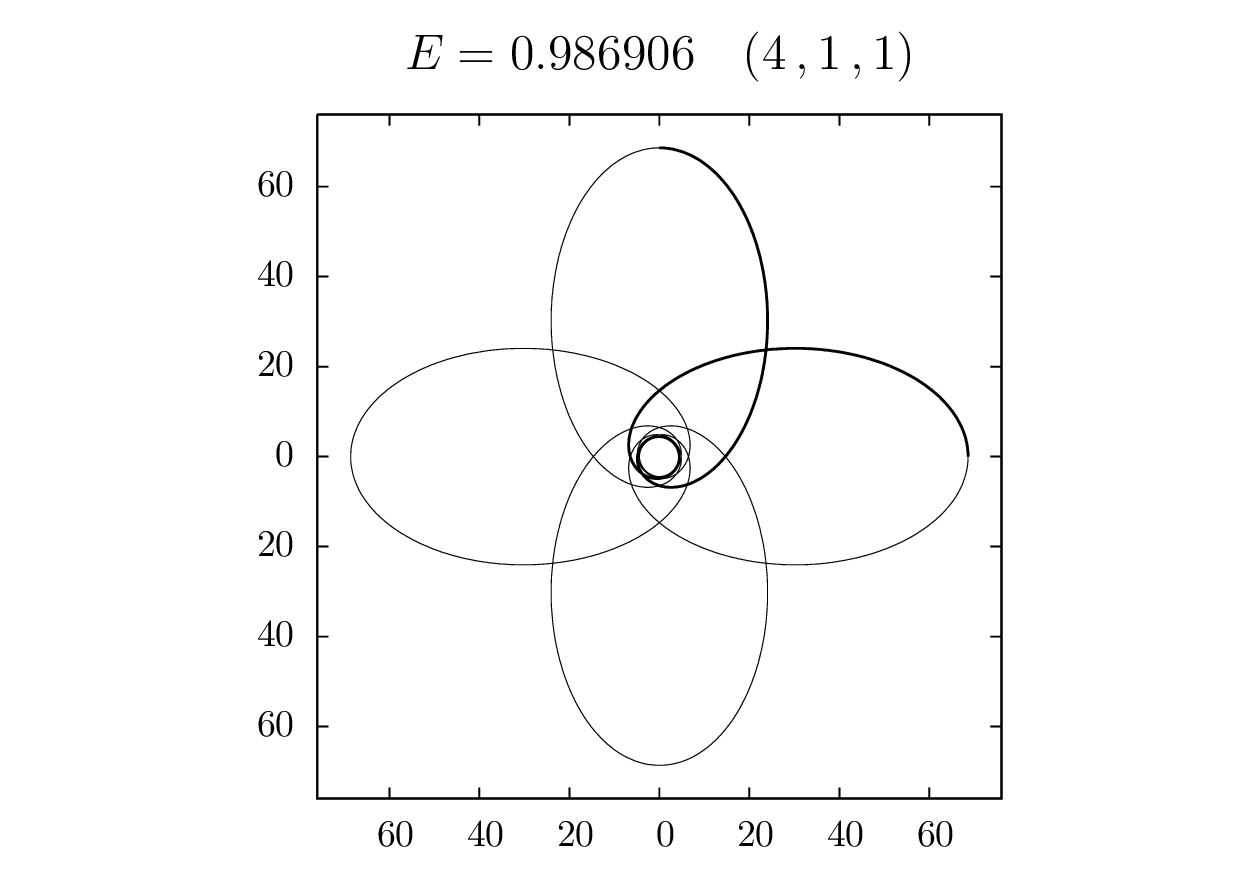}
\hspace{-80pt}
  \includegraphics[width=0.45\textwidth]{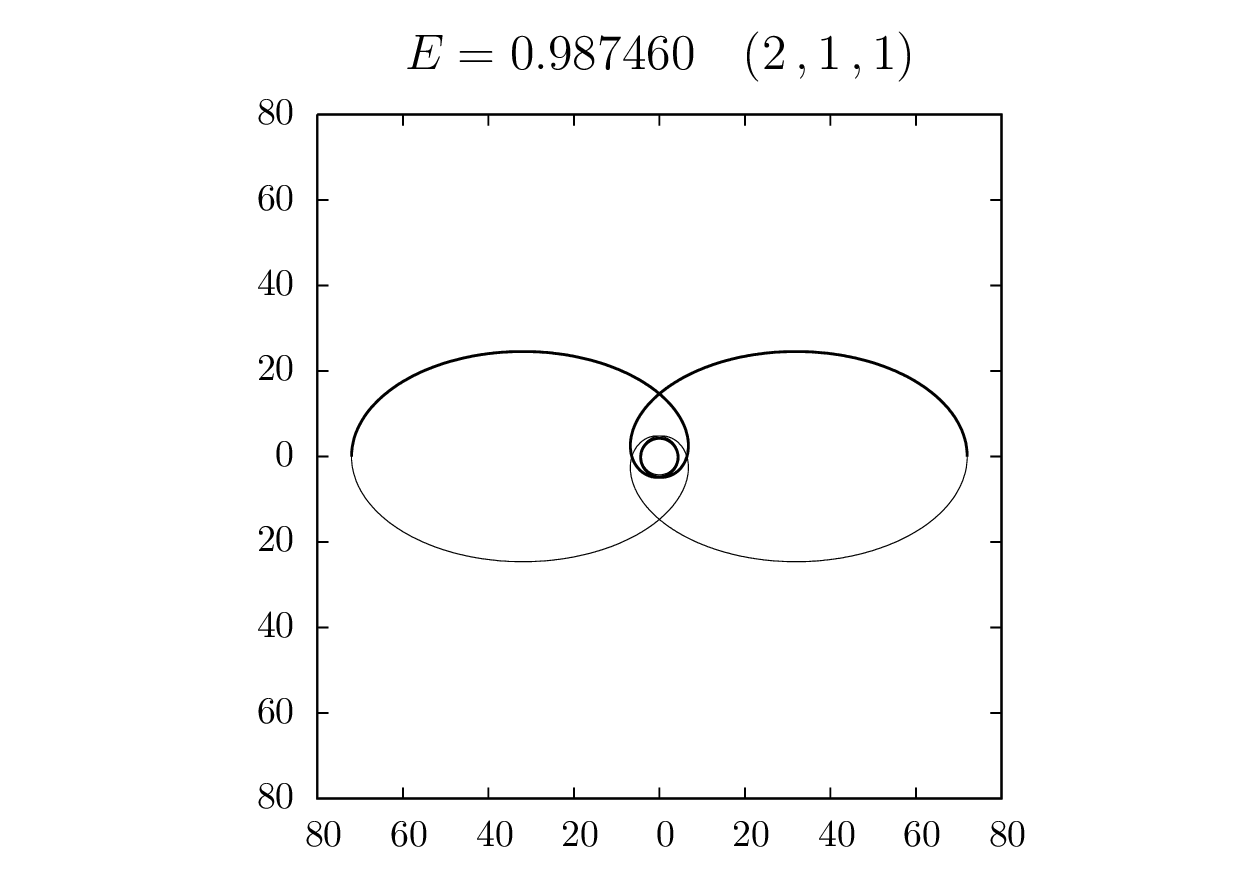}
\hspace{-80pt}
  \includegraphics[width=0.45\textwidth]{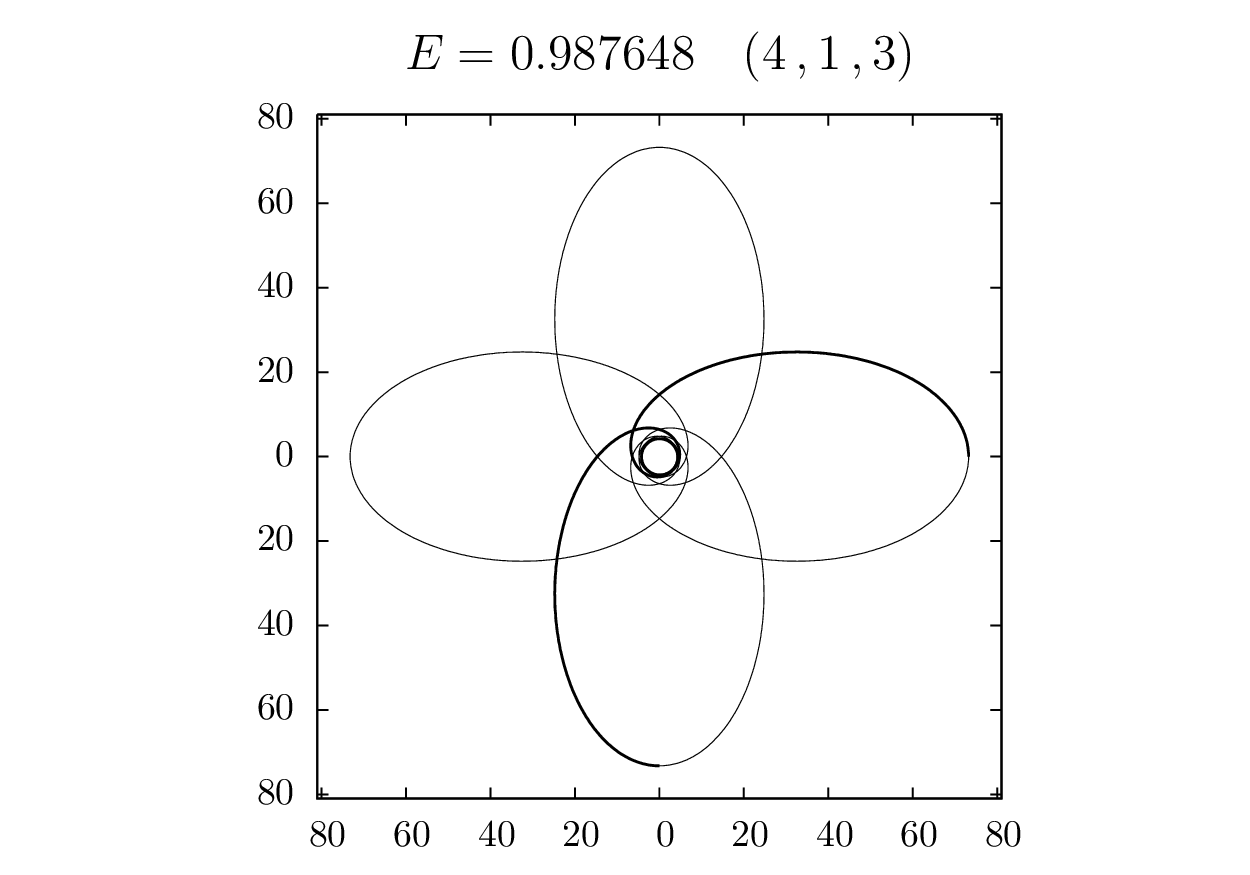}
\hfill
\\
\hspace{-22pt}
  \includegraphics[width=0.45\textwidth]{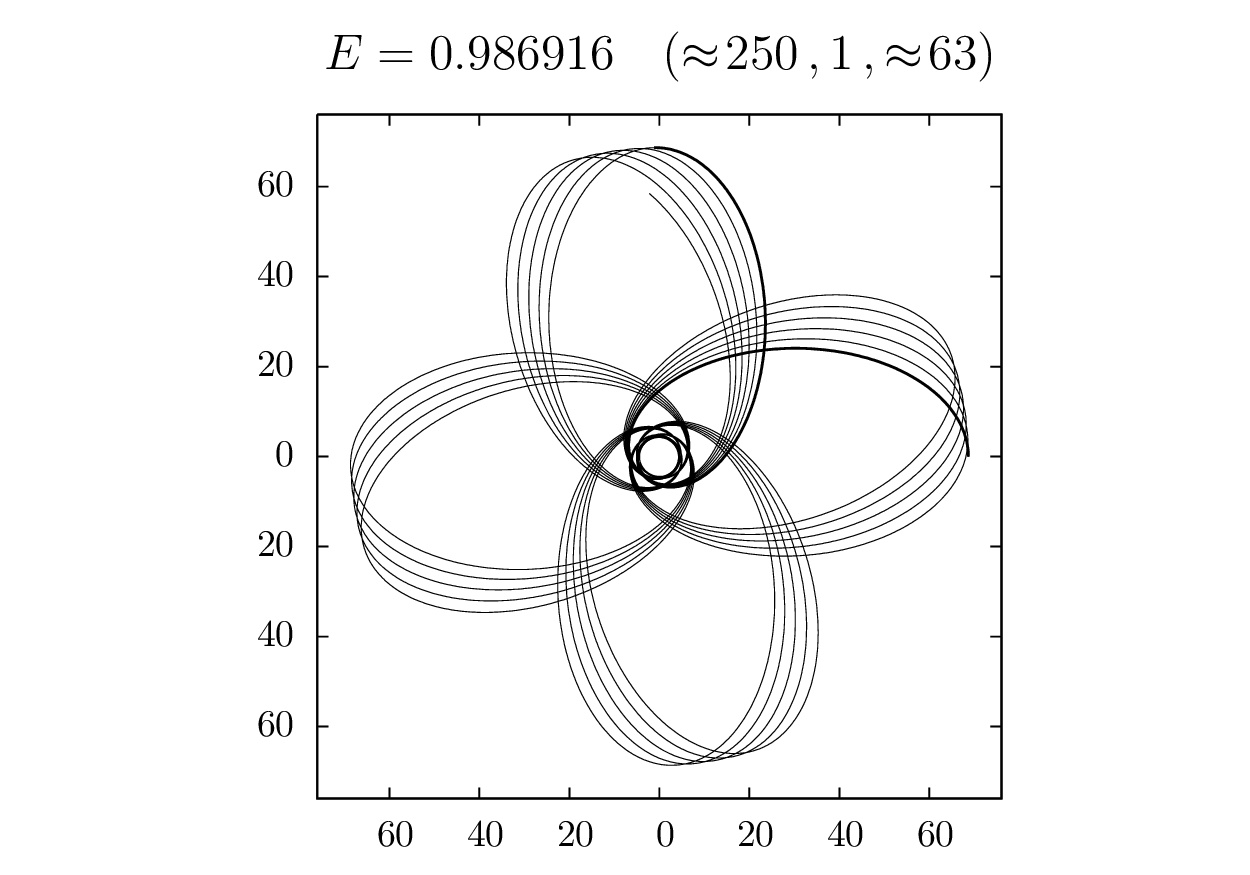}
\hspace{-80pt}
  \includegraphics[width=0.45\textwidth]{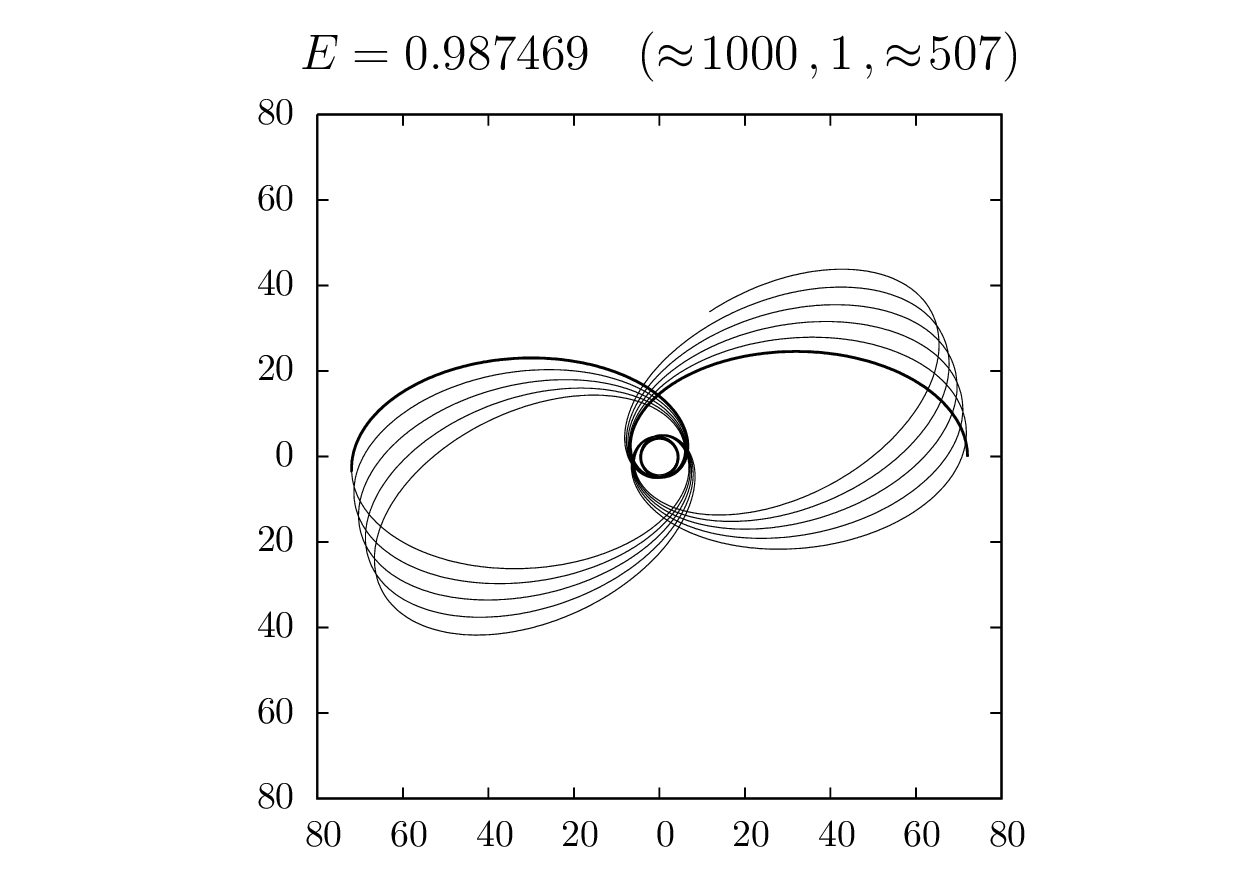}
\hspace{-80pt}
  \includegraphics[width=0.45\textwidth]{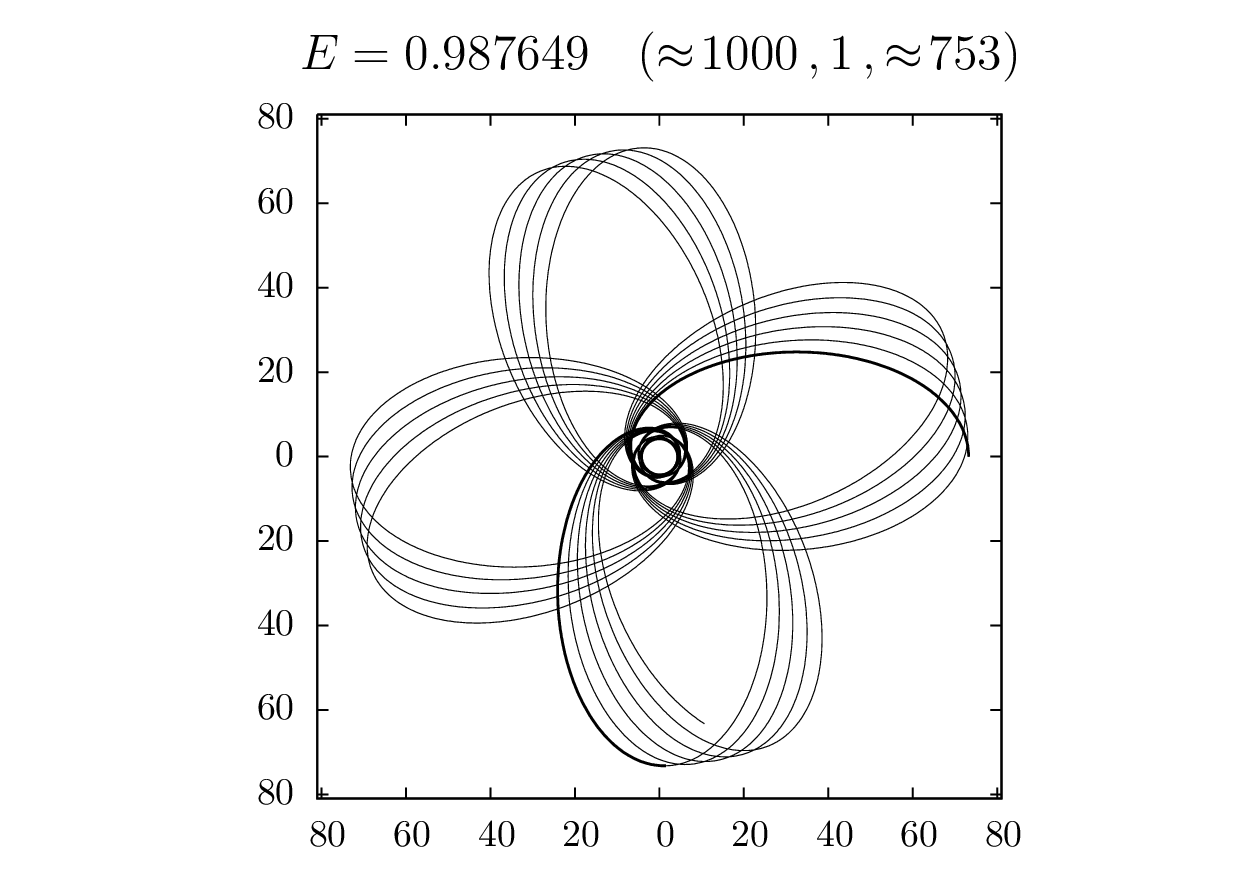}
\hfill
  \caption{A series of $w=1$ orbits for the $L=3.9$ line of {figure 
  \ref{sch}}. Orbits increase in energy from top to bottom and left to
  right. All $z=1,2,3,4$ orbits are shown. Also shown are randomly
  selected high $z$ orbits. Notice that the high $z$ orbits look like
  precessions of the energetically closest low $z$ orbit.}
  \label{twelveS}
\end{figure}

Finally, figure \ref{tenS} shows a periodic table of orbits (still for
$L=3.9$). For
illustration, we only inlcude orbits with $z$ up to $3$, although the
spectrum allows $z\rightarrow \infty$ orbits. The first column
corresponds to $w=0$, the middle column to $w=1$ and the last column to
$w=2$. The series could be continued through to $w\rightarrow
\infty$. The sequence should be read from top to bottom and then from
left to right to indicate 
increasing energy.
Notice that the first entry in the sequence of
periodic orbits is blank to indicate the complete absence of the
$(1,0,0)$ orbit as discussed above. The second entry in the sequence
is also blank indicating the complete absence of the $(3,0,1)$ orbit --
as figure \ref{sch} shows, $q$ never drops to $1/3$.

 Figure \ref{twelveS}, which was previewed in the introduction, shows
 the progression through orbits with higher  
numbers of zooms.
All orbits are drawn from the $w=1$ band -- the middle column 
of figure \ref{tenS}. As before they are arranged in order of increasing
 energy from top to bottom and left to right. All orbits with
 $z=1,2,3,4$ are drawn. Between each of these low leaf orbits, randomly
 selected high zoom orbits are shown as well. The high zoom orbits
 look like precessions of the low zoom orbits. 

Notice that for Schwarzschild the vast majority of low leaf orbits are
stacked at high 
eccentricity. 
Orbits around spinning black
holes, by contrast, show a 
wider spread of low $z$ orbits at low eccentricities as we now demonstrate.

\subsection{The Kerr zoo}
\label{zooK}

Kerr orbits do not admit a simple one-dimensional effective
potential description.\footnote{There is an unappetizing effective
  potential description for equatorial orbits that we do not want to
  rely upon here.}
Despite this complication, the generic Kerr system (appendix \S
\ref{kerreqs}) has all the key
features we need for our taxonomy. For any $L$, there is
always one stable circular orbit. At some critical $L$ a first bound
unstable circular orbit will appear bringing with it an associated
homoclinic orbit. The unstable circular orbit always has energy
higher than that of the stable circular orbit. For a given
$L_{ISCO}<L<L_{IBCO}$, the energy 
continues to specify uniquely orbits ranging from the 
stable circular to the homoclinc.
Therefore, the same graphical analysis that we used to determine the
population of the Schwarzschild zoo can be used to determine the
population of the Kerr zoo.

\begin{figure}
\centering
\hspace{-95pt}
\subfigure[$w+v/z$ versus energy.] 
{
 \label{wvzeK}
    \includegraphics[width=12.4cm]{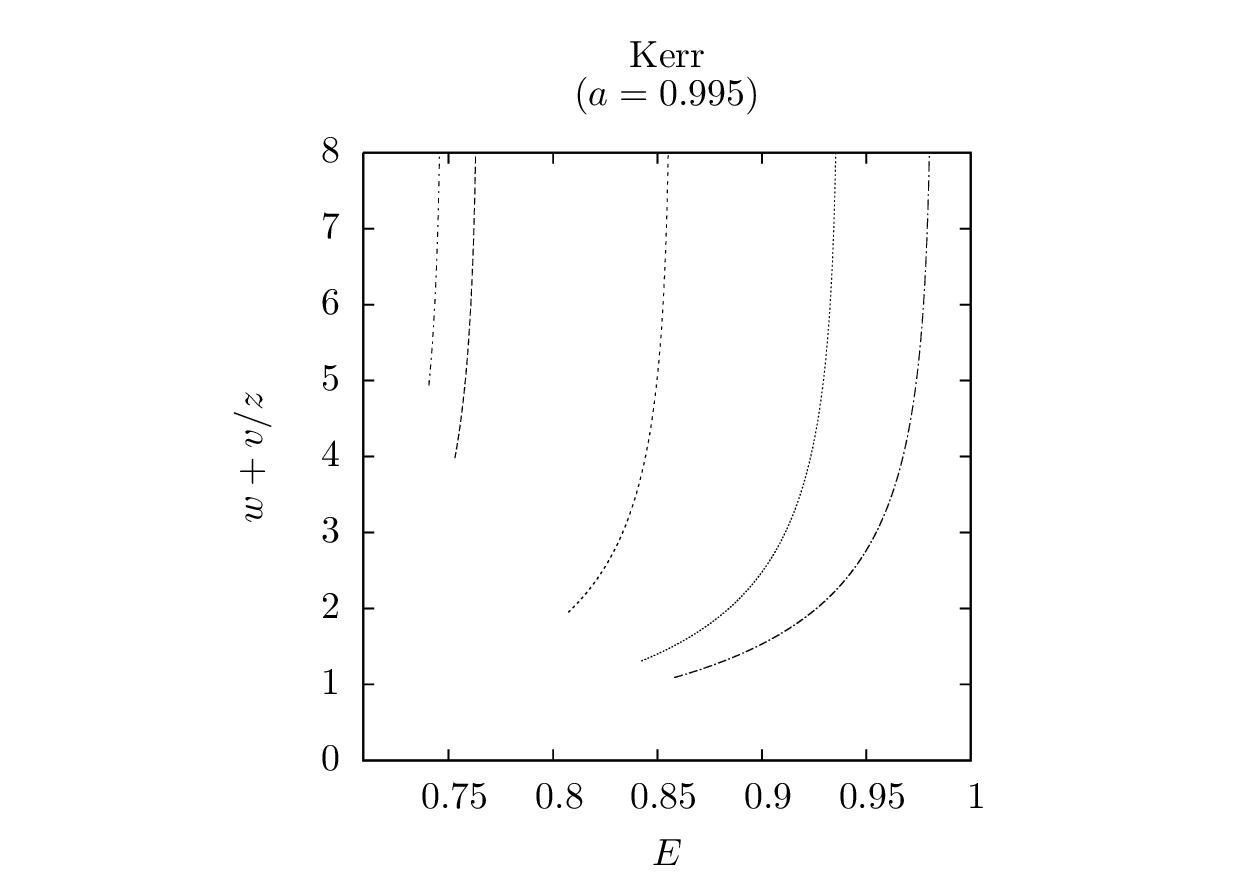}
}
\centering
\hspace{-165pt}
\subfigure[$w+v/z$ versus eccentricity.] 
{
 \label{wvzeccK}
    \includegraphics[width=12.4cm]{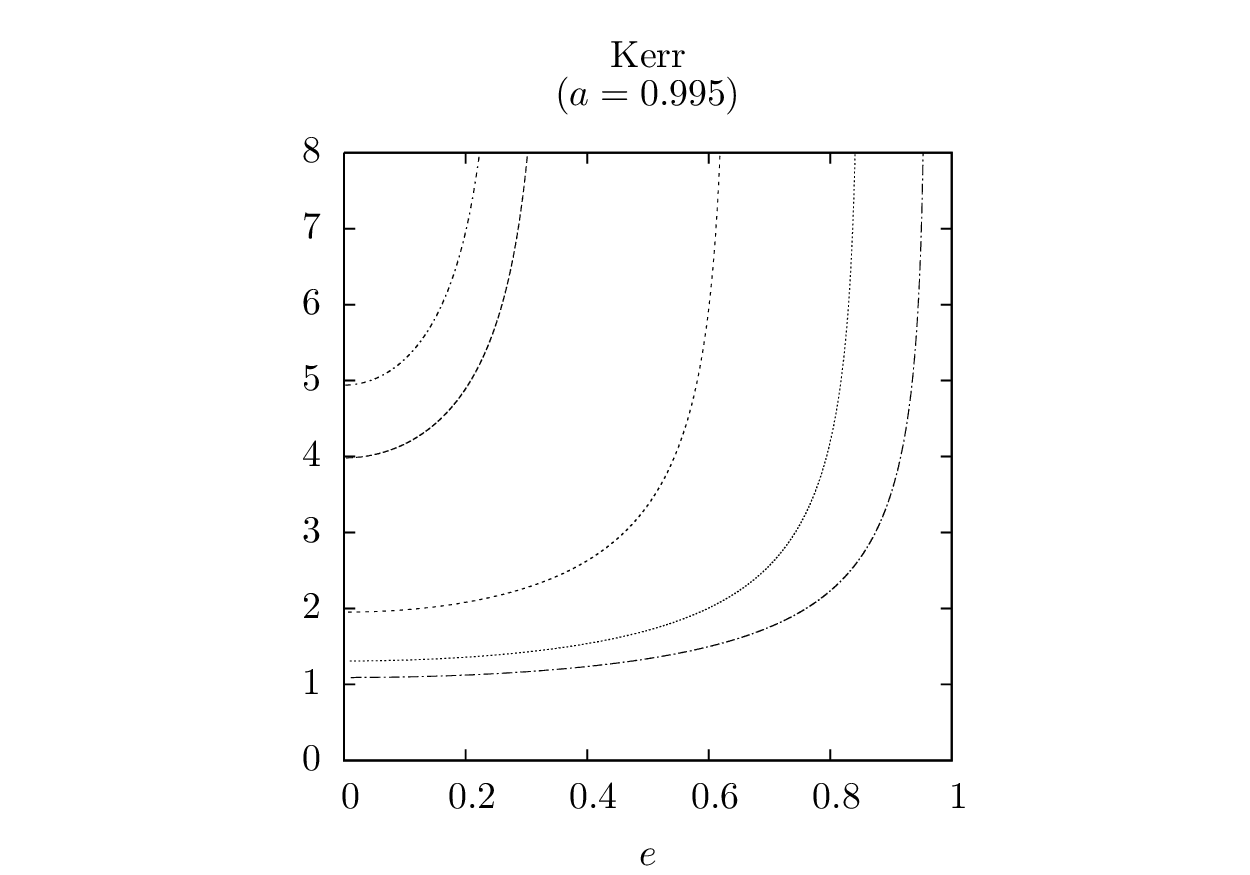}
}
\hspace{-60pt}
\caption{Prograde orbits around a spinning black hole with $a=0.995$.
The lines indicate increasing $L$ from left to right through the
values $L=1.57,1.61,1.82,2,2.1$
\label{kerr}}
\end{figure}

\begin{figure}
  \vspace{-20pt}
  \centering
\hspace{-30pt} 
 \includegraphics[width=0.45\textwidth]{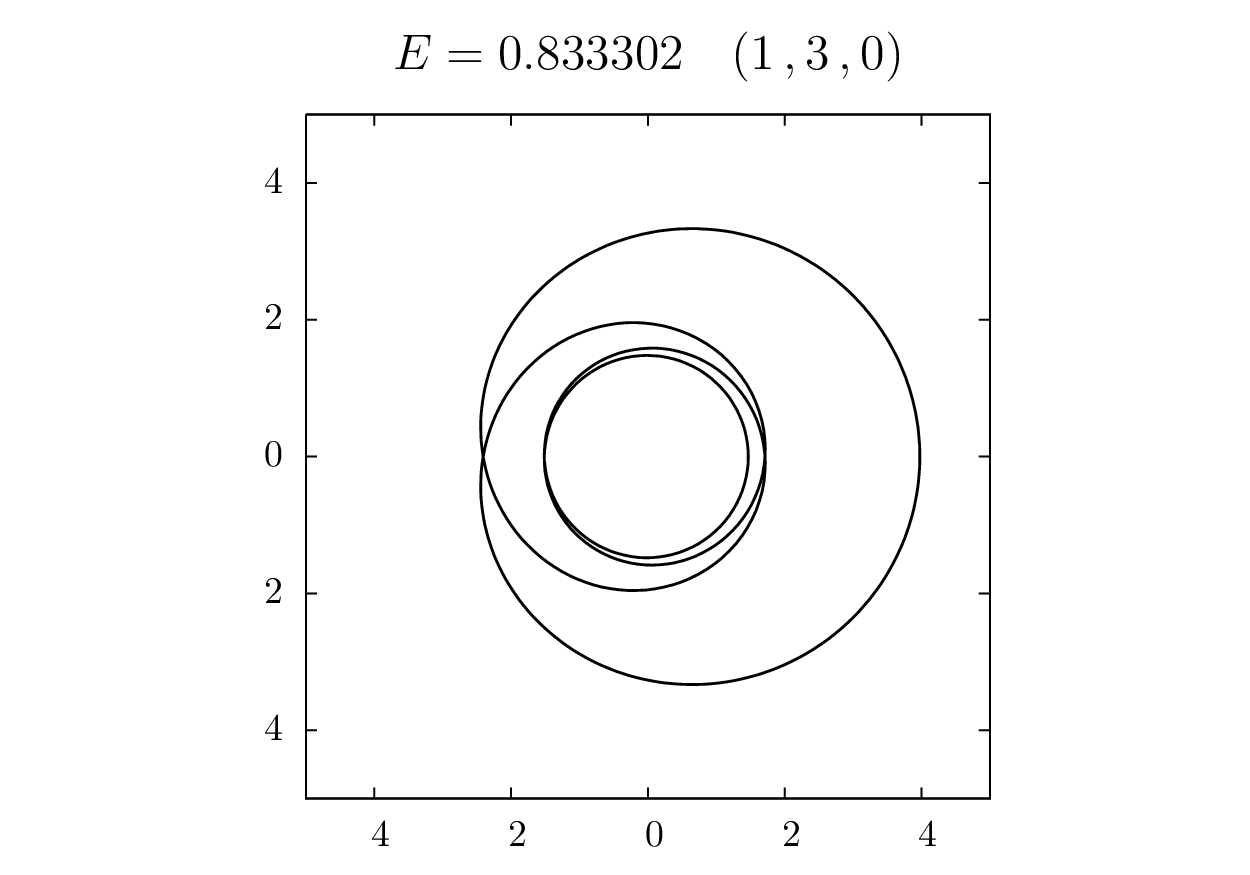}
\hspace{-80pt}
  \includegraphics[width=0.45\textwidth]{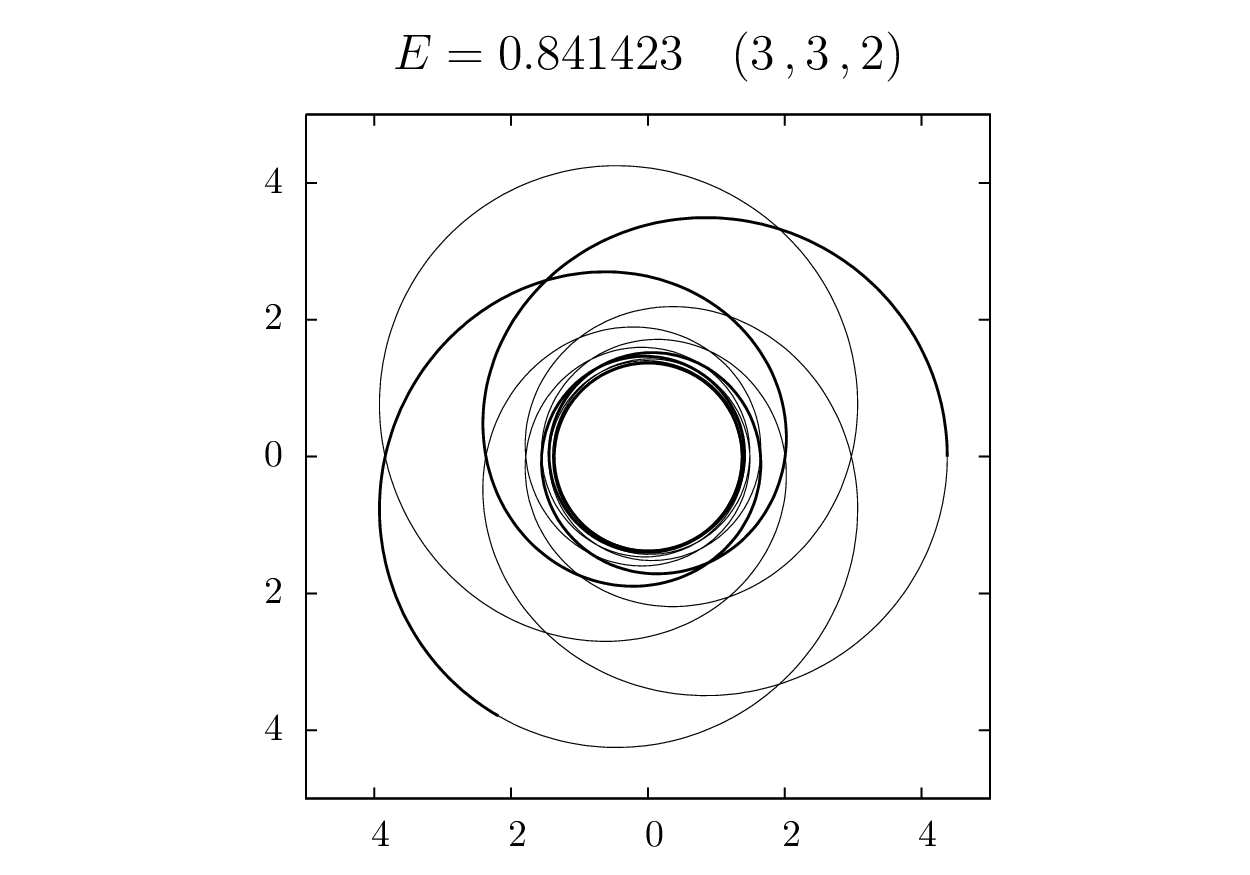} 
\hspace{-80pt}
  \includegraphics[width=0.45\textwidth]{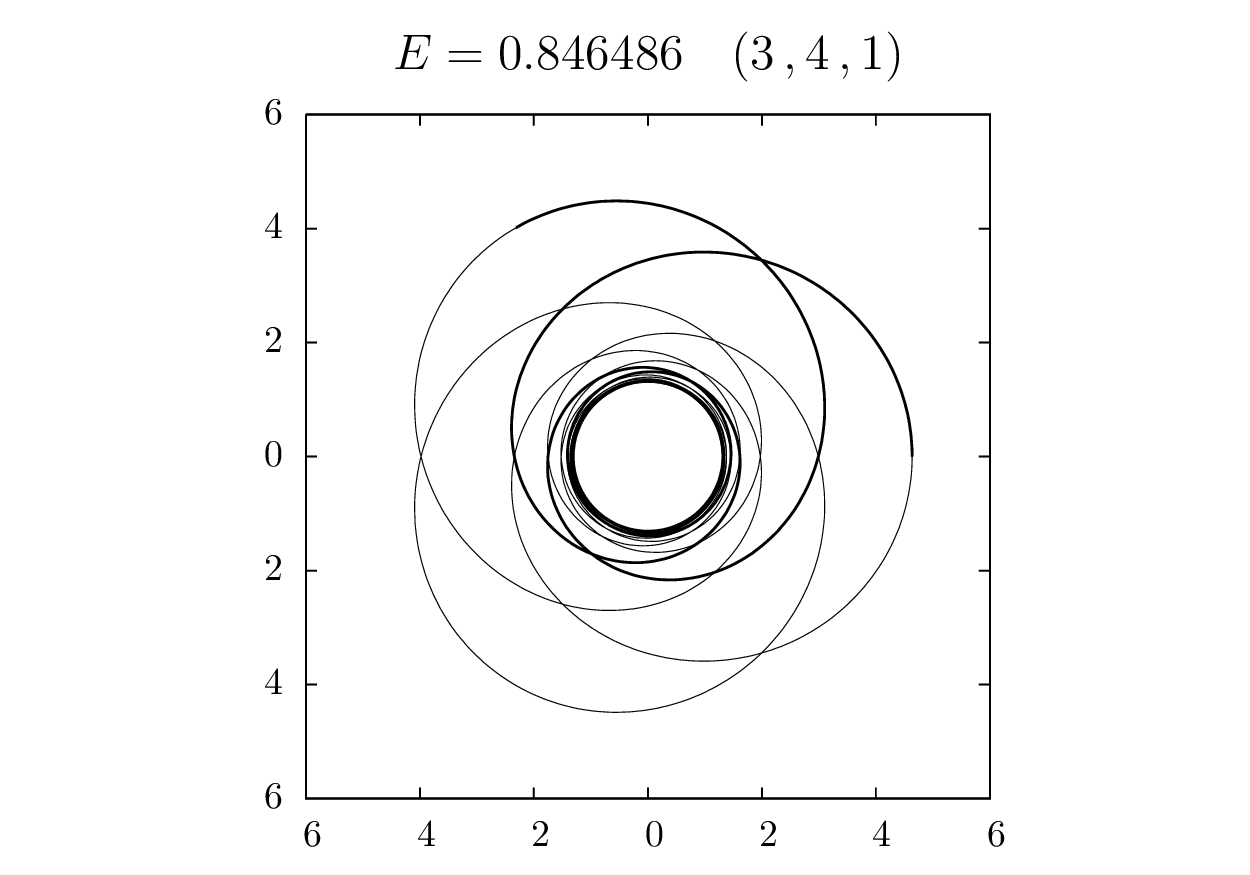}
\hfill
\\
\hspace{-26pt}
  \includegraphics[width=0.45\textwidth]{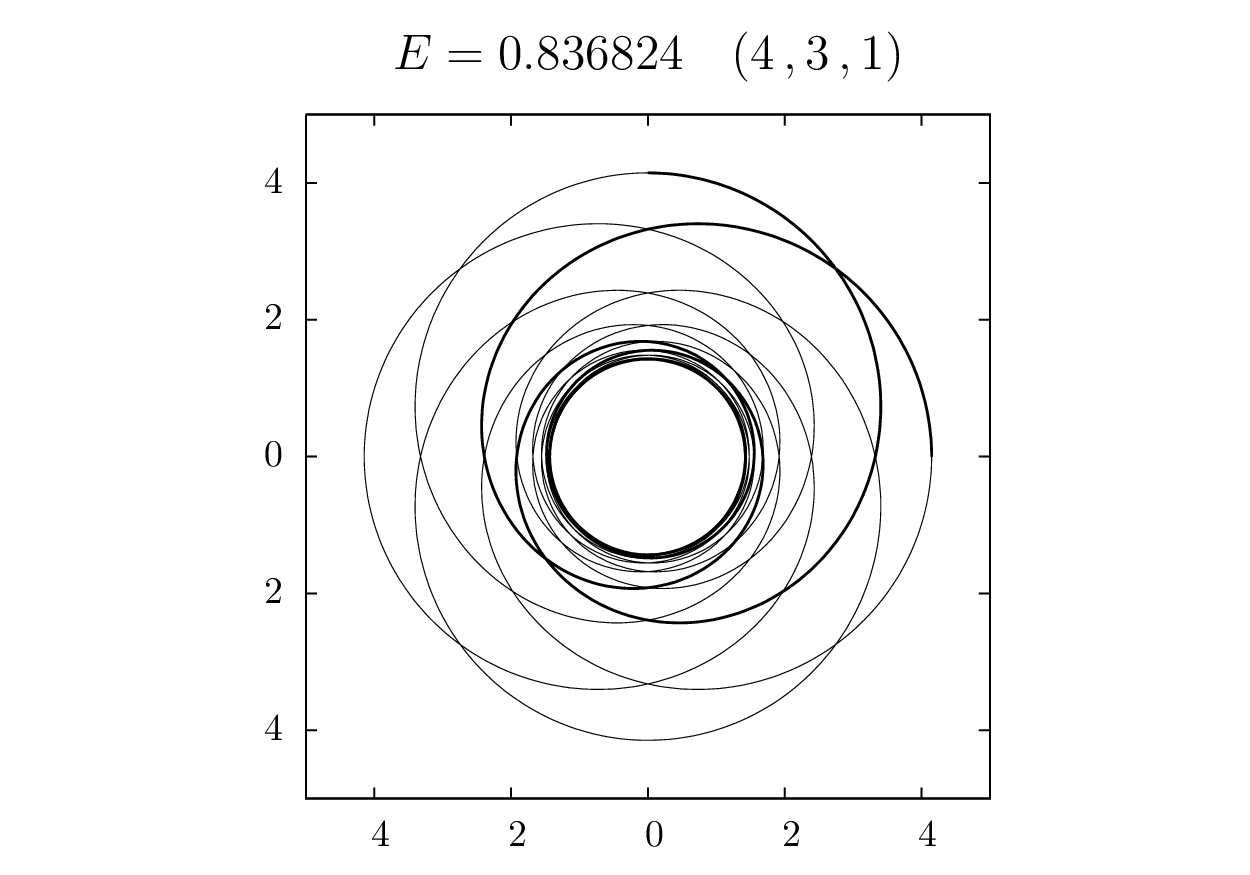}
\hspace{-80pt}
  \includegraphics[width=0.45\textwidth]{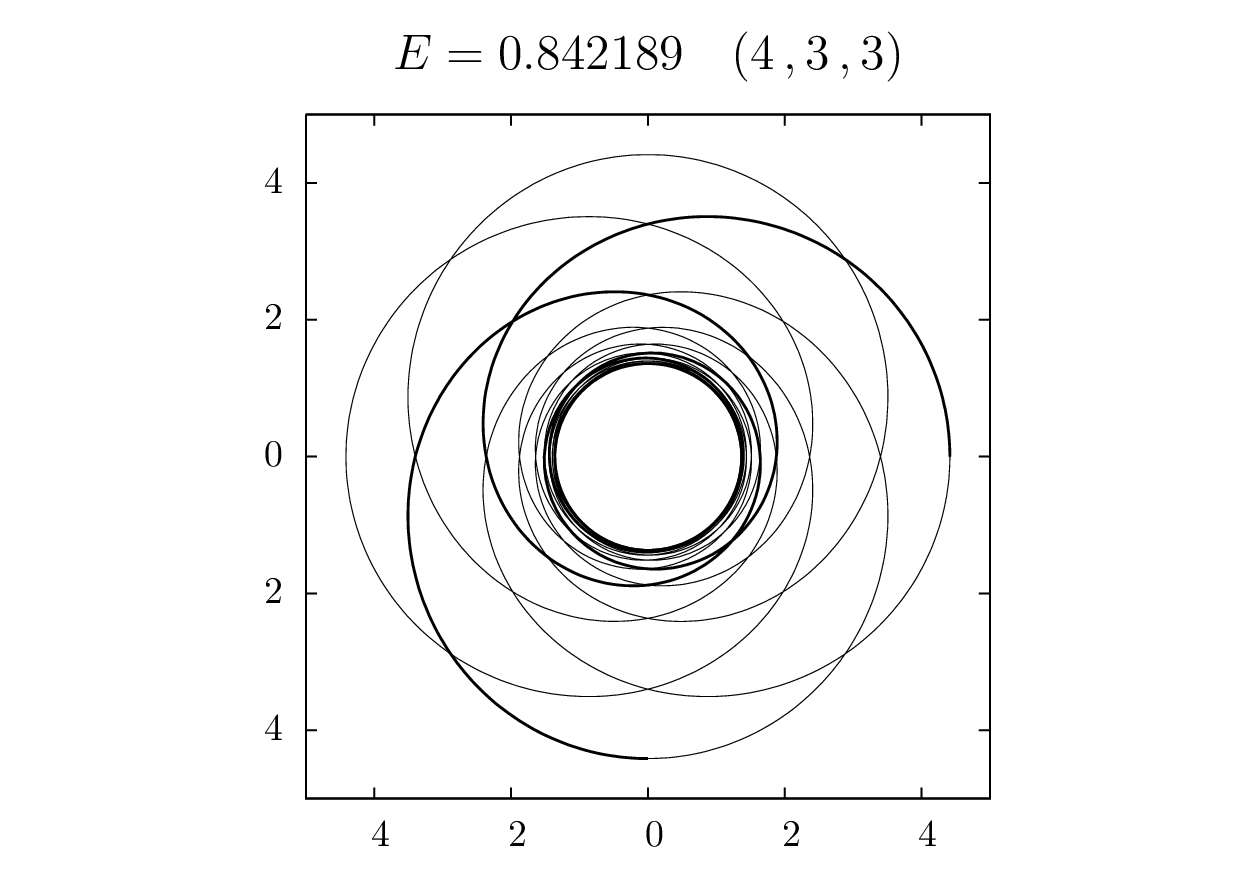}
\hspace{-80pt}
  \includegraphics[width=0.45\textwidth]{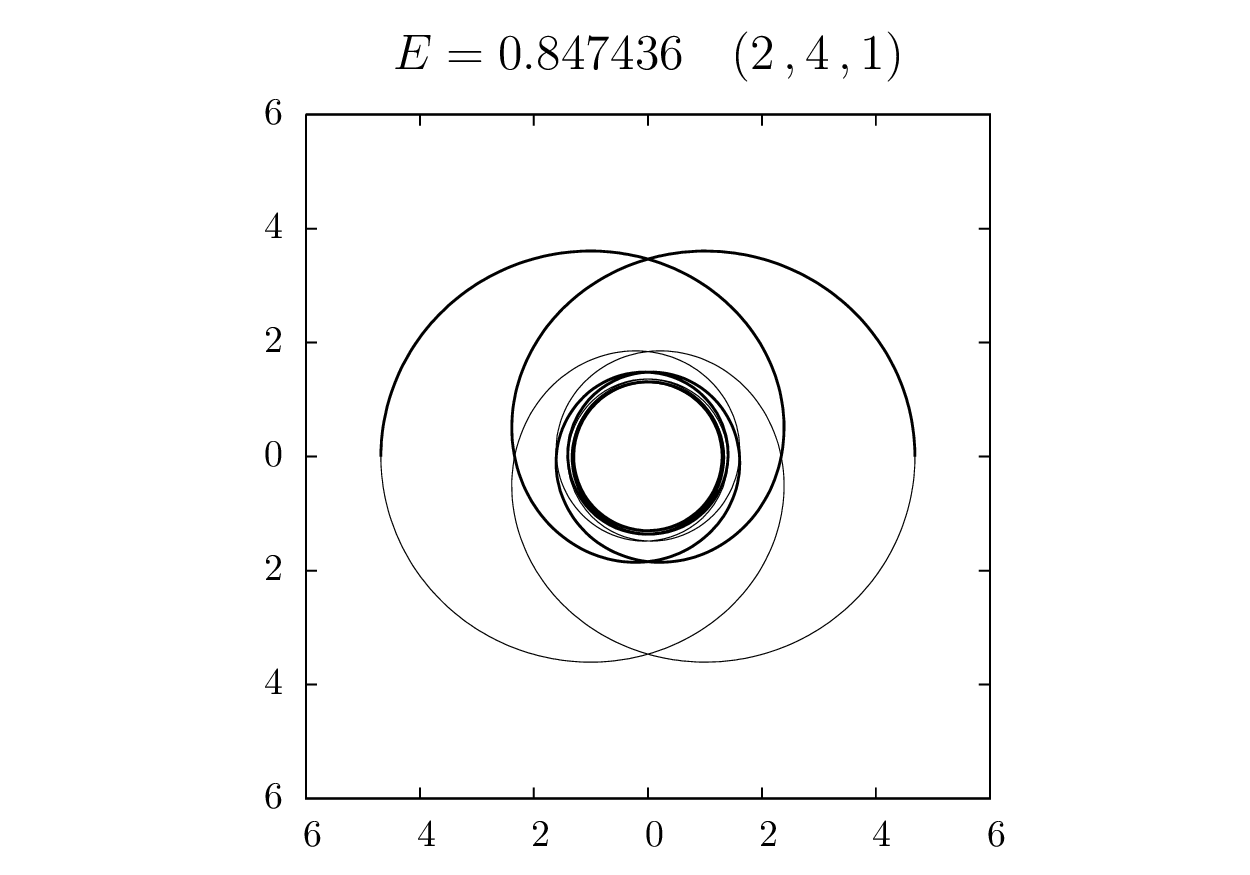}
\hfill
\\
\hspace{-24pt}
  \includegraphics[width=0.45\textwidth]{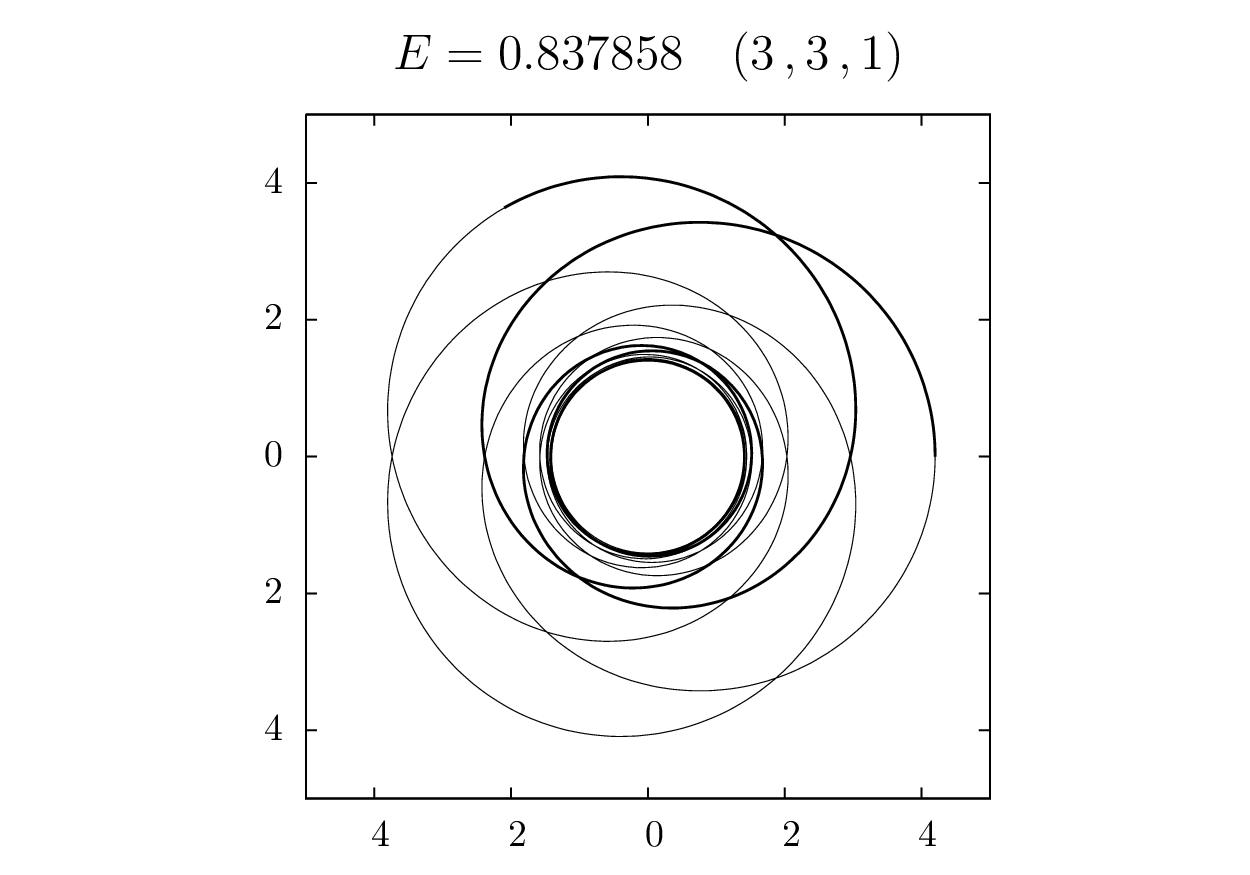}
\hspace{-80pt}
  \includegraphics[width=0.45\textwidth]{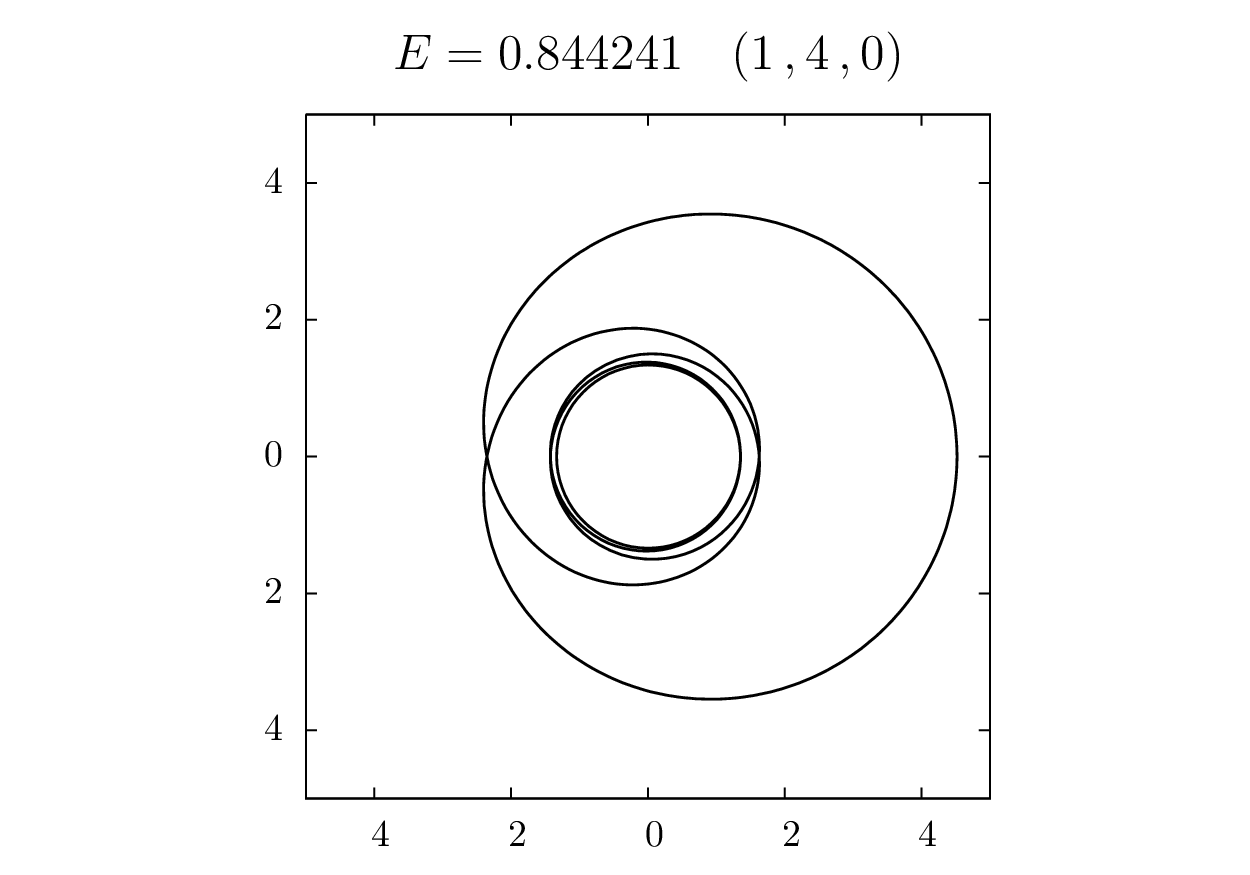}
\hspace{-80pt}
  \includegraphics[width=0.45\textwidth]{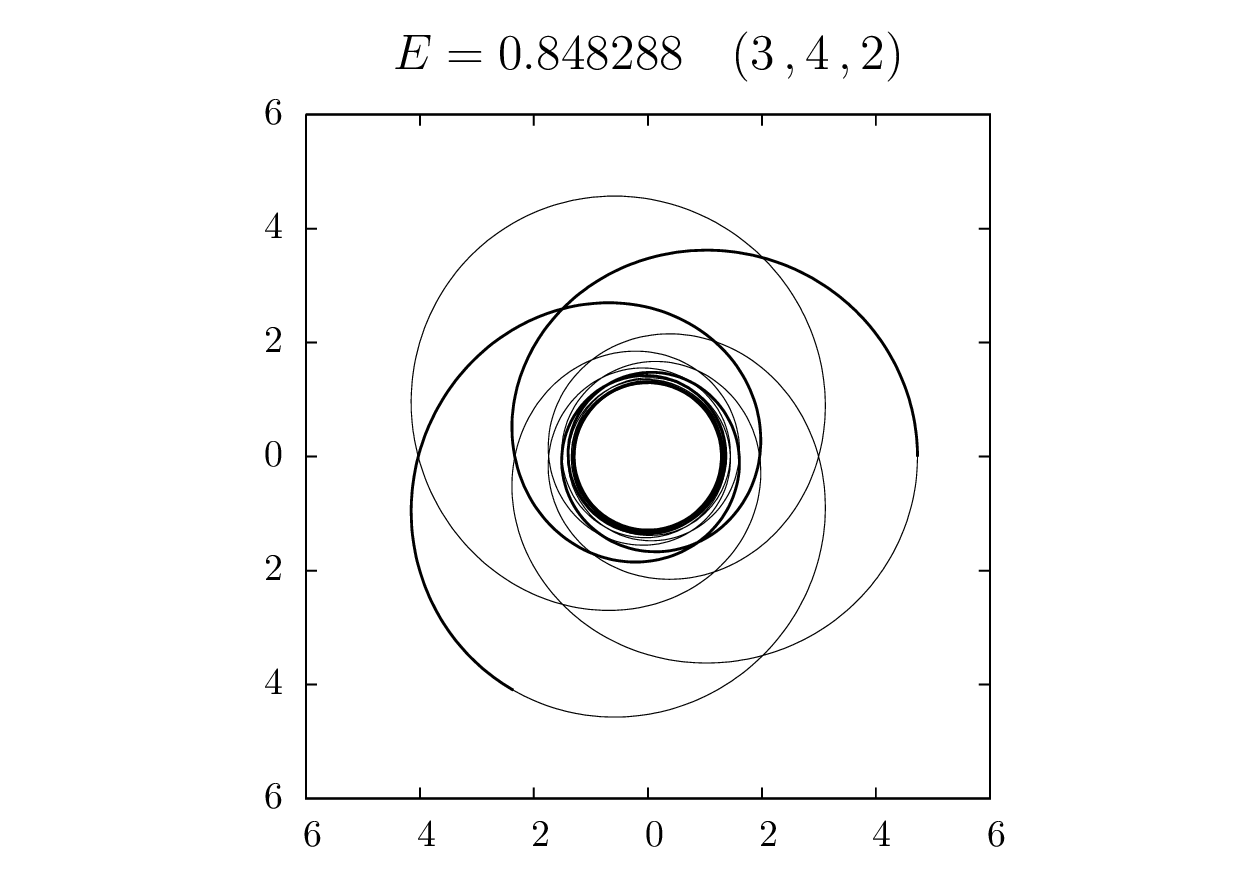}
\hfill
\\
\hspace{-22pt}
  \includegraphics[width=0.45\textwidth]{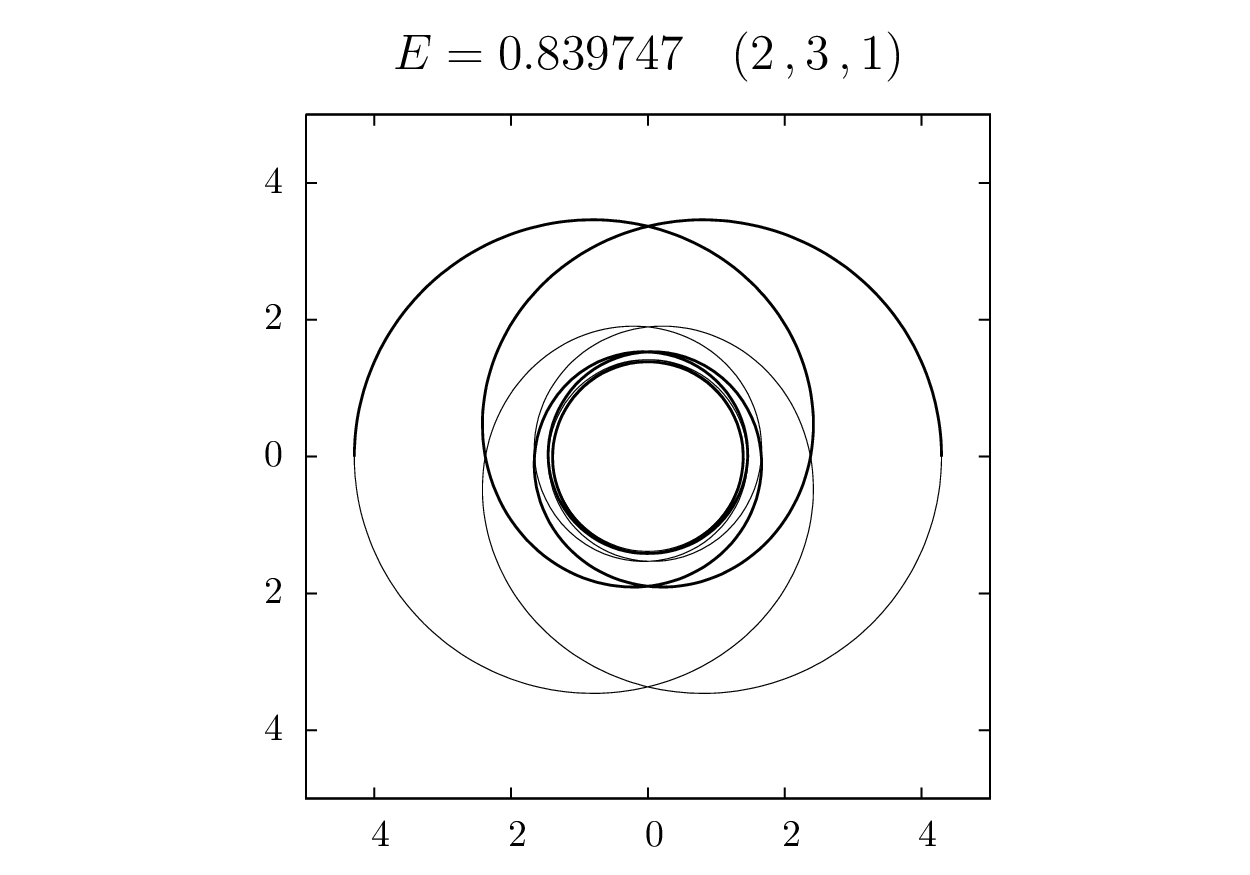}
\hspace{-80pt}
  \includegraphics[width=0.45\textwidth]{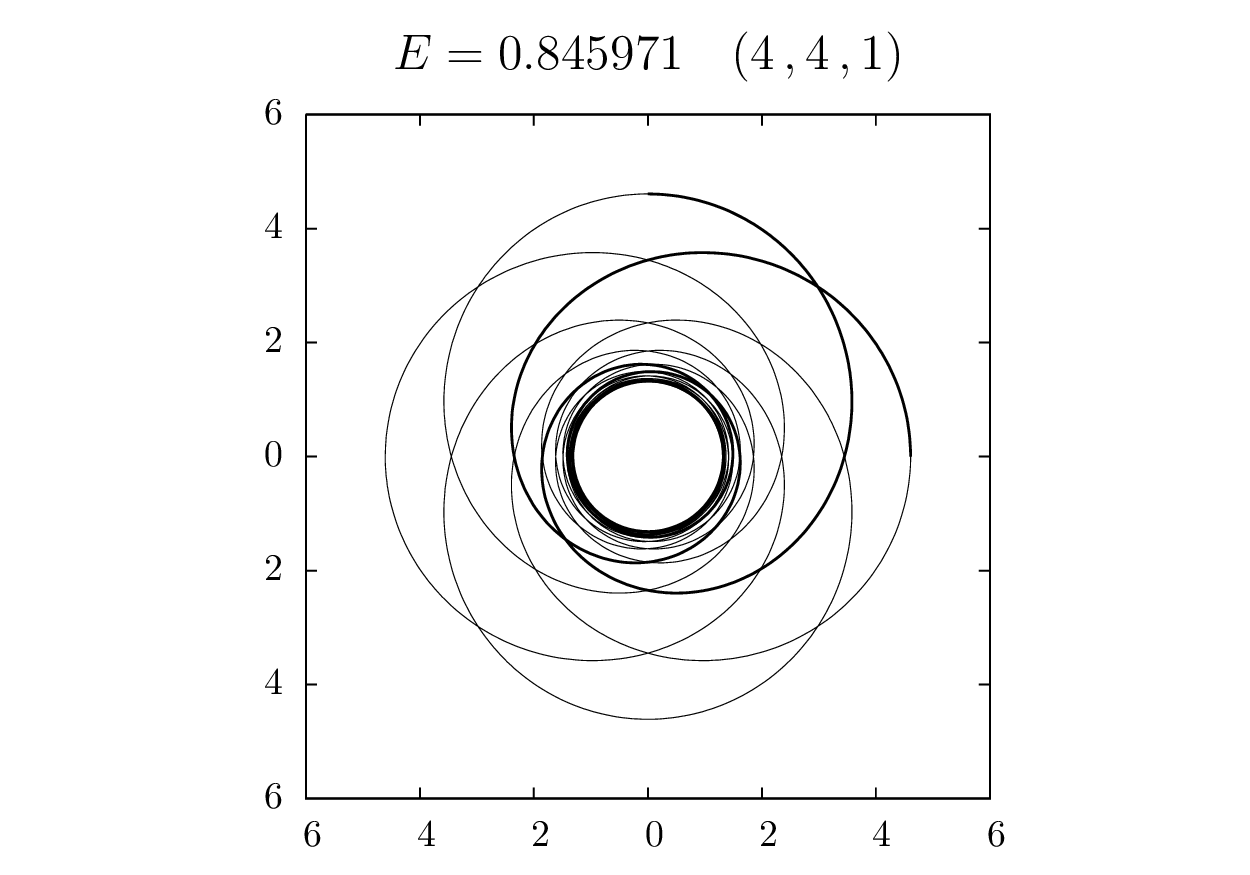}
\hspace{-80pt}
  \includegraphics[width=0.45\textwidth]{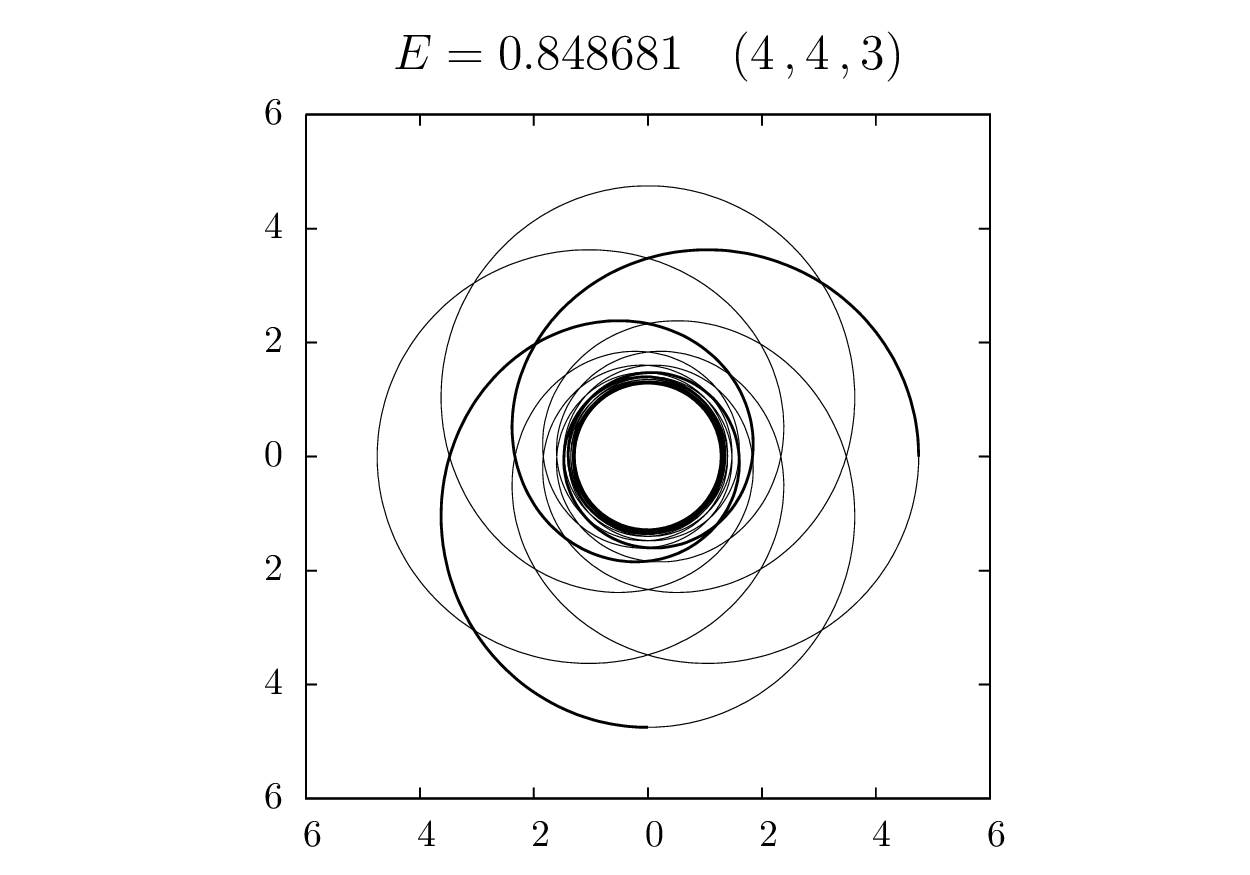}
\hfill
  \caption{A series of $w=3$ and $w=4$ orbits for the $L=1.82$ line of {figure 
  \ref{kerr}}. Orbits increase in energy from top to bottom and left to
  right. All $z=1,2,3,4$ orbits are shown.}
  \label{wK}
\end{figure}

\begin{figure}
  \vspace{-20pt}
  \centering
\hspace{-30pt} 
 \includegraphics[width=0.45\textwidth]{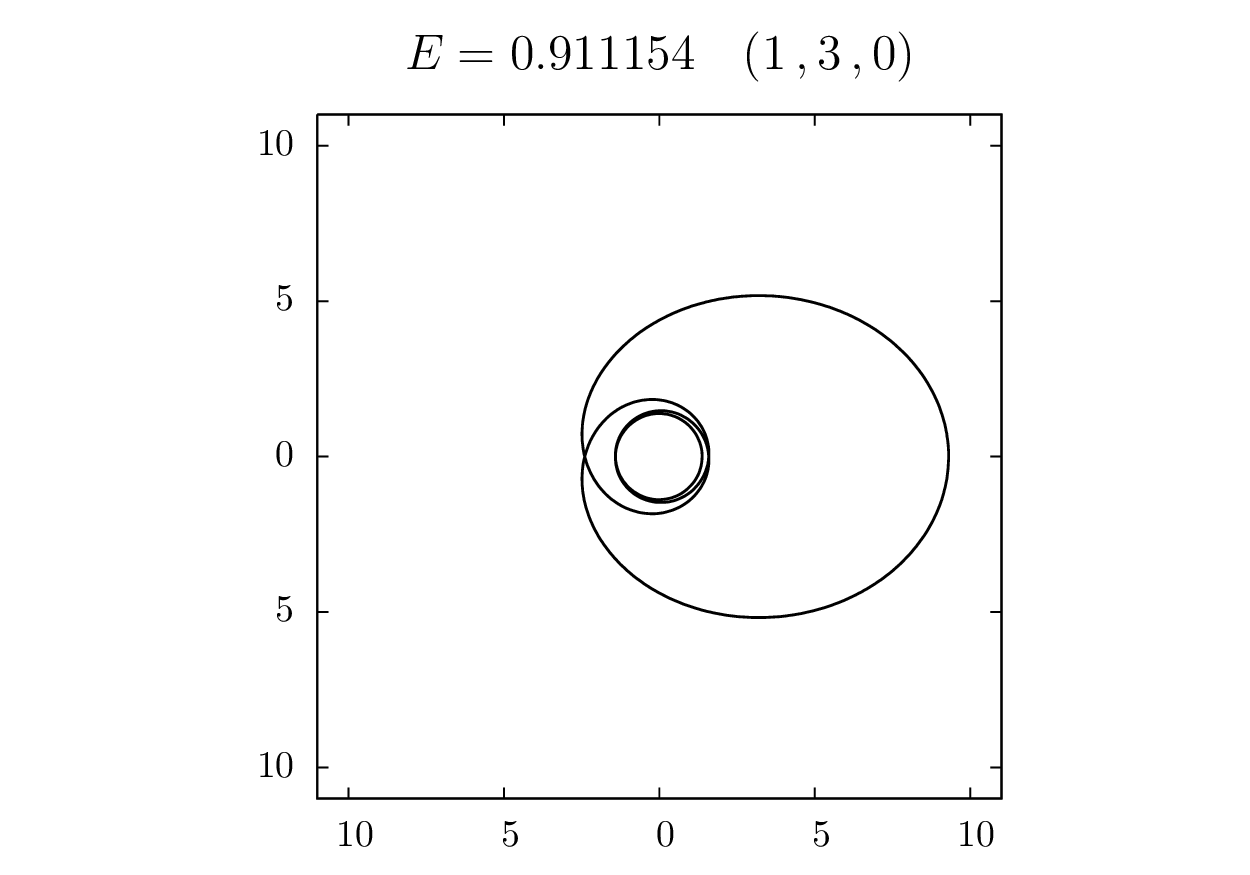}
\hspace{-80pt}
  \includegraphics[width=0.45\textwidth]{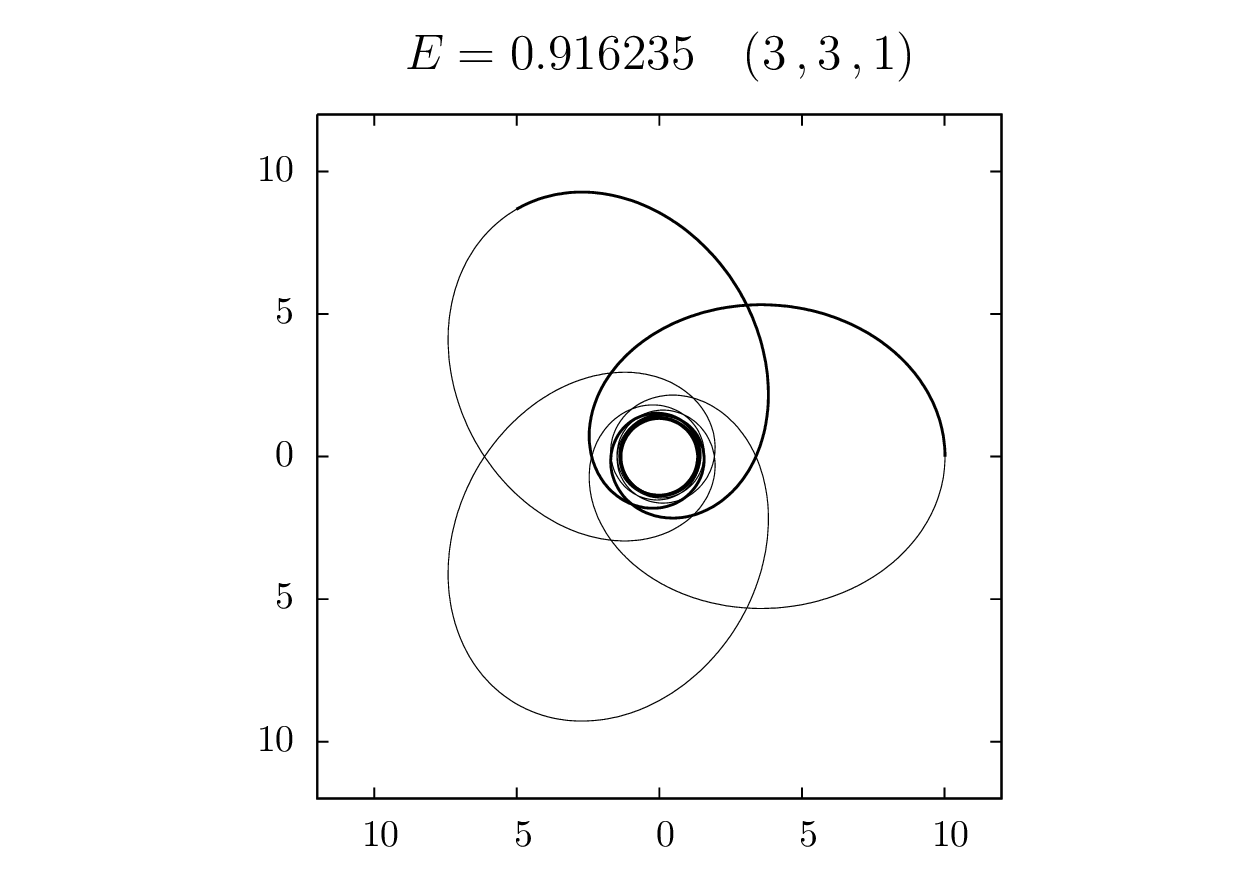} 
\hspace{-80pt}
  \includegraphics[width=0.45\textwidth]{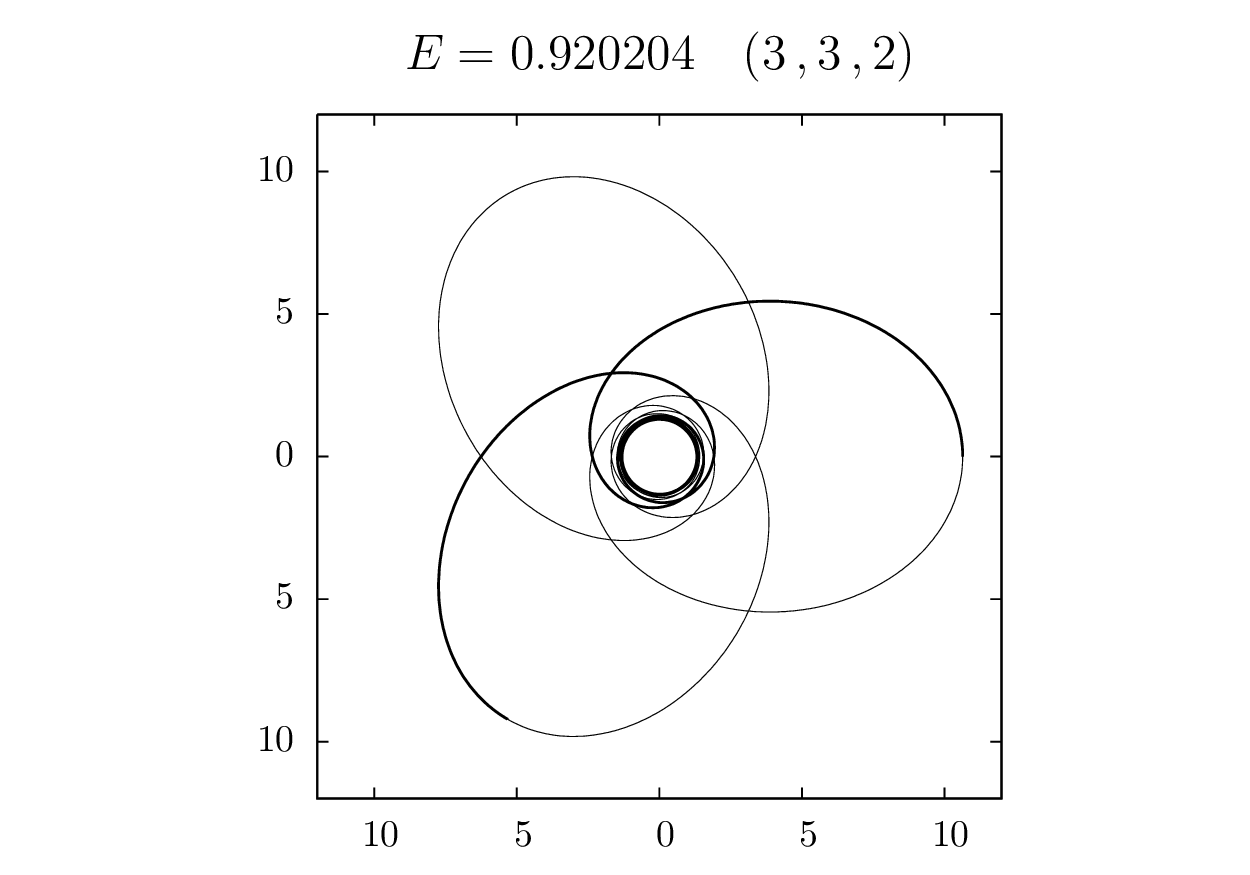}
\hfill
\\
\hspace{-26pt}
  \includegraphics[width=0.45\textwidth]{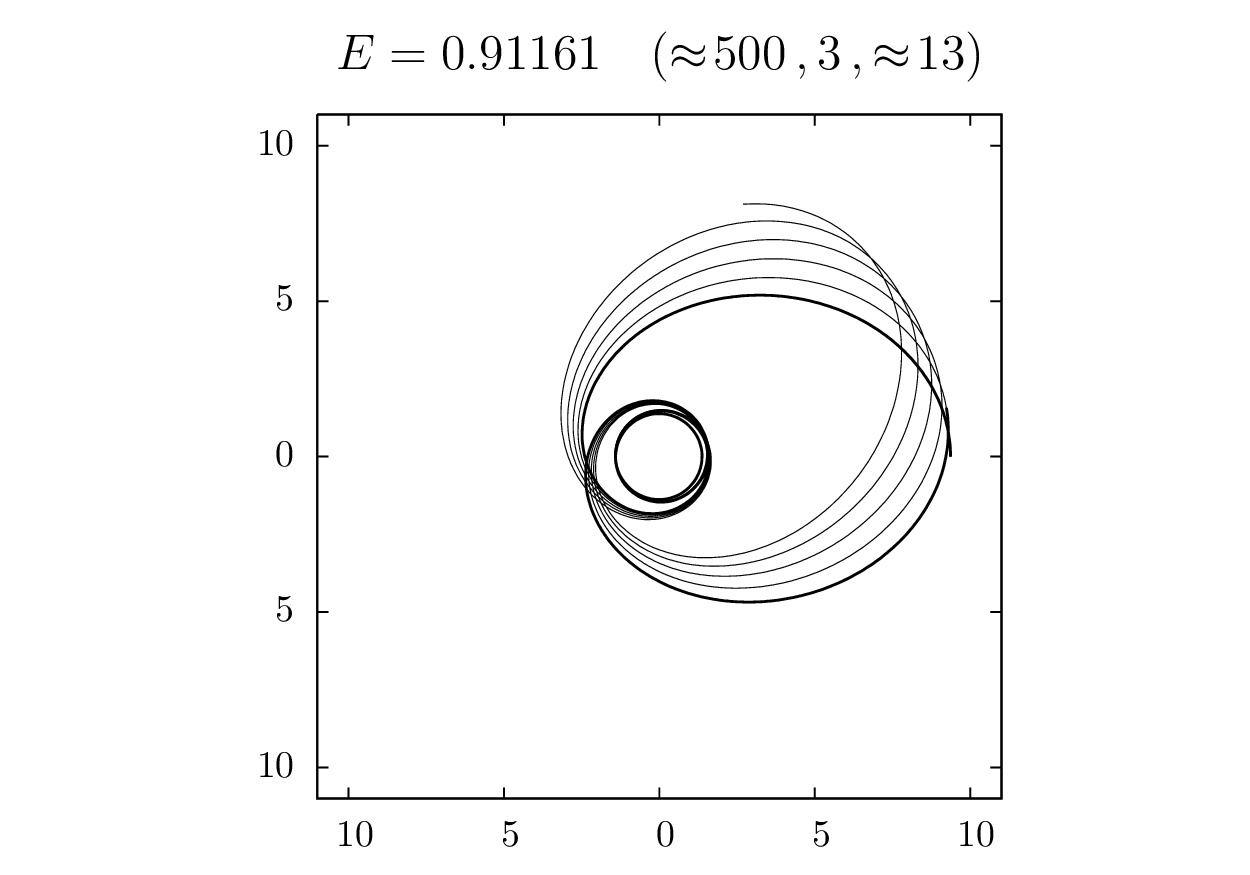}
\hspace{-80pt}
  \includegraphics[width=0.45\textwidth]{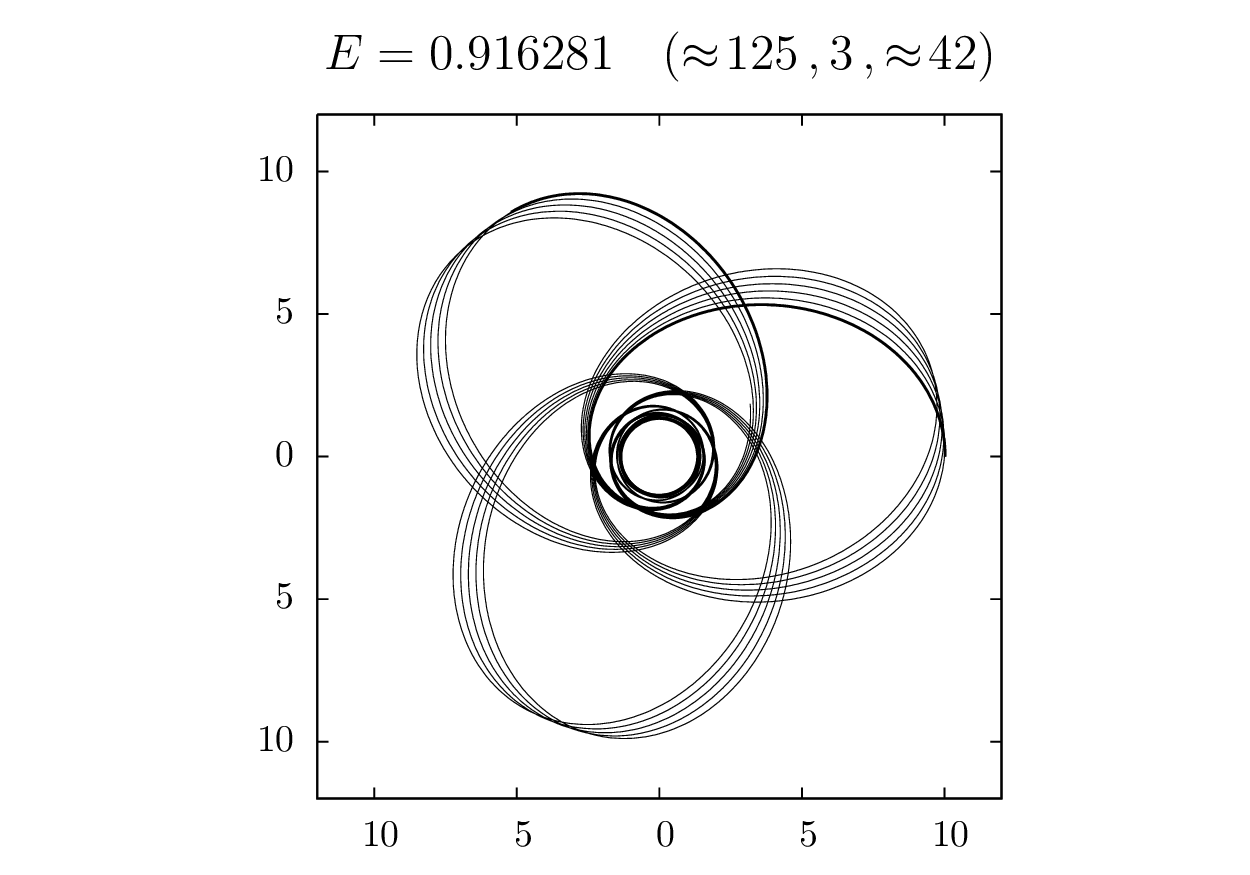}
\hspace{-80pt}
  \includegraphics[width=0.45\textwidth]{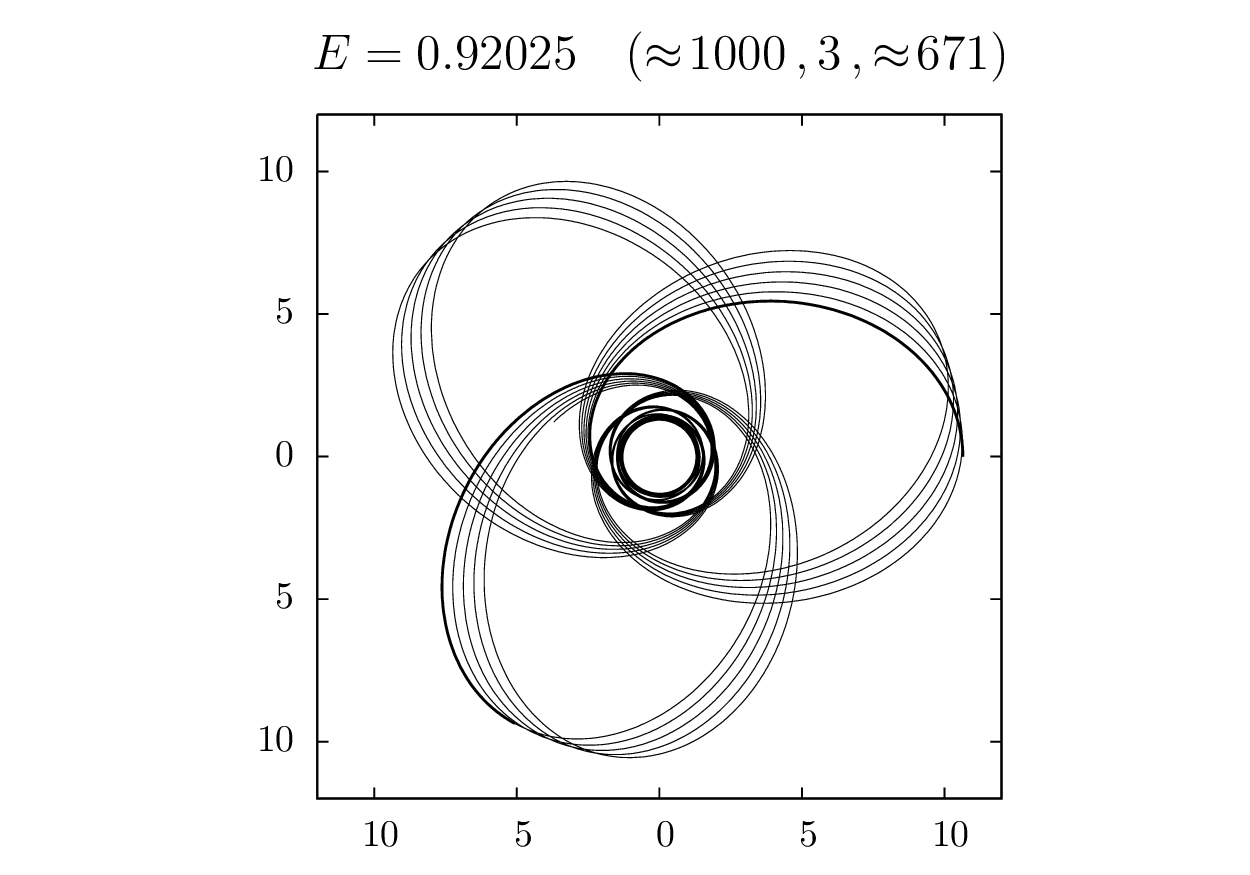}
\hfill
\\
\hspace{-24pt}
  \includegraphics[width=0.45\textwidth]{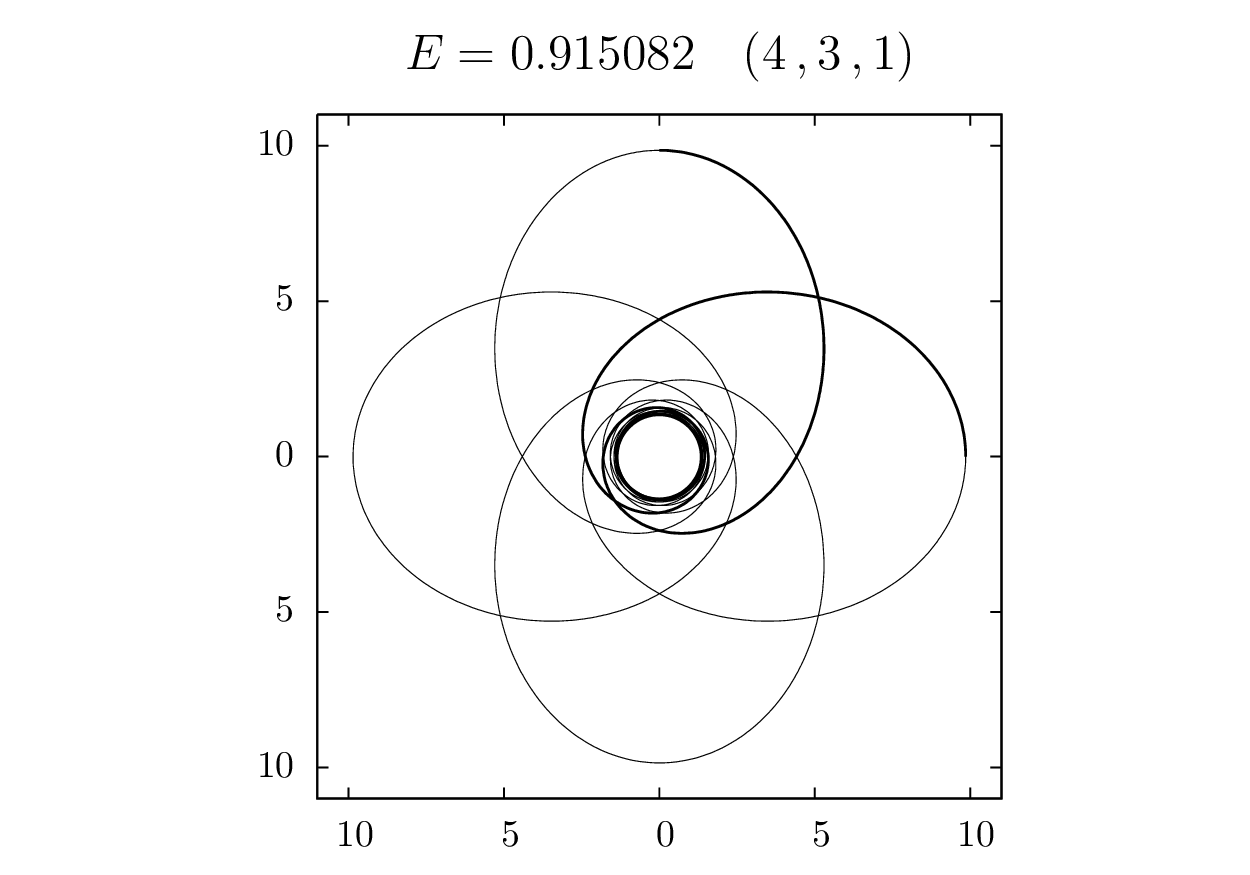}
\hspace{-80pt}
  \includegraphics[width=0.45\textwidth]{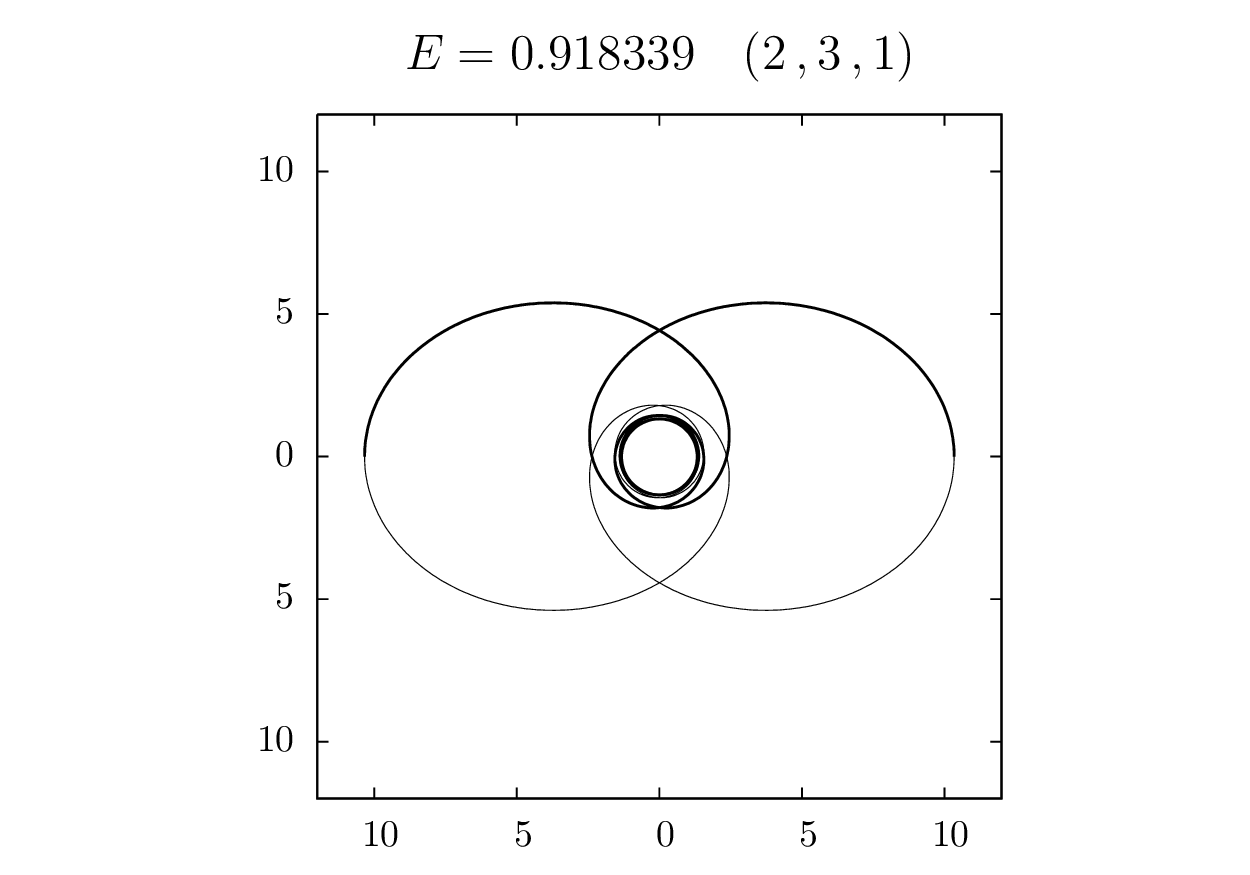}
\hspace{-80pt}
  \includegraphics[width=0.45\textwidth]{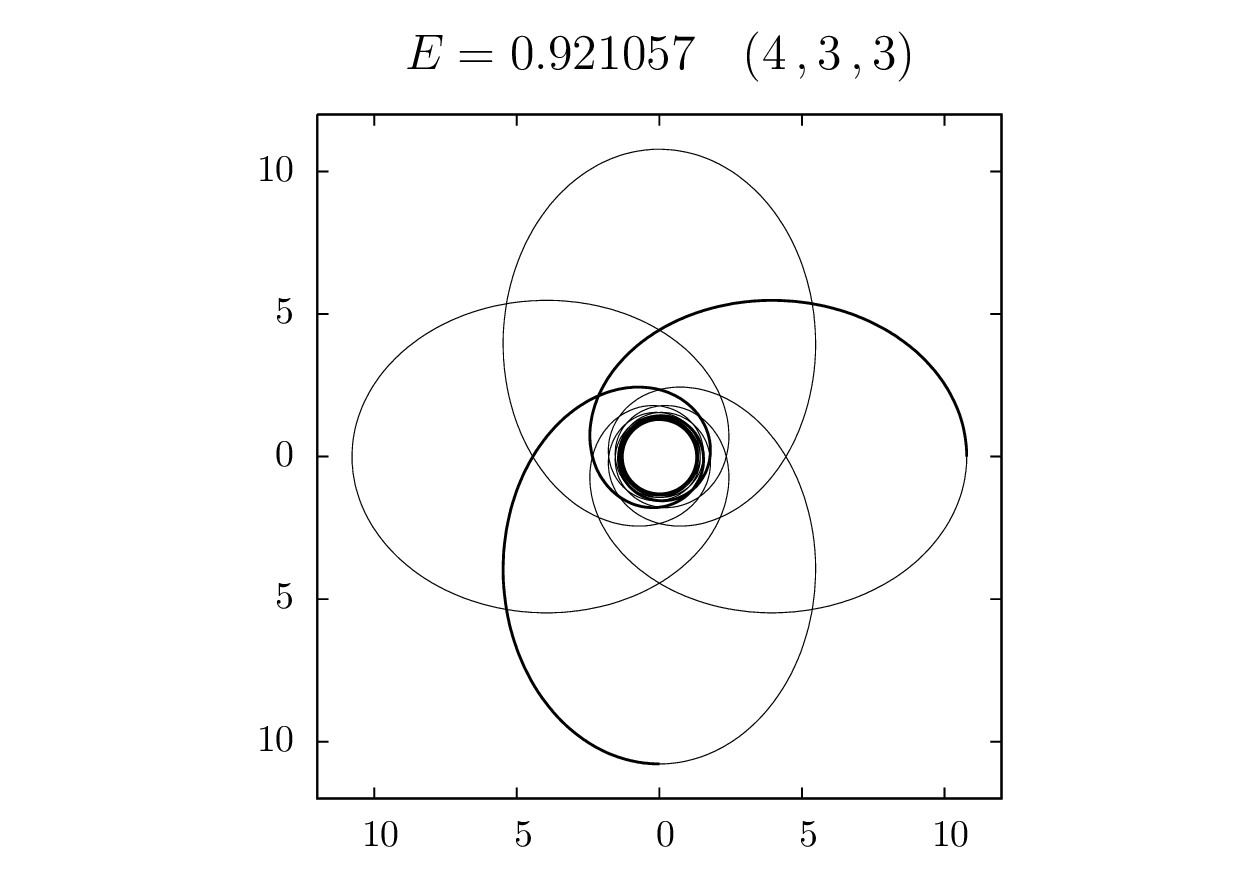}
\hfill
\\
\hspace{-22pt}
  \includegraphics[width=0.45\textwidth]{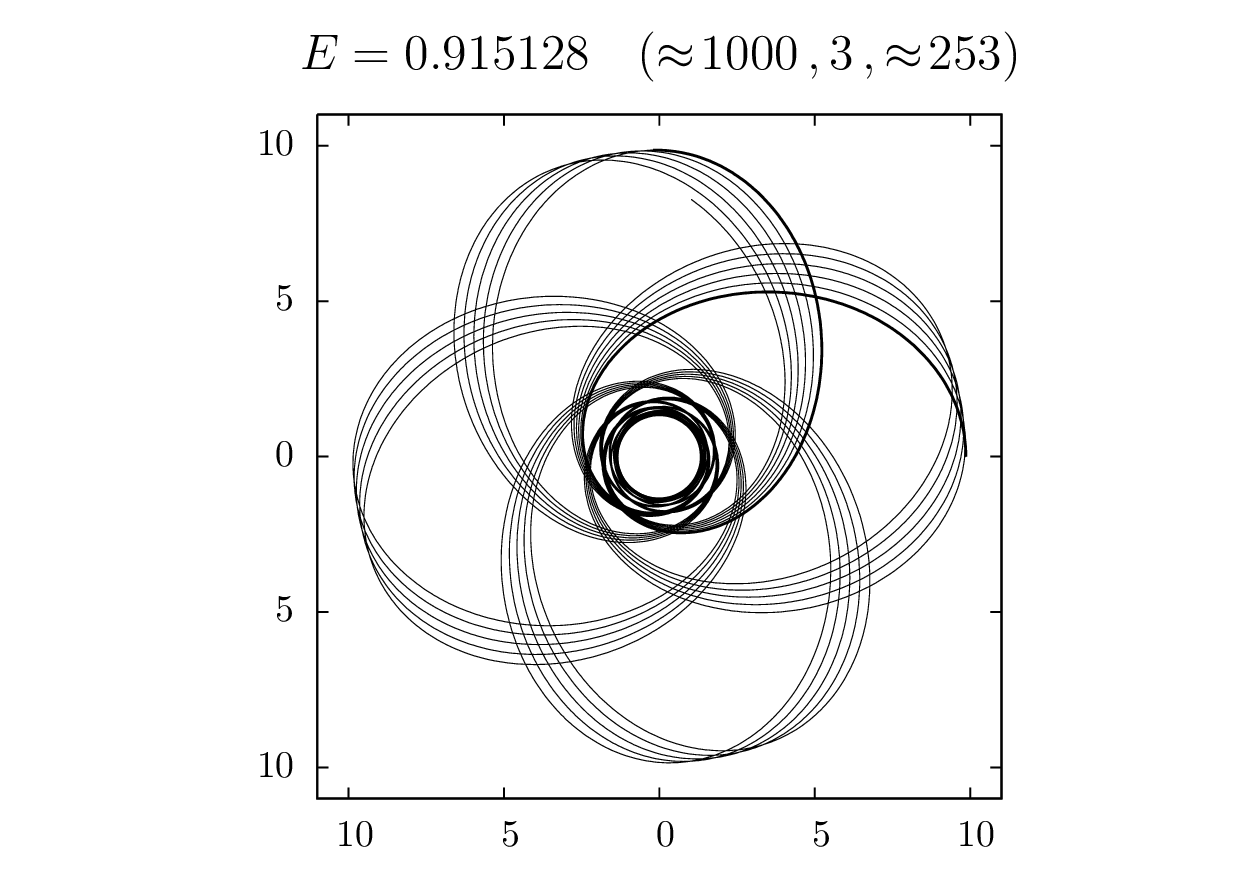}
\hspace{-80pt}
  \includegraphics[width=0.45\textwidth]{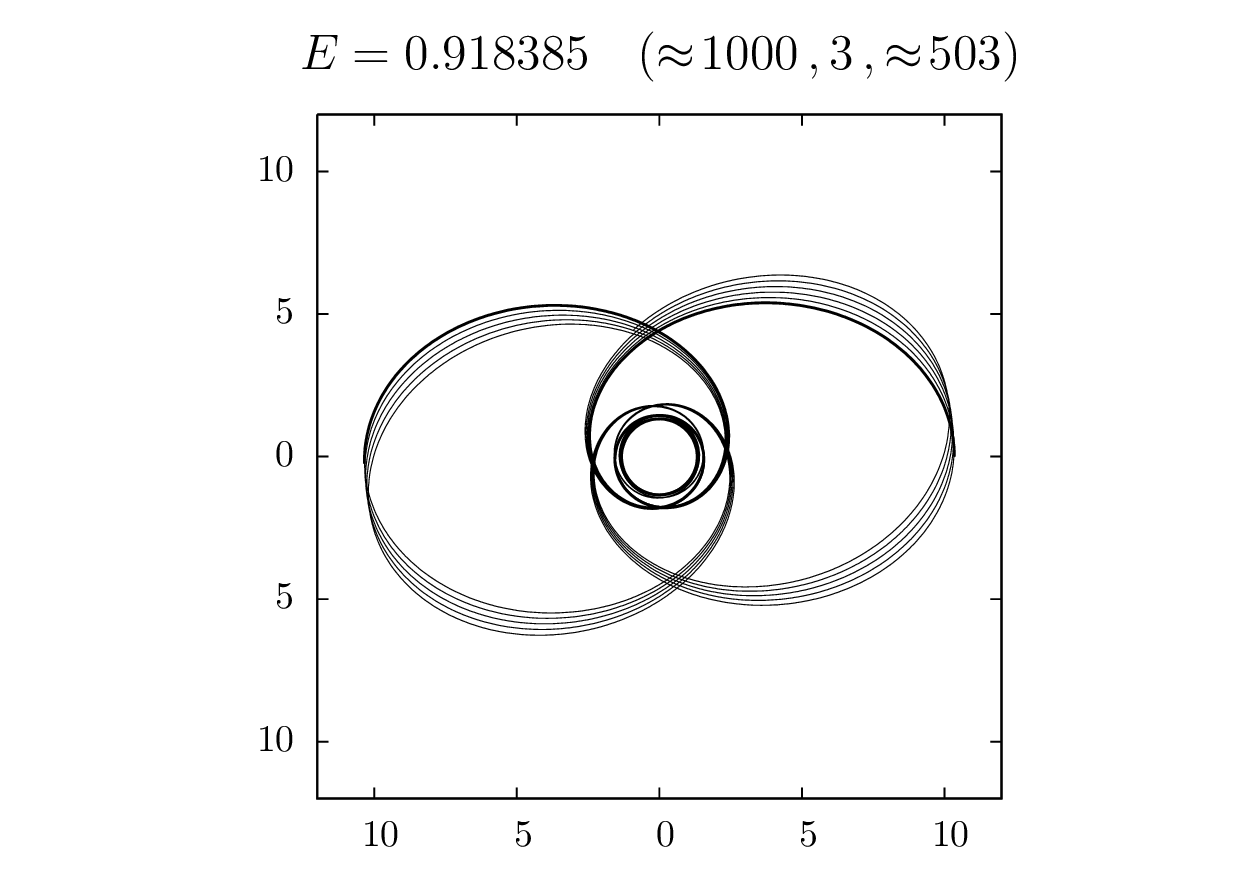}
\hspace{-80pt}
  \includegraphics[width=0.45\textwidth]{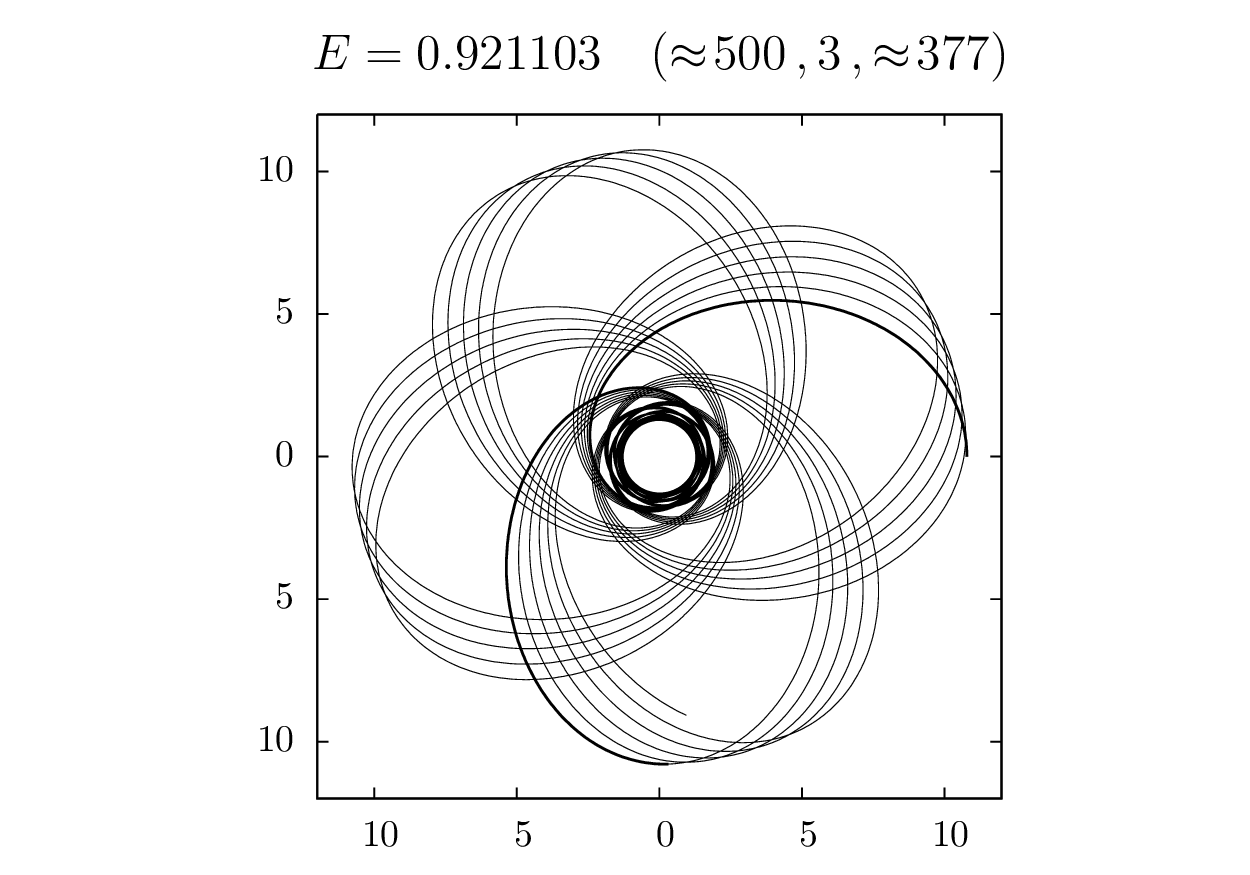}
\hfill
  \caption{A series of $w=3$ orbits for the $L=2$ line of {figure 
  \ref{kerr}}. Orbits increase in energy from top to bottom and left to
  right. All $z=1,2,3,4$ orbits are shown. Also shown are randomly
  selected high $z$ orbits. Notice that the high $z$ orbits look like
  precessions of the energetically closest low $z$ orbit.}
  \label{twelveK}
\end{figure}

Because the weak-field regimes look essentially the same for all $a$,
we consider 
only values of $L_{ISCO}<L<L_{IBCO}$.
As demonstrated for a Kerr black hole with
spin $a=0.995$ in figure \ref{wvzeK}, the
rational numbers corresponding to closed orbits increase monotonically with
energy. The rationals diverge at the homoclinic orbit for each $L$
since the number of whirls diverges. The rationals are bounded from
below by $q_c$ at the stable circular orbit:
\begin{equation}
 q_{c}\le q\le \infty
\label{limsK}
\end{equation}
Figure 
\ref{wvzeccK}. 
shows the increase of the rationals from the stable circular orbit to the
homoclinic orbit is shown versus eccentricity for $a=0.995$ and a
range of $L$'s.

In contrast to the non-spinning case of figure \ref{sch},
all eccentric orbits (with $L<L_{IBCO}$) have at least one whirl for
such high spin.  
Generally, $a\ne 0$ orbits exhibit more whirls
at low and intermediate
eccentricities than their $a=0$ counterparts.
As eccentricity (and energy) increase for a given $L$, the
number of whirls diverges as a homoclinic orbit is approached. 

Figure \ref{wK} displays a periodic table, energy increasing from
top to bottom
through the $w=3$ and $w=4$ bands for
$L=1.82$. Each entry has a periodic
orbit. All $z=1,2,3,4$ closed orbits are shown. The series shows the
passage through these low leaf, moderate eccentricity orbits as energy
-- and therefore 
$q$ increases. 

Finally, figure \ref{twelveK}
displays a periodic table with both low and high $z$ orbits, 
energy increasing from top to bottom for $L=2$ and $a=0.995$. 
The sequence takes you through low leaf orbits to high leaf orbits
that look very much like 
precessions of lower leaf periodics. 

The conclusion is that all eccentric Kerr orbits show
zoom-whirl behavior of some kind for this $a$ and $L<L_{IBCO}$. In the
strong-field regime, once $L<L_{IBCO}$, the pattern of
zoom-whirls at the lower energy bound 
looks like a precession of a low-leaf clover and at the upper energy
bound marches toward homoclinic -- a single leaf with an infinite
number of whirls.

\centerline{\bf RECAP OF DYNAMICAL RESULTS}

This concludes the primary results of the paper. Before turning to a
discussion of astrophysical applications of the taxonomy, we
briefly recap the results of this section:\par

$\bullet $ For a given $a$ and $L$, all periodic orbits corresponding
to rationals in the range
\begin{equation}
q_{c}
\le q
\le
q_{max}
\end{equation}
define the skeleton of the entire orbital dynamics, where $q_c$
is the limiting rational associated with the stable circular orbit and
$q_{max}$ is the approximate rational of the maximum energy bound orbit.

$\bullet$ In the Newtonian limit, the upper and lower bounds both
approach zero; Keplerian orbits are ellipses. 

$\bullet$ There is no
zoom-whirl behavior in the weak-field regime.

$\bullet $ At the ISCO, the upper and lower $q$ bounds both approach
$\infty $ because the ISCO is at once both a circular and a homoclinic
orbit and all 
homoclinic orbits have $q=\infty$.

$\bullet $ In the very strong-field regime, which we take  here to
correspond to ($L<L_{IBCO}$),
$q$ has no upper bound:
\begin{equation}
q_{c}
\le q
\le
\infty
\end{equation}

$\bullet $ In the strong-field regime, 
the simple precessing ellipse familiar from
planetary orbits is forbidden.

$\bullet $ {\it All} aperiodic eccentric orbits are precessions of
low-leaf periodics.

$\bullet $ As the ISCO is approached, {\it all} orbits whirl
as well as zoom for any $a$.

\section{Applications and future work}
\label{util}

Our approach -- representing the entire black hole
dynamics in terms of a 
periodic  
skeleton -- can be extended to non-equatorial orbits and then applied
to astrophysical  
problems such as gravitational wave
astronomy and a dynamical analysis of spinning black hole pairs.
We
briefly describe each of these applications in turn.

{\bf Non-equatorial Orbits}:
Equatorial orbits yield especially well to a visual rendering of the
properties of periodic orbits and their relationship to aperiodic
ones. We focused here exclusively on equatorial orbits for precisely
those pedagogic reasons. Already underway, our next order of business
is to extend the periodic orbit analysis to non-equatorial Kerr
geodesics.
Generic non-equatorial trajectories can look complex particularly for
high eccentricities and incliniations in the strong-field \cite{dh}.
Still, periodic orbits must exist. 
Just as
in the equatorial case periodic orbits will correspond to rational
numbers. Identifying the ordering of
those rationals with respect to those orbital parameters should offer
an elegant and illuminating description of the dynamics.

Framing the non-equatorial periodic skeleton is not a solely academic
enterprise. The modeling of 
extreme mass ratio inspirals (EMRIs) of compact objects into rotating
supermassive black holes relies heavily on non-equatorial Kerr
dynamics.

{\bf Gravitational Wave Astronomy:}  Compact objects orbiting
supermassive black holes in galactic
centers are expected to be plentiful sources for gravitational waves
in the LISA bandwidth \cite{{hopman},{hughesconf2}}.
Tens of thousands of such compact objects populate our own galactic
center. 
Through gravitational scattering, a stellar mass compact object can be
thrown into a highly eccentrc orbit in the strong-field regime
\cite{bigref}. 
We have shown that all eccentric orbits show zoom behavior of some
sort -- that is to say, a pattern of discernible leaves -- and large
ranges of $L$ and $E$ generate
whirl behavior 
(figures \ref{sch} and \ref{kerr}). In particular, in the strong-field
regime, we should not expect eccentric orbits to be characterized by
precessing ellipses, but rather by the precession of multi-leaf orbits.
These EMRIs
 will therefore show 
the zoom-whirl behavior
taxonomized in this paper.

So far we have only described the
conservative dynamical system. Astrophysical objects orbiting near a
massive black hole will radiate gravitational waves and veer off
geodesics under the effects of dissipation. A study of adiabatic
inspiral through periodic orbits might offer computational advantages.
For instance, periodic orbits should have more rapidly converging
Fourier series, than aperiodic ones. Additionally, the evolution of
the conserved quantities $E$, $L$, and $Q$ may not be independent,
thereby cutting down dramatically on the computational expenses. 

Closed orbits also show promise as a benefit to signal extraction from the
gravitational wave observatories in a time-frequencey
analysis. Furthermore, the templates of aperiodic orbits, where
required \cite{poisson}, might be suitably replaced by periodic tables. 

{\bf Spinning Pairs:} 
The Kerr orbits are only integrable in the case of non-spinning test
particles. If the companion spins, a constant of integration is lost
(namely $Q$) and the dynamics is no longer integrable
\cite{{sm},{jl1}}. While many of the other techniques for
studying the dynamics may not generalize to spinning pairs, periodic
orbits still define the 
skeleton of the phase space. Even if no clean taxonomy for those
orbits exists, a search for and analysis of periodic orbits should
reveal general properties of the dynamics.

Further, it is already known that spinning pairs can exhibit chaos 
\cite{{sm},{jl1},{jl2},{jlnc},{hb},{jl3},{hartl1},{hartl2}}.
Treating spin of the companion as a perturbation to Kerr
motion, the progression of closed oribts under that perturbation could
track the transition to chaos. 
As the companion spin increases, the
periodic orbits proliferate 
and the homoclinic orbit is replaced with a homoclinic tangle -- a
fractal set of unstable periodic orbits
\cite{{bc},{cornishfrankel}}. A fractal in phase space of such 
rapidly spinning systems was revealed in Ref.\ \cite{jl1} for comparable
mass black hole binaries in the Post-Newtonian expansion by
numerically scattering
black holes. An analysis centered on periodic orbits should
reveal the fractal set
more readily, thereby removing the need to search and
survey phase space blindly for chaotic behavior.

\section{Conclusion}
\label{conc}

Periodic orbits structure the entire equatorial
dynamics around black holes. In the spirit of 
Poincar\'e's approach, we have defined a
taxonomy of all periodic orbits based on zooms, whirls, and
vertices and forged an explicit connection with the set of rational
numbers. Under this scheme,
all generic aperiodic orbits can always be approximated
as near one of the periodics.

Our method reveals that 
no eccentric orbits will trace out planetary type elliptical precessions in the
strong-field regime. Instead, all eccentric orbits are
precessions of multi-leaf orbits with whirls multiplying as the black
hole spins. The implication for gravitational wave astronomy is
intriguing: all eccentric orbits will transition through these nearly
periodic orbits as they inspiral to the plunge. 

\vfill\eject

\noindent{*Acknowledgments*}

We are especially grateful to Becky Grossman for her valuable and generous
contributions to this work and to Jamie Rollins for his careful
reading of the manuscript, inspired suggestions, and for coining our
figures ``periodic tables''. We also thank
Szabi Marka for valuable discussions of this work. 
JL and GP-G
acknowledge financial support from a Columbia 
University ISE grant. This material is based in part upon work
supported under a National Science Foundation Graduate Research
Fellowship. 

\vfill\eject

\appendix

\section{THE KERR EQUATIONS}
\label{kerreqs}

We now present the explicit equations integrated to generate the
results for \S \ref{zooS} and \ref{zooK}. We will also derive the
association between periodic orbits and rational 
numbers from the dynamical systems perspective.  Since the natural
language for that discussion is Hamiltonian mechanics, we begin with
the Hamiltonian formulation of Kerr geodesic motion.

\subsection{The Kerr Metric and Geodesic equations}
\label{geods}

In Boyer-Lindquist coordinates and geometrized units ($G = c =1$), the
Kerr metric is
\begin{eqnarray}
ds^2 =-\mu^2 d\tau^2=
-\left (1-\frac{2Mr}{\Sigma}\right )dt^2
&-&\frac{4Mar\sin^2\theta}{\Sigma}dtd\varphi 
+\frac{\Sigma}{\Delta}dr^2+\Sigma d\theta^2 \nonumber \\
&+&\sin^2\theta\left
(r^2+a^2+\frac{2Ma^2r\sin^2\theta}{\Sigma}\right ) d\varphi^2 \quad .
\label{metric}
\end{eqnarray}
Here
\begin{eqnarray}
\Sigma &\equiv & r^2+a^2\cos^2\theta \nonumber \\
\Delta &\equiv & r^2-2Mr+a^2 \quad\quad ,
\end{eqnarray}
and $M, a$ denote the central black hole mass and spin angular
momentum per unit mass, respectively.  Besides the test particle mass
$\mu$, three additional quantities are conserved along Kerr geodesics:
the energy $E$ and $z$-component of angular momentum $L$ as measured
by observers at infinity, and the Carter constant $Q$.
Following a useful and now common convention, we hereafter set both
$M=1$ and $\mu=1$.  This choice leaves the overall appearance of the
metric and other Kerr expressions otherwise unchanged provided all
Boyer-Lindquist coordinates, the proper time $\tau$, the spin
parameter $a$ and all conserved quantities are now interpreted as
dimensionless quantities.

Since they possess as many constants of motion ($\mu, E, L, Q$) as
degrees of freedom ($t, r, \theta, \varphi$), the usually second order
geodesic equations can be integrated to yield a set of 4 first order
equations of motion for the coordinates \cite{carter}, 
\begin{eqnarray}
\Sigma \dot r &=& \pm \sqrt{R} \nonumber \\
\Sigma \dot \theta &=& \pm \sqrt{\Theta} \nonumber \\
\Sigma \dot \varphi &=&-\left
(aE-\frac{L}{\sin^2\theta}\right )+\frac{a}{\Delta}P\nonumber \\
\Sigma \dot t&=& -a\left (aE\sin^2\theta -L\right
)+\frac{r^2+a^2}{\Delta}P \quad\quad ,
\label{dimcarter}
\end{eqnarray}
where an overdot denotes
differentiation
with respect to the particle's proper time $\tau$ and
\begin{eqnarray}
\Theta(\theta) &=& Q - \cos^2\theta \left
\{a^2(1- E^2)+\frac{L^2}{\sin^2\theta}\right \} \nonumber \\
P(r) &=& E(r^2+a^2)-aL \nonumber \\
R(r) &=& P^2-\Delta \left\{  r^2+(L-aE)^2 + Q \right\} \quad .
\label{dimpots}
\end{eqnarray}

We note that while these equations are concise and appealing in some
ways, during numerical integration they tend to accumulate error at
the turning points due to the explicit square roots in the $r$ and
$\theta$ equations, not to mention the nuisance of having to change
the signs of the $r$ and $\theta$ velocities by hand at every turning
point.  While other authors circumvent this problem with a rather
involved reparametrization of the equations in (\ref{dimcarter}), we
will see that in a Hamiltonian formulation the numerical difficulties
are avoided naturally.

\subsection{Hamiltonian Formulation of Kerr Geodesic Motion}
\label{hamsec}

We begin with a Lagrangian for a free particle in the Kerr 
spacetime:
\begin{equation}
{\cal L} = \frac{1}{2} g_{\alpha \beta} \dot{q^\alpha} \dot{q^\beta}
\quad\quad .
\label{lagrangian}
\end{equation}
Because the $q^\alpha$ are dimensionless coordinates ($\mu = 1$), and
because $g_{\alpha \beta} \dot{q^\alpha} \dot{q^\beta} = -1$ for
timelike trajectories, eqn.\ (\ref{metric}) implies that ${\cal L}$ is
identically equal to $-1/2$ along any trajectory.  The particle's 4-momentum
\[
p^\alpha \equiv \dot{q}^\alpha
\]
is also dimensionless and is identical to the 4-velocity.  Defining
the canonical momentum $p_\alpha$ for the system in the standard way,
we see that it coincides with the 4-momentum one-form,
\begin{eqnarray}
  p_\alpha & \equiv & \frac{\partial {\cal L}}{\partial\dot{q}^\alpha} \\
  & = & g_{\alpha \beta} \dot{q}^\beta = g_{\alpha \beta} p^\beta \quad.
\end{eqnarray}
Explicitly, the components of the momentum\footnote{``Momentum'' will
always refer to the canonical momentum, i.e. the momentum
one-form on the spacetime, unless otherwise noted.} are
\begin{eqnarray}
\label{momenta}
  p_t &=& -\left(1-\frac{2r}{\Sigma}\right ) \dot t
  -\frac{2ar\sin^2\theta}{\Sigma}\dot \varphi \label{e} \\
  p_r &=& \frac{\Sigma}{\Delta} \dot r \label{pr}\\
  p_\theta &=& \Sigma \dot{\theta} \label{ptheta}\\
  p_\varphi &=& \sin^2\theta\left
  (r^2+a^2+\frac{2a^2r\sin^2\theta}{\Sigma}\right ) \dot\varphi \\
  & & -\frac{2ar\sin^2\theta}{\Sigma}\dot t \quad .\label{pphi}
\end{eqnarray}

We finally define the Hamiltonian in the standard way,
\begin{equation}
H = p_\mu \dot{q}^\mu - {\cal L}
= \frac{1}{2}g^{\alpha \beta} p_\alpha p_\beta \quad\quad .
\label{ham}
\end{equation}
Note that the Hamiltonian is numerically identical to our Lagrangian
($- 1/2$) along any orbit because each quantity is just half the
contraction of the 4-momentum.  This is not surprising since, for
geodesic motion, the Lagrangian and Hamiltonian contain only
kinetic terms.

To get an explicit expression for the Hamiltonian in terms of our
dimensionless coordinates and momenta, we could compute the inverse
metric $g^{\alpha \beta}$ and grind through eqn.\ (\ref{ham}) by brute
force. However, a simple observation yields a nice expression for
$H(q,p)$ with much less effort. Using eqns.\ (\ref{pr})-(\ref{ptheta})
in eqn.\ (\ref{lagrangian}), we know the Hamiltonian must have the
form
\begin{equation}
H(\vctr{q}, \vctr{p}) =
\frac{\Delta}{2\Sigma}p_r^2+\frac{1}{2\Sigma}p_\theta^2+f(r,\theta,p_t,p_\varphi)
\quad\quad ,
\end{equation}
where $f$ is a function yet to be determined that depends only on
$p_t$ and $p_\varphi$ and, via the $g^{\alpha \beta}$, on $r$ and
$\theta$.  Eliminating the $p_r$ and $p_\theta$ above with the use of
(\ref{momenta}) and (\ref{dimcarter}) and noting that $H$ must
identically equal $-1/2$, we see that
\begin{equation}
H(\vctr{q}, \vctr{p}) =
\frac{R}{2\Delta\Sigma} + \frac{\Theta}{2\Sigma} + f(r,\theta,p_t,p_\varphi)=
-1/2 
\quad\quad .
\end{equation}
This fixes $f$ and allows us finally to write $H$ as
\begin{equation}
H(\vctr{q}, \vctr{p}) =
\frac{\Delta}{2\Sigma}p_r^2+\frac{1}{2\Sigma}p_\theta^2
-\frac{R + \Delta\Theta}{2\Delta \Sigma} -\frac{1}{2}
\quad\quad ,
\label{niceham}
\end{equation}
where $R$ and $\Theta$ are the functions listed in eqns.\
(\ref{dimpots}) with every $E$ or $L$ therein interpreted as a $-p_t$ or
$p_\varphi$, respectively, and with every $Q$ treated as a constant.

With the Hamiltonian written as in (\ref{niceham}), Hamilton's
equations
\begin{equation}
{\dot q_i} =\frac{\partial H}{\partial p_i} \quad , \quad
{\dot p_i}=-\frac{\partial H}{\partial q_i}
\end{equation}
for the test particle motion become, explicitly,
\begin{eqnarray}
\label{eom}
\dot{r} & =& \frac{\Delta}{\Sigma}p_{r}  \\
 \dot{p}_{r} & = &
-\left (\frac{\Delta}{2\Sigma}\right )'p_{r}^{2} -
\left (\frac{1}{2\Sigma}\right )'p_{\theta}^{2} + \left (\frac{R +
 \Delta\Theta}{2\Delta\Sigma}\right )' \nonumber \\
\dot{\theta}& = & \frac{1}{\Sigma}p_{\theta} \nonumber
 \\
 \dot{p}_{\theta} & = &
-\left (\frac{\Delta}{2\Sigma}\right )^{\theta}p_{r}^{2} -
 \left (\frac{1}{2\Sigma}\right )^{\theta}p_{\theta}^{2} + \left (\frac{R +
 \Delta\Theta}{2\Delta\Sigma}\right )^{\theta}  \nonumber \\
  \dot{t} & = & \frac{1}{2\Delta\Sigma} \pd{E}{} \left(R +
  \Delta\Theta \right) \\ 
  \dot{p_t} & = & 0 \nonumber \\
\dot{\varphi} & = & -\frac{1}{2\Delta\Sigma}
  \pd{L}{} \left(R + \Delta\Theta \right) \nonumber \\
  \dot{p_\varphi} & = & 0 \, ,
\end{eqnarray}
where the superscripts $'$ and $\theta$ denote differentiation with
respect to $r$ and $\theta$, respectively.  Note that since none of
the equations contain square roots any more, they constitute a
smoothly differentiable system of ODEs, even at turning points, and
they can be integrated directly without resorting to an intricate
change of variable.  Since the $\dot{p_t}$ and $\dot{p_\varphi}$
equations vanish, the only added cost of retaining analytical
transparency in the equations is that we must integrate 6 equations in
(\ref{eom}) instead of 4 in (\ref{dimcarter}). These are the quations
we integrate numerically to identify the periodic orbits. 

We isolate the periodic orbits by inputing $(z,w,v)$ for a given $a$
and $L$ to locate the $E$, apastron $r_a$ and perihelion $r_p$ of the corresponding
periodic orbit.
Numerically the periodics are
extracted by explicitly integrating
\begin{equation}
\Delta \varphi_r=2\int_{t(r_p)}^{t(r_a)}\frac{d \varphi}{dt} dt = 2\int_{r_p}^{r_a}
\frac{d\varphi}{dr} dr \quad\quad 
\label{represent}
\end{equation}
to find those orbits for which the integral is a rational multiple of $2\pi$.

As a final check of our numerical results, we run an
independent code that computes the orbits directly from the geodesic
equation and find the two independent computations
in complete accord.

\vfill\eject
\section{The Action-Angle Picture}
\label{actionangle}

In addition to deriving numerically supple equations of motion, we can
use Hamiltonian dynamics to relate the periodic orbits to rationally
related canonical frequencies. 
We now relate those frequencies to the taxonomy we have developed.

It is a classic result of Hamiltonian dynamics that motion in a system with $N$
coordinates and $N$ conserved momenta will be confined to
$N$-dimensional tori. 
In this case we do have 4 coordinates
$(t,r,\theta,\varphi)$ and 4 conserved quantitites
$(\mu,E,L,Q)$. However, this classic result only
applies to orbits bound in phase space. 
In our relativistic system, the
$t$ direction throws a wrench in the works since every orbit
increases without bound in the timelike direction
-- worldlines, in both phase space and
configuration space, are not compact.  
To circumvent this problem, we work in a formally \emph{reduced phase
  space}.\footnote{A more detailed exposition of which we leave to the
  nonequatorial 
  case.}  
Any canonical coordinate can be chosen as the time parameter and its
canonical momentum will become the Hamiltonian for the reduced phase
space \cite{textbooks}. In our case, we choose to parameterize the
orbits by $t$ and use $p_t=E$ as the Hamiltonian of the reduced phase space.

In this paper we are only dealing with equatorial motion, for which
motion lies in the 4D hypersurface defined by $\theta \equiv \pi/2,
p_\theta \equiv 0$ and on which $Q \equiv 0$.  On that hypersurface
$r, \varphi, p_r$ and $p_\varphi$ still form a canonical set, so the
hypersurface corresponding to equatorial orbits constitutes a
\textit{bona fide} 4D phase space.  

On this space, all phase space trajectories corresponding to orbits
that are bound and non-plunging in configuration space lie on compact
hypersurfaces topologically equivalent to 2D tori.  Thus, we can make
the transformation to action-angle variables for those orbits, to
which we now turn.

A canonical transformation to action variables allows
us to set each canonical
momentum $J_i$ equal to a function of the constants of motion. 
The reduced Hamiltonian can be rewritten as a function of the $\vctr{J}$ only, 
\begin{equation}
\label{HAA}
  E = E(\vctr{J}) \quad \quad ,
\end{equation}
so that it is cyclic in all the $\psi_i$, and the $(\psi_i,
J_i)$ form 
a canonical set whose evolution is still determined by Hamilton's
equations
\begin{eqnarray}
\label{eomAA}
\frac{d J_i }{dt}&=& -\frac{\partial E}{\partial \psi_i}\nonumber \\
\frac{d \psi_i}{dt} &=& \frac{\partial E}{\partial J_i} \quad ,
\end{eqnarray}
($i=r,\varphi$).
Together, (\ref{HAA}) and (\ref{eomAA}) imply that each $J_i$ is a
constant of the motion and that each $\psi_i$ is linear in time,
\begin{equation}
  \psi_i = \omega_i t + \psi_i(0) \quad ,
\end{equation}
where each
\begin{equation}
\label{omegadefs}
   \omega_i(\vctr{J}) \equiv \dot{\psi_i} =
  \pd{J_i}{E(\vctr{J})}
\end{equation}
is a function only of the $J_i$ and thus also a constant of the
motion.  These constants are the orbital frequencies used
as a basis in which to Fourier decompose functions of generic Kerr orbits,
including the instantaneous adiabatic gravitational waveforms they
emit \cite{{schmidt},{drasco}}.

We define the radial and azimuthal actions in the standard way,
as
\begin{eqnarray}
\label{Jdefs}
J_r  &=& \frac{1}{2\pi}\oint_{C_r}p_r dr \\
J_\varphi  &=& \frac{1}{2\pi}\oint_{C_\varphi}p_\varphi d\varphi 
\quad \quad,
\end{eqnarray}
where the curves $C_r$ and $C_\varphi$ are projections of the orbit into
the $r-p_r$ and $\varphi-p_\varphi$ planes, respectively.  
Because $p_\varphi$ is
a constant we see that
\begin{equation}
  J_\varphi \equiv p_\varphi = L \quad .
\end{equation}
We are more interested in the frequencies $\omega_r$ and
$\omega_\varphi$.  An explicit expression for the Hamiltonian in terms of
the $J$'s requires inverting the $J_r$ integral in (\ref{Jdefs}) to
solve for $E(J_r, L) = E(J_r, J_\varphi)$ and thus is not
analytically accessible.  Still, expressions for the $\omega_i$ in
terms of $E, L$ rather than in terms of $J_r, J_\varphi$ \emph{can} be
computed as follows.

Since a trajectory can alternately be specified by constants of the
motion $(E, L)$ or $(J_r, J_\varphi)$, there exists some transformation
from the $(E,L)$ set to the $J_r, J_\varphi$ set.
Since
\begin{equation}
  J_\varphi(E, L) \equiv L \quad ,
\end{equation}
the Jacobian of the transformation is
\begin{equation}
\left|\begin{array}{cc}
  \pd{E}{J_r} & \pd{L}{J_r} \\
  \pd{E}{J_\varphi} & \pd{L}{J_\varphi}
\end{array}\right| = 
\left|\begin{array}{cc}
  \pd{E}{J_r} & \pd{L}{J_r} \\
  0 & 1
\end{array}\right| = 
\pd{E}{J_r} \quad ,
\end{equation}
which, since $J_r$ increases monotonically with $E$ at a given $L$, 
never vanishes.  There thus exists an inverse transformation $E(J_r,
J_\varphi), L(J_r, J_\varphi)$ from the actions back to the usual orbital
constants.  The product of those two transformations should be
identity:
\begin{equation}
\left(\begin{array}{cc}
  \pd{E}{J_r} & \pd{L}{J_r} \\
  \pd{E}{J_\varphi} & \pd{L}{J_\varphi}
\end{array}\right) 
\left(\begin{array}{cc}
  \pd{J_r}{E} & \pd{J_\varphi}{E} \\
  \pd{J_r}{L} & \pd{J_\varphi}{L}
\end{array}\right) = 
\left(\begin{array}{cc}
  \pd{E}{J_r} & \pd{L}{J_r} \\
  0 & 1
\end{array}\right) 
\left(\begin{array}{cc}
  \pd{J_r}{E} & \pd{J_\varphi}{E} \\
  0 & 1
\end{array}\right) = 
\left(\begin{array}{cc}
  1 & 0 \\
  0 & 1
\end{array}\right) \quad ,
\end{equation}
which is equivalent to the two independent equations
\begin{eqnarray}
  \pd{E}{J_r}\pd{J_r}{E} & = & 1 \\
  \pd{E}{J_r}\pd{J_\varphi}{E} & = & -\pd{L}{J_r} \quad .
\end{eqnarray}
But in our reduced phase space, $E$ \emph{is} the reduced Hamiltonian, so
each $\txtpd{J_i}{E} \equiv \omega_i$.  The equations above then
yield
\begin{eqnarray}
  \omega_r & = & \frac{1}{\txtpd{E}{J_r}} \nonumber \\
  \omega_\varphi & = & -\frac{\txtpd{L}{J_r}}{\txtpd{E}{J_r}} \quad .
\end{eqnarray}
Combining equations (\ref{momenta}), (\ref{dimpots}) and
(\ref{Jdefs}), we see that
\begin{equation}
  \oint_{C_r}p_r dr =  2 \int_{r_p}^{r_a} dr \frac{\sqrt{R}}{\Delta}
\end{equation}
and, combining this with the equations of motion, that
\begin{eqnarray}
  -\pd{L}{J_r} & = & -\frac{1}{2\pi} 2 \frac{\partial}{\partial L}\int_{r_p}^{r_a} dr
  \frac{\sqrt{R}}{\Delta} \nonumber \\
&=& \frac{1}{2\pi} 2 \int_{r_p}^{r_a} dr
  \frac{1}{2\sqrt{R}\Delta}\left (-\pd{L}{R}\right ) \nonumber 
\end{eqnarray}
where the term proportional to the partial of the limits vanishes
  since the integrand vanishes at the limits. Continuing,
\begin{eqnarray}
  -\pd{L}{J_r}  & = & \frac{1}{2\pi} 2 \int_{r_p}^{r_a} dr \frac{\Sigma}{\sqrt{R}}
  \frac{-\txtpd{L}{R}}{2\Delta\Sigma} \nonumber \\
  &=& \frac{1}{2\pi} 2 \int_{r_p}^{r_a} dr \frac{\dot{\varphi}}{\dot{r}}
\nonumber \\
  &=& \frac{1}{2\pi} \Delta\varphi_r \quad ,
\end{eqnarray}
where the last equality comes from equation (\ref{dphir}).
Additionally,
\begin{eqnarray}
  \pd{E}{J_r} & = & \frac{1}{2\pi} 2 \pd{E}{}\int_{r_p}^{r_a} dr
  \frac{\sqrt{R}}{\Delta} \nonumber \\
&=& \frac{1}{2\pi} 2 \int_{r_p}^{r_a} dr
  \frac{1}{2\sqrt{R}\Delta}\left (\pd{E}{R}\right ) \nonumber \\
  & = & \frac{1}{2\pi} 2 \int_{r_p}^{r_a} dr \frac{\Sigma}{\sqrt{R}}
  \frac{\txtpd{E}{R}}{2\Delta\Sigma} \nonumber \\
  &=& \frac{1}{2\pi} 2 \int_{r_p}^{r_a} dr \frac{\dot{t}}{\dot{r}}
\nonumber \\
  &=& \frac{1}{2\pi} T_r \quad ,
\end{eqnarray}
where $T_r$ is the radial period of an orbit.

The physical interpretation of the $\omega$'s is now clear.
As expected, $\omega_r$ is
\begin{equation}
  \omega_r = \frac{2\pi}{\int_{r_p}^{r_a} dr \frac{1}{\Delta\sqrt{R}}}
  \pd{E}{R} = \frac{2\pi}{T_r} \quad ,
\label{omegar}
\end{equation}
where $T_r$ as measured in coordinate time $t$. The rate
at which azimuth accumulates averaged over the radial period is $\omega_\varphi$:
\begin{equation}
  \omega_\varphi = \frac{-\int_{r_p}^{r_a} dr \frac{1}{\Delta\sqrt{R}} \pd{L}{R}}
	{\int_{r_p}^{r_a} dr \frac{1}{\Delta\sqrt{R}} \pd{E}{R}} =
	\frac{\Delta\varphi_r}{T_r} \quad .
\label{omegaphi}
\end{equation}
Because for a periodic orbit, the total orbital period is an integer
multiple of $T_r$ (in fact, $z T_r$), we can equivalently say that
$\omega_\varphi$ is the rate of increase of $\varphi$ averaged over the
entire orbital period (this latter interpretation is what will
generalize when we leave the equatorial plane).

Finally, to make the connection with the \S \ref{find}, note that the
frequency ratio 
\begin{equation}
  \frac{\omega_\varphi}{\omega_r} = \frac{\Delta\varphi_r}{2\pi} =
  1 + w + \frac{v}{z} =1+q\quad .
\end{equation}
Thus, in the action-angle picture, the periodic orbits are those with
rationally related fundamental frequencies.

A typical orbit will fill out the
torus because the canonical angular frequencies are not
commensurate. In contrast, closed orbits do not.
It is precisely this
relation we have exploited to explicitly locate the periodic orbits in
our taxonomy.

\vfill\eject 
\section{Circular Orbits Revisited}
\label{circs}

As discussed briefly in \S \ref{circs1}, circular orbits,
while clearly periodic, are somewhat anomalous in 
the geometric picture and our $(z, w, v)$ scheme.
After all, our taxonomy hinges on quantities calculated per radial
period.  Since the radial period of a circular orbit is zero, it is
not clear what rational number to associate with circular orbits.
Nevertheless, it turns our that in a useful and enlightening sense,
some stable circular orbits (a measure zero set of them, to be precise)
\emph{are} mappable to the rational numbers.  
Naively, we might expect them to correspond to the rational number 1,
just like Keplerian circular orbits, but this turns out to be a mistake.

The action-angle picture is significantly more informative here.  Our
difficulty stems from the fact that all circular orbits have a
vanishing radial action $J_r$.  In a loose sense, the actions
correspond to the two circumferences that characterize a
two-dimensional torus.  When one of those circumferences is zero,
the 2-torus $T^{2}$ just becomes a 1-torus $T^{1} \equiv S^{1}$: a
circle.  Orbits that are circular in configuration space thus live on
surfaces in the phase space that also have the topology of
circles.  In that sense, circular orbits only really have one
fundamental frequency, $\omega_\varphi$, and there is no rational
frequency ratio to speak of.  They are all nonetheless periodic, for
any value of $\omega_\varphi$, because any curve that fills out
$S^{1}$ with a constant velocity necessarily closes on itself and
becomes periodic.
\footnote{Note that for spherical orbits, i.e. constant $r$ orbits not
confined to the equatorial plane, which other authors usually refer to
as ``non-equatorial circular orbits'', this will no longer be the
case.  Non-equatorial orbits in general have 3 associated actions
$J_r, J_\theta$ and $J_\varphi$ and are confined to surfaces with the
topology of $T^{3}$.  On spherical orbits, $J_r$ vanishes, but the
remaining actions do not, leaving spherical orbits to occupy surfaces
with the topology of $T^2$.  Much like generic equatorial orbits,
then, spherical orbits will only be exactly periodic only when the
ratio of their frequencies $\omega_\theta/\omega_\varphi$ is rational.}

Nonetheless
some stable circular orbits
can be mapped to the rational numbers.  We can construct that
map in two equivalent ways.  First, for a given $L$, we can take the
zero eccentricity limit of the values of $\omega_r, \omega_\varphi,$ and
their ratio and define the corresponding values for the
circular orbit with that same $L$ to be those limits. This was the attitude
taken in \S \ref{circs1}. 
 
Alternately, we
can perform a linear stability analysis 
of the circular orbits. We find the frequencies of small
oscillations of the $r$ and $\varphi$ motions of low eccentricity orbits
around the $r$ and $\varphi$ motions of the reference circular orbits
and examine the ratios of those frequencies\footnote{Often
called epicycle frequencies, or just
epicycles for short}.

\begin{figure}
  \vspace{-20pt}
  \centering
\includegraphics[width=80mm]{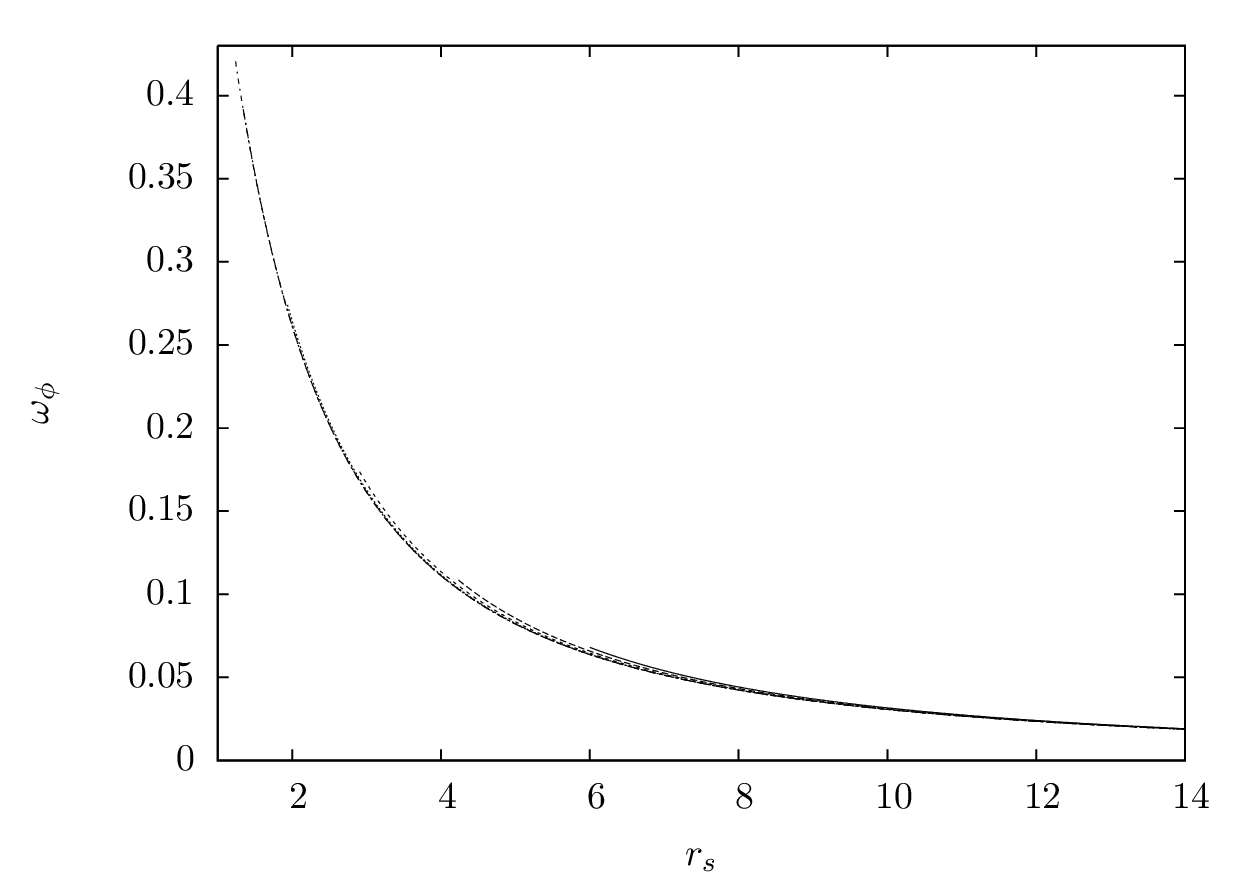}
\hspace{-10pt}
\includegraphics[width=80mm]{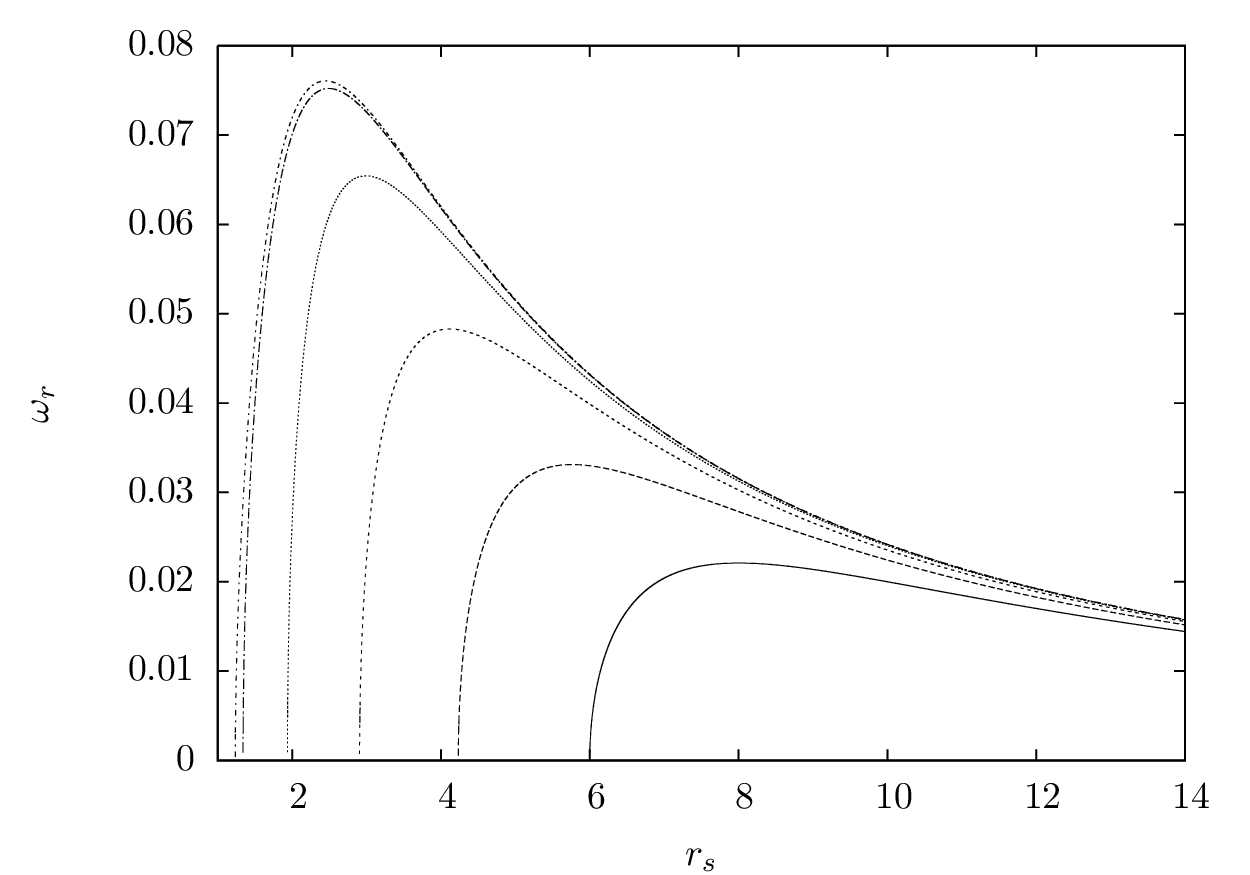}
\hfill
  \caption{Left: $\omega_\varphi$ versus the radii of stable circular
  orbits. Right: $\omega_r$ versus the radii of stable circular
  orbits. Increasing from right to left the spin values are $a=0, 0.5,
  0.8, 0.95, 0.995, 0.998$. All orbits are prograde.} 
  \label{omegafig}
\end{figure}

\begin{figure}
  \vspace{-20pt}
  \centering
  \includegraphics[width=0.5\textwidth]{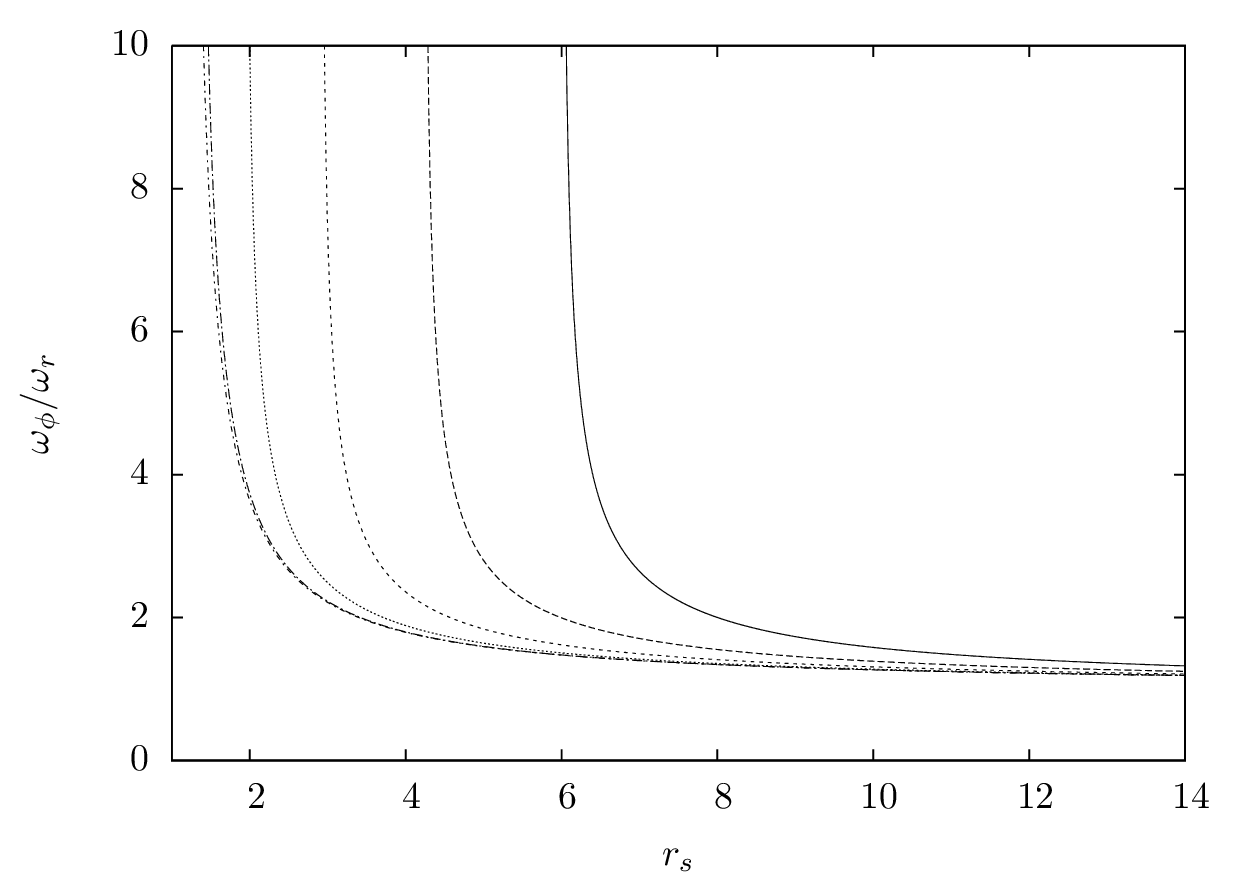}
  \caption{The ratio $\omega_\varphi/\omega_r$, which equals $1+w+v/z$, as
    a function of the radii of stable circular orbits.
Increasing from right to left the spin values are $a=0, 0.5,
  0.8, 0.95, 0.995, 0.998$. All orbits are prograde.} 
  \label{omegaphitor}
\end{figure}

Whether derived from a stability analysis or as the limiting values of
low eccentricity orbits, 
figure \ref{omegafig} shows the
resulting values of $\omega_\varphi$ and 
$\omega_r$ while figure \ref{omegaphitor} shows $\omega_\varphi/\omega_r$
for stable circular orbits as a 
function of their radial coordinate $r$ for various values of
$a$. 
Some features are worth noting.  First, $\omega_\varphi$ for a circular
orbit is just the coordinate angular velocity and thus satisfies
\begin{equation}
  \omega_\varphi = \D{t}{\varphi} = \pm\frac{1}{r^{3/2} \pm a} \quad ,
\end{equation}
the well-known relativistic generalization of Kepler's third law 
\cite{BPT}.
Second, the $\omega_r$'s
differ from the Keplerian in that the increase with decreasing $r$
until they hit some maximum (in the Keplerian case, the $\omega_r$
continue to increase and diverge at $r = 0$), at which point they
decrease, reaching zero at the ISCO.  The ISCO, then, will have a
diverging $\omega_\varphi/\omega_r$ ratio, an observation that was
relevant in our discussion of homoclinic orbits. All homoclinic orbits
have $q=w+v/z=\omega_\varphi/\omega_r=\infty$, including the ISCO is (the
eccentricity zero homoclinic orbit).

Third, $\omega_\varphi/\omega_r$ for circular orbits is a continuous and
monotonic function of $r$, increasing from an asymptotic value of $1$
in the $r\to\infty$ limit and diverging as $r \to r_{ISCO}$.  The
continuity of this ratio means that although a measure zero subset of
the circular orbits corresponds to rational numbers or, equivalently,
to $(z,w,v)$ triplets, most do not.  Nevertheless, even the irrational
circulars are arbitrarily close to some rational.  We can thus
characterize every circular orbit either exactly (the rational
circulars) or approximately (the irrational circulars) by a $(z, w,
v)$ triplet.

Finally,
dissipative dynamical systems with
multiple coupled frequencies, can sometimes attractors 
(or at
least transient attractors).
Those attractors typically correspond to rationals $q$
with low denominator fractional parts (low $z$).
The $1:1$ and $3:2$ spin-orbit
frequency ratios of the moon-earth and Mercury-sun systems,
respectively, are solar system examples of where this sort of behavior
takes place.  The possibility that an orbit decaying adiabatically
under radiation reaction might become trapped in a resonance en route
to plunge is thus open.  At the time of submission of this article, we
are still 
investigating that possibility.

\end{document}